\documentclass[journal]{IEEEtran}
\usepackage{epsf,exscale,times}
\usepackage{graphicx}
\usepackage{amssymb}
\usepackage{mathtools}
\usepackage{subcaption}
\usepackage{multirow}
\usepackage{multicol}
\usepackage{glossaries}
\usepackage{cite}
\usepackage{algorithm}
\usepackage{nccmath}
\usepackage{commath}
\usepackage{comment}
\usepackage{color}

\newacronym{FNR}{FNR}{Luxembourg National Research Fund}
\newacronym{SNR}{SNR}{Signal to Noise Ratio}
\newacronym{INR}{INR}{Interfernece to Noise Ratio}
\newacronym{SINR}{SINR}{Signal to Interference plus Noise Ratio}
\newacronym{AF}{AF}{Ambiguity Function}
\newacronym{MIMO}{MIMO}{Multiple-Input Multiple-Output}
\newacronym{SISO}{SISO}{Single Input Single Output}
\newacronym{CD}{CD}{Coordinate Descent}
\newacronym{BCD}{BCD}{Block Coordinate Descent}
\newacronym{GD}{GD}{Gradient Descent}
\newacronym{MM}{MM}{Majorization-Minimization}
\newacronym{FMCW}{FMCW}{Frequency Modulated Continuous Wave}
\newacronym{DFT}{DFT}{Discrete Fourier Transform}
\newacronym{FFT}{FFT}{Fast Fourier Transform}
\newacronym{MVDR}{MVDR}{Minimum Variance Distortionless Response}
\newacronym{MBI}{MBI}{Maximum Block Improvement}
\newacronym{RFPA}{RFPA}{Radio Frequency Power Amplifier}
\newacronym{BPSK}{BPSK}{Binary Phase Shift Keying}
\newacronym{QPSK}{QPSK}{Quadrature Phase Shift Keying}
\newacronym{ULA}{ULA}{Uniform Linear Array}
\newacronym{DOF}{DOF}{Degrees of Freedom}
\newacronym{PSK}{PSK}{Phase Shift Keying}
\newacronym{PSL}{PSL}{Peak Sidelobe Level}
\newacronym{PSLR}{PSLR}{Peak Sidelobe Level Ratio}
\newacronym{ISL}{ISL}{Integrated Sidelobe Level}
\newacronym{ISLR}{ISLR}{Integrated Sidelobe Level Ratio}
\newacronym{LFM}{LFM}{Linear Frequency Modulation}
\newacronym{CPI}{CPI}{Coherent Pulse Interval}
\newacronym{RCS}{RCS}{Radar Cross Section}
\newacronym{CNR}{CNR}{Clutter to Noise Ratio}
\newacronym{MTI}{MTI}{Moving Target Indicator}
\newacronym{ROC}{ROC}{Receiver Operating Characteristic}
\newacronym{MPSK}{MPSK}{$M$-ary Phase Shift Keying}
\newacronym{PAR}{PAR}{Peak-to-Average Ratio}
\newacronym{GFP}{GFP}{Generalized Fractional Programming}
\newacronym{PRI}{PRI}{Pulse Repetition Interval}
\newacronym{PRF}{PRF}{Pulse Repetition Frequency}
\newacronym{MMM}{MM}{Majorization Minimization or Minorization Maximization}
\newacronym{QCQP}{QCQP}{Quadratic Constraint Quadratic Programming}
\newacronym{SDP}{SDP}{Semi-definite Programming}
\newacronym{CCL}{CCL}{Cross-Correlation Level}
\newacronym{TDMA}{TDMA}{Time-Division Multiple Access}
\newacronym{FDMA}{FDMA}{Frequency-Division Multiple Access}
\newacronym{CDMA}{CDMA}{Code-Division Multiple Access}
\newacronym{DDMA}{DDMA}{Doppler-Division Multiple Access}
\newacronym{TDM}{TDM}{Time Division Multiplexing}
\newacronym{FDM}{FDM}{Frequency Division Multiplexing}
\newacronym{CDM}{CDM}{Code Division Multiplexing}
\newacronym{DDM}{DDM}{Doppler Division Multiplexing}
\newacronym{SDR}{SDR}{Semi-definite Relaxation}
\newacronym{QSDR}{QSDR}{Quantized Semi-definite Relaxation}
\newacronym{CA}{CA}{Cyclic Algorithm}
\newacronym{ADMM}{ADMM}{Alternating Direction Method of Multipliers}
\newacronym{PDR}{PDR}{Projection, Descent, and Retraction}
\newacronym{SQP}{SQP}{Semidefinite Quadratic Programming}
\newacronym{CM}{CM}{Constant Modulus}
\newacronym{MPS}{MPS}{Minimum Peak Sidelobe}
\newacronym{BiST}{BiST}{Binary Sequences seTs}
\newacronym{ESA}{ESA}{Effective Simulated Annealing}
\newacronym{CAN}{CAN}{Cyclic Algorithm New}
\newacronym{SRR}{SRR}{Short-Range Radar}
\newacronym{MRR}{MRR}{Mid-Range Radar}
\newacronym{LRR}{LRR}{Long-Range Radar}
\newacronym{EDDB}{EDDB}{Environmental Dynamic Database}
\newacronym{DoA}{DoA}{Direction of Arrival}
\newacronym{STTC}{STTC}{Space-Time Transmitting Code}
\newacronym{MIA}{MIA}{Majorized Iterative Algorithm}
\newacronym{MIACMC}{MIA-CMC}{Majorized Iterative Algorithm - Constant Modulus Constraint}
\newacronym{MIAPC}{MIA-PC}{Majorized Iterative Algorithm - PAR Constraint}
\newacronym{NSGA}{NSGA}{Non-Dominated Sorting Genetic Algorithm}

\newcommand{\bA}{\mbox{\boldmath{$A$}}}
\newcommand{\ba}{\mbox{\boldmath{$a$}}}

\newcommand{\bJ}{\mbox{\boldmath{$J$}}}

\newcommand{\bS}{\mbox{\boldmath{$S$}}}
\newcommand{\bs}{\mbox{\boldmath{$s$}}}

\newcommand{\bw}{\mbox{\boldmath{$w$}}}

\newcommand{\RN}[1]{%
  \textup{\uppercase\expandafter{\romannumeral#1}}%
}
\newcommand{\Rn}[1]{%
  \textup{\lowercase\expandafter{\romannumeral#1}}%
}
\newtheorem{remark}{Remark}

\graphicspath{ {figures/} }

\begin{document}

\title{Spatial- and Range- ISLR Trade-off in MIMO Radar via Waveform Correlation Optimization}

\author{Ehsan~Raei,~\IEEEmembership{Student Member,~IEEE,}
        Mohammad~Alaee-Kerahroodi,~\IEEEmembership{Member,~IEEE,}
        and~M.R.~Bhavani~Shankar,~\IEEEmembership{Senior~Member,~IEEE}% <-this % stops a space
% \thanks{This work was supported by \gls{FNR} through the BRIDGES project AWARDS under Grant CPPP17/IS/11827256/AWARDS and CORE project SPRINGER}% <-this % stops a space
% \thanks{J. Doe and J. Doe are with Anonymous University.}% <-this % stops a space
% \thanks{Manuscript received April 19, 2005; revised August 26, 2015.}
}

% \markboth{IEEE TRANSACTIONS ON SIGNAL PROCESSING,~Vol.~~, No.~~, July~~}%
% {Shell \MakeLowercase{\textit{et al.}}: Bare Demo of IEEEtran.cls for IEEE Journals}

\maketitle

\begin{abstract}
This paper aims to design a set of transmit waveforms in cognitive colocated Multi-Input Multi-Output (MIMO) radar systems considering the simultaneous minimization of spatial- and the range- Integrated Sidelobe Level Ratio (ISLR). The design problem is formulated as a bi-objective Pareto optimization under practical constraints on the waveforms, namely total transmit power, peak-to-average-power ratio (PAR), constant modulus, and discrete phase alphabet. A Coordinate Descent (CD) based approach is proposed, in which at every single variable update of the algorithm we obtain the solution of the uni-variable optimization problems. The novelty of the paper comes from deriving a flexible waveform design problem applicable for 4D imaging MIMO radars which is optimized directly over the different constraint sets. The simultaneous optimization leads to a trade-off between the two ISLRs and the simulation results illustrate significantly improved trade-off offered by the proposed methodologies.
%compared to literature.
%
%iteration of the devised method requires the solution of a non-convex NP-hard problem. It is handled either through a novel direct solution based on gradient under the power and PAR constraints, derivative under constant modulus constraint, or an FFT-based method for the discrete phase constraint. Simulation results illustrate that the proposed methodologies can outperform some counterparts providing sequences that trade-offs between spatial- and range-ISLRs. 
\end{abstract}


\IEEEpeerreviewmaketitle

\section{Introduction}
%Waveform design is a key factor in \gls{MIMO} radar systems that enhances the spatial resolution, detection, localization, and classification performance \cite{LImimo}.  
Transmit beampattern shaping and orthogonality have been the key waveform design aspects influencing the performance of colocated \gls{MIMO} radar systems \cite{LImimo}. Beampattern shaping involves steering the radiation power in a spatial region of desired angles while reducing interference from sidelobe returns to improve target detection \cite{skolnik2008radar}. There exists a rich literature on waveform design for beampattern shaping following different approaches with regards to the choice of the variables, the objective function and the constraints; kindly refer to \cite{7435338, 7126203, 8387476, 7955071, 8713914, 4276989} for details. An interesting approach to enhance detection of weak targets in the vicinity of strong ones is the design of waveforms with a small \gls{ISLR} \cite{7435338, 7126203} in the beam domain or spatial-\gls{ISLR}. This can be achieved by {\em imparting appropriate correlation among the waveforms} transmitted from the different antennas  \cite{4516997}.  Waveform orthogonality, on the other hand, aims to enhance spatial resolution through the concept of virtual array. Similar to the beampattern design, there is a rich literature on orthogonal waveform design; kindly refer to \cite{7060251, 5960679, 6178055, 6882135} for details. Waveforms with low \gls{ISLR} in time domain, also known as range-\gls{ISLR}, are typically sought \cite{8706639, he2012waveform}, to enable an effective virtual array. This is achieved by designing a {\em set of waveforms that are uncorrelated} with each other (within and across antennas).
% $-$ ideally  exhibiting  orthogonality in the spatial and temporal domains. 
%determines the characteristic of the transmit waveforms in the temporal domain, that requires to orthogonality in terms of auto- and cross-correlation functions.
%While orthogonality enables recovering the transmitted signals at the receive array but  precludes  transmit beamforming.  
%Note that since each element in the transmit array radiates independently, there is no transmit beamforming, so the transmit pattern is broad and covers a large field of view (FOV). 
Thus, a contradiction arises in  achieving small spatial- and range-\gls{ISLR} simultaneously, leading to a waveform design trade-off between spatial- and range-\gls{ISLR}. This trade-off necessitates a dedicated waveform design approach \cite{he2012waveform}, a subject pursued in this paper.
%
%\vspace*{-0.01in}
%\subsection{Waveform design achieving i}
%\label{ssec:SoA}

{\sl Spatial-\gls{ISLR} minimization:} In the spatial-\gls{ISLR} the approach is to maximize/ minimize the response of beampattern on desired/ undesired angles respectively. In \cite{7126203}, a waveform covariance design based on \gls{SDR} under a constraint on the $3$ dB main-beam is proposed to minimize the spatial-\gls{ISLR}. In \cite{7435338}, robust waveform covariance matrix designs through the worst case transmit beampattern optimization are considered to minimize the spatial-\gls{ISLR} and -\gls{PSLR}. Unlike the aforementioned methods, \cite{8387476} proposes a direct design of the waveform entries based on \gls{ADMM} to minimize the spatial-\gls{PSLR} under constant modulus constraint.  
% Some of papers does not deal with spatial-\gls{ISLR} directly. These papers aims to maximize the \gls{SINR} while considering the target and multiplier interference \gls{DoA} information. For example the authors \cite{8239836} propose \gls{MIA} based on \gls{MM} method for joint waveform and filter design under similarity, constant modulus and \gls{PAR} constraints. While \gls{STTC} \cite{8401959} is proposed based on \gls{CD} to solve the problem under similarity, uncertain steering matrices, continuous or discrete phase constraints. The authors propose a Dinkelbach based method and exhaustive search for continuous and discrete phase constraints respectively. However the both paper just consider one target in their problem formulations, therefore nor \gls{MIA} neither \gls{STTC} can steer the beampattern in several directions.
In \cite{8239836} \gls{MIA} approach was proposed based on \gls{MM} for joint waveform and filter design under similarity, constant modulus and \gls{PAR} constraints. In \cite{8401959} a \gls{CD} based method (\gls{STTC}) was proposed to design space-time codes  under similarity, uncertain steering matrices, continuous or discrete phase constraints. The authors propose a Dinkelbach based method and exhaustive search for continuous and discrete phase constraints, respectively. In \cite{8239836, 8401959}, steering the beampattern to several desired angles is not considered in the problem formulation.

{\em Range-\gls{ISLR} minimization:} Unlike aforementioned spatially correlated designs, set of waveforms  having low {\em cross-correlation} for all lags have been investigated in \cite{LImimo, he2012waveform, 4383615, 5072243, 4602540}  to exploit the virtual array in \gls{MIMO} radar systems. Further, low {\em auto-correlation sidelobes} is a requirement \cite{5072243,7967829, 8239862,7707413,8267026,8239836, S20132096}, to avoid masking of the weak targets by the range sidelobes of a strong target \cite{barker1953group, 68151}, and to mitigate the harmful effects of distributed clutter returns close to the target of interest \cite{4517015}. These two requirements naturally lead to the use of \gls{ISLR}/ \gls{PSLR} minimization as the metric which is pursued through several approaches including,
\gls{CAN}, \gls{MM}, \gls{ADMM} and \gls{CD}. The authors in \cite{5072243,he2012waveform} proposed the \gls{CAN} algorithm to optimize sequence with good \gls{ISL} using the alternating minimization technique. However, instead of directly solving the \gls{ISL} minimization, they solved its approximation. To solve the \gls{ISL} minimization problem the authors in \cite{7420715} proposed the MM-Corr algorithm and the authors in \cite{8239862} proposed the ISL-NEW algorithm, both using the majorization-minimization technique. The authors in \cite{7529179} used the \gls{ADMM} technique to solve an approximation of the ISL minimization problem. The authors in \cite{8706639} used the \gls{CD} technique, and not only solve the \gls{ISL} minimization, but also solved the \gls{PSL} minimization problem under discrete phase constraint. They have reported superior performance comparing with the state-of-the art by using the \gls{CD} approach. 
% {\bf \gls{CAN} approach}: For example Multi-CAN, Multi-PeCAN \cite{5072243,he2012waveform} were introduced for constant modulus aperiodic and periodic sequence designs and 1bCAN and 1bPeCAN \cite{8926388} are the binary versions of Multi-CAN and Multi-PeCAN respectively.

% {\bf \gls{MM} approach} Such as MM-Corr \cite{7420715} and ISLNew \cite{8239862} for constant modulus sequences.

% {\bf \gls{CD} approach} Iterative Direct Search \cite{cui2017constant} proposed an exhaustive search to minimize the range-\gls{ISLR} under continuous and discrete phase while \gls{BiST} \cite{8706639} proposed and iterative algorithm based on \gls{FFT} to minimize the range\gls{ISL}/\gls{PSL} under discrete phase (binary) constraint.

% including Multi-CAN/ Multi-PeCAN \cite{5072243,he2012waveform}, Iterative Direct Search \cite{cui2017constant}, ISLNew \cite{8239862}, MM-Corr \cite{7420715} and \gls{BiST} \cite{8706639}. 
%

{\em Simultaneous range and spatial-\gls{ISLR} designs:}
It is clearly evident that simultaneous minimization of range- and spatial-\gls{ISLR} would be essential to achieve high performance in both range and spatial domains while minimizing the interfering radiation  or clutter reflections. In addition, simultaneous minimization provides a new design perspective offering novel waveforms. In this context, there are a few works even on the general topic of waveform design considering simultaneous waveform orthogonality and beampattern shaping. The same holds for the case of \gls{ISLR} minimization. The authors in \cite{4516997} bring out the contradictory nature of the two \gls{ISLR} designs and propose a method for beampattern matching under particular constraints on the waveform cross-correlation matrix. In \cite{7511868}, the authors present an algorithm which, at first, minimizes the difference between desired and designed beampattern responses for one sub-pulse. Subsequently, other sub-pulses are obtained through random permutation. The waveforms obtained exhibit quasi-Dirac auto-correlation and the different waveforms are quasi-orthogonal. Since the spatial-\gls{ISLR} is the ratio of beampattern response on undesired and desired angles, the approach in \cite{7511868} is not equivalent to minimizing the spatial-\gls{ISLR}. In \cite{8645373}, the authors introduce a beampattern matching by including orthogonality requirement as a penalty in the objective function. and using the \gls{PDR} approach for the solution. In \cite{8835646}, the authors propose a method based on \gls{ADMM} to design a beampattern with good cross-correlation property but they do not consider the need for a good auto-correlation in their formulation. The aforementioned papers design constant modulus waveforms with continuous phase alphabets. However, they do not consider simultaneous minimization of range- and spatial-\gls{ISLR} metrics in designing the waveform set; nor do they consider discrete-phase designs.

Another approach considering both orthogonality and beampattern shaping is the phased-\gls{MIMO} technique where the transmit array is divided into a number of sub-arrays and each sub-array coherently transmits a waveform which is orthogonal to those transmitted by the other sub-arrays. For instance, \cite{5419124} considers designing a weight vector for each sub-array to form a beam in a desired direction. In order to obtain the orthogonality, \cite{5419124} allocates non-overlapping bandwidth to each sub-array, where the bandwidth is greater than the \gls{PRF} of the system (similar to  \gls{DDMA} technique). In this case, the radar system may occupy large bandwidth leading to inefficient spectrum allocation. On the other hand, the authors in \cite{5393290} first, generate correlated waveform to achieve arbitrary beampattern subsequently the matrix waveform is  permuted to achieve a pseudo noise like quasi-orthogonal waveform. However, phased-\gls{MIMO} radars tend to be effective for large antenna systems and may not be suitable for applications with few transmit antennas.
\subsection{Contributions}
\begin{figure}
\centering
    \begin{subfigure}{0.49\textwidth}
    \centering
        \includegraphics[width=0.95\linewidth]{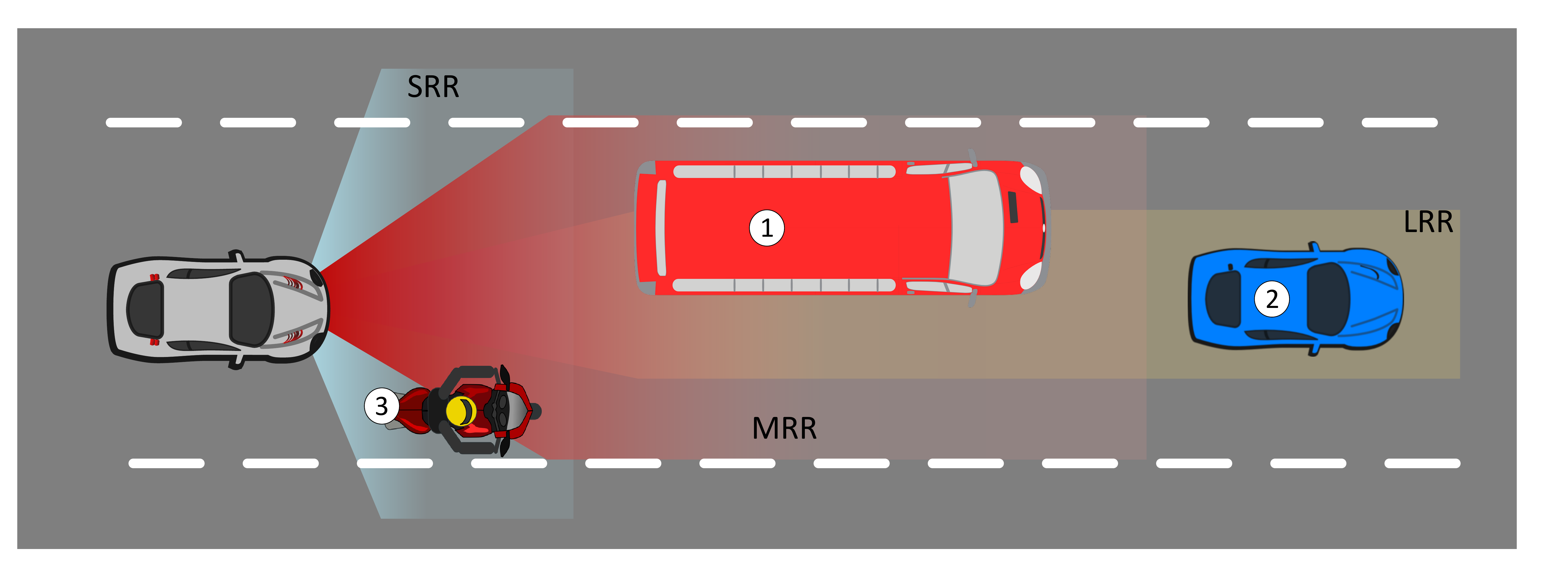}
    	\caption[]{Current automotive \gls{MIMO} radar systems.}    
    \end{subfigure}
    \begin{subfigure}{0.49\textwidth}
    \centering
        \includegraphics[width=0.95\linewidth]{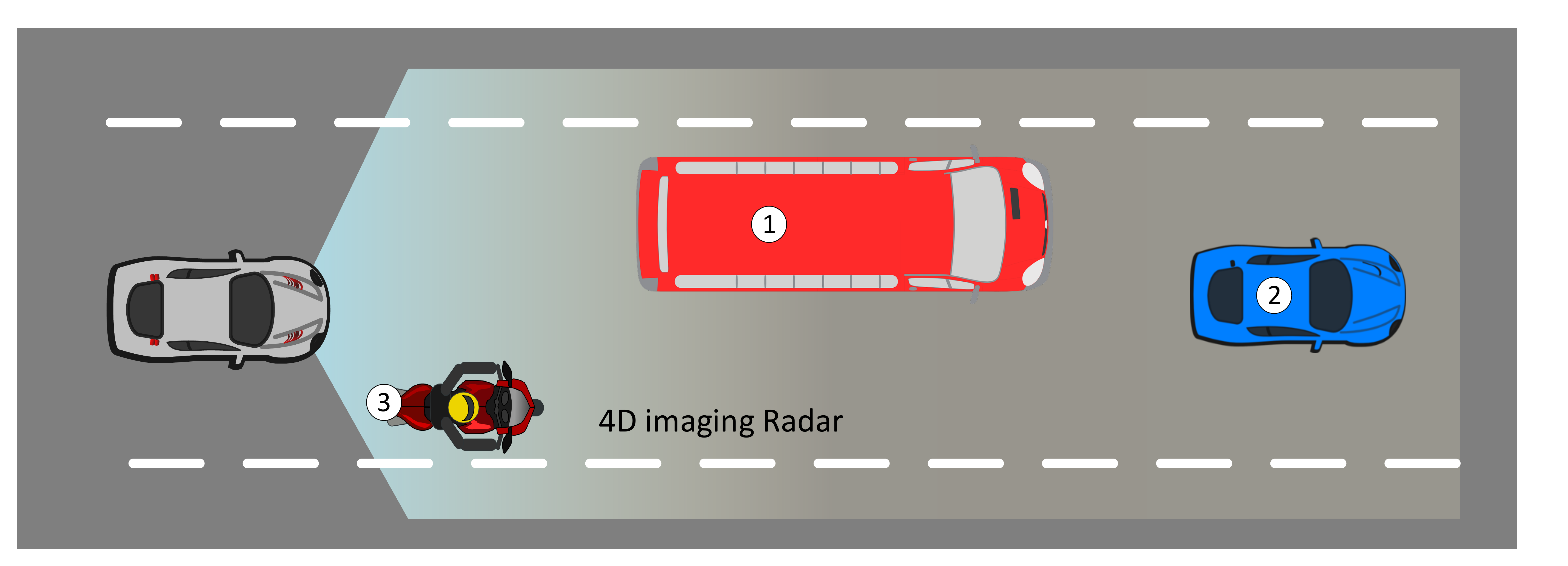}
		\caption[]{4D imaging automotive \gls{MIMO} radar systems.}
    \end{subfigure}
    \caption{A comparison between the conventional and 4D imaging \gls{MIMO} radar system. }
    \label{fig:4Dimaging}
\end{figure}

In the emerging 4D-imaging automotive \gls{MIMO} radar systems, the \gls{SRR}, \gls{MRR}, and \gls{LRR} applications are planned to be merged, to provide unique and high angular resolution in the entire radar detection range, as depicted in \figurename{~\ref{fig:4Dimaging}}. In this application, both long range property and fine angular resolution are required. To achieve the long range property, the \gls{MIMO} radar system should have the capability of beampattern shaping to enhance the received \gls{SINR} and the detection performance while, the orthogonality is required to build the \gls{MIMO} radar virtual array in the receiver and obtaining fine angular resolution.

The novel problem in this paper is aimed to address the above practical requirements, by considering an \gls{CD} framework subsuming the key objectives and constraints while offering an elegant design methodology.
This motivation drives the following contributions of the paper:
%In order to solve the problem, we propose an effective iterative algorithm based on \gls{CD} which minimizes the objective function monotonically in each iteration \cite{8706639, 7967829, 9054519, 9078952, 9052442, 9082109, 9093027, 9103620, doi:10.1137/120891009, boyd2004convex, wright2015coordinate}. 
%The numerical results show that the proposed method can make a good trade-off between beampattern shaping and orthogonality. 
%In a nutshell, the current study aims at:
\begin{itemize}
\item  {\sl Use of both range- and spatial-\gls{ISLR}:} 
% The paper offers a novel flexible  waveform design formulation for minimizing range-\gls{ISLR} and spatial-\gls{ISLR} 
Since these two aspects are important in \gls{MIMO} radar systems, we exploit the well-known weighting to propose a flexible framework enabling a trade-off between spatial- and range-\gls{ISLR} in a cognitive \gls{MIMO} radar paradigm.  This is considered by resorting to a scalarization of the multi-objective problem through its weighted sum. The weight offers a trade-off between spatial- and range- \gls{ISLR}. This property is very useful for cognitive radars where the system can set the operation levels for the two \gls{ISLR}s  based on the scenario. The proposed  optimization problem is then augmented with different sets of practical constraints, i.e., limited energy, \gls{PAR}, constant modulus and discrete phase. This novel exercise of consolidation eases design and achieves higher design efficiency.
\item  
% {\sl \gls{CD} based approach:} 
{\sl Optimization framework:}
The problem formulation leads to an objective function comprising a weighted sum of fractional quadratic (spatial-\gls{ISLR}) and quartic (range-\gls{ISLR}) functions; together with the constraints, the formulation leads to a non-convex, multi-variable, and NP-hard optimization problem. The paper proposes a unified framework based on the \gls{CD} approach to solve the optimization problem under the different sets of constraints. An effective iterative algorithm based on \gls{CD}, which minimizes the objective function monotonically, in each iteration is devised. While the \gls{CD} approach is well-known \cite{8706639, 7967829, 9054519, 9078952, 9052442, 9082109, 9093027, 9103620, doi:10.1137/120891009, boyd2004convex, wright2015coordinate, 8890867} challenges lie in deriving an efficient solution to each of the single variable optimization problems. A key analytical contribution of this paper is to specialize the single variable objective functions and obtain closed-form or numerically efficient design methodologies based on the constraints. Particularly, the paper considers the following approaches to derive the global optimum at each single variable update (\Rn{1}) gradient based approach for limited power and \gls{PAR} constraints wherein the minimization problems are reformulated to enable derivation of gradients efficiently using real computations, (\Rn{2}) a traditional calculus approach for continuous phase followed by simplification, (\Rn{3}) solving the problem to yield an efficient \gls{FFT} based solution for discrete phase problems.
% Particularly, a novel direct solution using the gradient under limited power and \gls{PAR}, derivative under continuous phase, and \gls{FFT} under discrete phase constraints are provided. 
%
\item {\sl Discrete Phase Design:} A systematic approach to the design of discrete phase sequences, generally not addressed in the literature, is considered in this paper. The design of discrete phase sequences is important since its allows for the efficient utilization of the limited transmitter power. Further, the phases of these sequences are chosen from a limited alphabet, lending it attractive for radar engineers/designers from the point of view of hardware implementation. A \gls{FFT} based methodology is considered to handle \gls{CD} for such sequences.
\item {\sl Trade-off and Flexibility:} Extensive simulations comparing the proposed method with literature are provided to illustrate the superior trade-off obtained by the proposed solutions in minimizing the spatial- and range- \gls{ISLR}.  The flexibility of the framework is also illustrated by reporting superior performance when minimizing only the spatial-\gls{ISLR} or the range-\gls{ISLR}.
\end{itemize}
\subsection{Organization and Notations}
The rest of this research is organized as follows. In Section \ref{sec:System Model}, the system model and the design problem is formulated. We develop the \gls{CD} framework to solve the problem in Section \ref{sec:Proposed method} and provide numerical experiments to verify the effectiveness of proposed algorithm in Section \ref{sec:Numerical Results}.
\paragraph*{Notations} This paper uses lower-case and upper-case boldface for vectors ($\ba$) and matrices ($\bA$) respectively. The conjugate, transpose and the conjugate transpose operators are denoted by the $(.)^*$, $(.)^T$ and $(.)^H$ symbols respectively. Besides the Frobenius norm, $l_2$ norm and absolute value are denoted by $\norm{.}_F$, $\norm{.}_2$ and $|.|$ respectively. For any complex number $a$, $\Re(a)$ and $\Im(a)$ denotes the real and imaginary part respectively. The letter $j$ represents the imaginary unit (i.e., $j=\sqrt{-1}$), while the letter $(i)$ is use as step of a procedure. Finally $\odot$ denotes the Hadamard product. 

\section{System Model and Problem Formulation}\label{sec:System Model}
We consider a colocated narrow-band \gls{MIMO} radar system, with $M_t$ transmit antennas, each transmitting a sequence of length $N$ in the fast-time domain. Let the matrix $\bS \in \mathbb{C}^{M_t \times N}$ denotes the transmitted set of sequences in baseband.
% \begin{equation*}\label{eq:S}
% 	\bS \triangleq 
% 	\begin{bmatrix}
% 	s_{1,1} & s_{1,2} & \dots  & s_{1,N} \\
% 	s_{2,1} & s_{2,2} & \dots  & s_{2,N} \\
% 	\vdots  & \vdots  & \vdots & \vdots  \\
% 	s_{M,1} & s_{M,2} & \dots  & s_{M,N}
% 	\end{bmatrix},
% \end{equation*}
Let us assume that $\bS \triangleq [\bar{\bs}_1, \dots, \bar{\bs}_N] \triangleq [\tilde{\bs}_1^T, \dots, \tilde{\bs}_N^T]^T$, where the vector $\bar{\bs}_n \triangleq [s_{1,n}, s_{2,n}, \ldots, s_{M_t,n}]^T \in \mathbb{C}^{M_t}$ ($n=\{1,\dots,N\}$) indicates the $n^{th}$ time-sample across the $M_t$ transmitters (the $n^{th}$ column of matrix $\bS$) while the $\tilde{\bs}_m \triangleq [s_{m,1}, s_{m,2}, \dots, s_{m,N}]^T \in \mathbb{C}^N$ ($m=\{1,\dots,M\}$) indicates the $N$ samples of $m^{th}$ transmitter (the $m^{th}$ row of matrix $\bS$).
% The $m^{th}$ row indicates the $N$ samples of $m^{th}$ transmitter while the $n^{th}$ column indicates the $n^{th}$ time-sample across the $M_t$ transmitters. 
In this paper, we deal with the spatial- and range- related \gls{ISLR}. To this end, in the following, we introduce the \gls{ISLR} model in these domains.
\subsection{System Model in Spatial Domain}\label{subsec:System_Model_in_Spatial_Domain}
% At time-instance $n$, the probing signal through the $M_t$ transmit antennas is, $\bar{\bs}_n=[s_{1,n}, s_{2,n}, \ldots, s_{M_t,n}]^T \in \mathbb{C}^{M_t}$,
% % \begin{equation}\label{eq:hat{s}}
% % 	\bar{\bs}_n=[s_{1,n}, s_{2,n}, \ldots, s_{M,n}]^T \in \mathbb{C}^{M},
% % \end{equation}
% where $s_{m,n}$ is the $n^{th}$ sample of $m^{th}$ transmitter. 
We assume a \gls{ULA} structure for the transmit array and the transmit steering vector takes the from \cite{LImimo},
\begin{equation}\label{eq:a(theta)}
	\ba(\theta)=[1, e^{j\frac{2\pi d_t}{\lambda}\sin(\theta)}, \ldots, e^{j\frac{2\pi d_t(M_t-1)}{\lambda}\sin(\theta)}]^T \in \mathbb{C}^{M_t}.
\end{equation}
In \eqref{eq:a(theta)}, $d_t$ is the distance between the transmitter antennas and $\lambda$ is the signal wavelength. 
%The baseband signal in the direction $\theta$ is $\ba^H(\theta)\bar{\bs}_n$ \cite{LImimo}. Therefore, 
The power of transmitted signal (beampattern) in the direction $\theta$ can be written as \cite{LImimo,7435338,4276989},
\begin{equation*}\label{eq:beampattern}
	P(\bS, \theta) = \textstyle \frac{1}{N}\sum_{n=1}^{N} \left|\ba^H(\theta) \bar{\bs}_n\right|^2 = \frac{1}{N}\sum_{n=1}^{N}\bar{\bs}_n^H\bA(\theta)\bar{\bs}_n
\end{equation*}
where, $\bA(\theta)=\ba(\theta)\ba^H(\theta)$. 
%It can be shown that the recent equation  can be recast as function of $\bS$ (see Appendix \ref{app:1}),
%\begin{equation}\label{eq:P(theta)}
%	P(\bS, \theta) =  \frac{1}{N}{\rm Tr}(\bA(\theta)\bS\bS^H).
%\end{equation}
Let $\Theta_d = \{\theta_{d,1}, \theta_{d,2}, \dots, \theta_{d,M_d}\}$ and $\Theta_u = \{\theta_{u,1}, \theta_{u,2}, \ldots, \theta_{u,M_u}\}$ denote the sets of $M_d$ desired and $M_u$ undesired angles in the spatial domain, respectively. 
%Their information can be obtained from a cognitive paradigm.
This information can be obtained from a cognitive paradigm.
We define the spatial-\gls{ISLR}, $\bar{f}(\bS)$, as the ratio of beampattern response on the undesired directions (sidelobes) to those on the desired angles (mainlobes) by the following equation,
\begin{equation}\label{eq:f_bar}
	\bar{f}(\bS) \triangleq \frac{\frac{1}{M_u}\sum_{r=1}^{M_u}P(\bS,\theta_{u,r})}{\frac{1}{M_d}\sum_{r=1}^{M_d}P(\bS,\theta_{d,r})} = \frac{\sum_{n=1}^{N}\bar{\bs}_n^H\bA_u\bar{\bs}_n}{\sum_{n=1}^{N}\bar{\bs}_n^H\bA_d\bar{\bs}_n},
\end{equation} 
where $\bA_u \triangleq \frac{\sum_{r=1}^{M_u}\bA(\theta_{u,r})}{NM_u}$ and $\bA_d \triangleq \frac{\sum_{r=1}^{M_d}\bA(\theta_{d,r})}{NM_d}$.  Note that $\bar{f}(\bS)$ is a fractional quadratic function.

\subsection{System Model in Fast-Time Domain}\label{subsec:System_Model_in_Fast_Time_Domain}
% The transmitted signal from $m^{th}$ transmit antenna is, $\tilde{\bs}_m=[s_{m,1}, s_{m,2}, \dots, s_{m,N}]^T \in \mathbb{C}^{N}$.
% \begin{equation}\label{eq:s_r}
% 	\tilde{\bs}_m=[s_{m,1}, s_{m,2}, \dots, s_{m,N}]^T \in \mathbb{C}^{N},
% \end{equation}
The aperiodic cross-correlation of $\tilde{\bs}_m$ and $\tilde{\bs}_l$ is defined as,
\begin{equation}\label{eq:cross_correlation}
	%r_{m,l}(k) = (\tilde{\bs}_m \circledast \tilde{\bs}_l)_k =  \textstyle \sum_{n=1}^{N-k} s_{m,n}s_{l,n+k}^*,
	r_{m,l}(k) = \textstyle \sum_{n=1}^{N-k} s_{m,n}s_{l,n+k}^*,
\end{equation}
where $m,l \in \{1,\dots,M_t\}$ are the transmit antennas indices and $k \in \{-N+1,\dots,N-1\}$ denotes the lag of cross-correlation. If $m = l$, \eqref{eq:cross_correlation} represents the aperiodic auto-correlation of signal $\tilde{\bs}_m$. The zero lag of auto-correlation represents the mainlobe of the matched filter output and contains the energy of sequence, while the other lags ($k \neq 0$) are referred to the sidelobes. The range-\gls{ISL} can therefore be expressed by \cite{8706639,8239862},
\begin{equation}\label{eq:ISL}
	\textstyle \sum_{\substack{{m,l=1}\\{l \neq m}}}^{M_t}\sum_{k=-N+1}^{N-1}|r_{m,l}(k)|^2 + \sum_{m=1}^{M_t}\sum_{\substack{{k=-N+1}\\{k \neq 0}}}^{N-1}|r_{m,m}(k)|^2,
\end{equation}
where the first and second terms represent the cross- and auto-correlation sidelobes, respectively. For the sake of convenience, \eqref{eq:ISL} can be written as,
% follow (see Appendix \ref{app:1}),
\begin{equation}\label{eq:ISL2}
	\text{ISL} = \textstyle \sum_{m,l=1}^{M_t}\sum_{k=-N+1}^{N-1}|r_{m,l}(k)|^2 - \sum_{m=1}^{M_t}|r_{m,m}(0)|^2.
\end{equation}
% \begin{equation}\label{eq:ISL2}
% 	\text{ISL} = \textstyle \sum_{m=1}^{M}\sum_{l=1}^{M}\sum_{k=-N+1}^{N-1}|r_{m,l}(k)|^2 - \sum_{m=1}^{M}|r_{m,m}(0)|^2.
% \end{equation}
%where \gls{ISL} is expressed as the subtraction of the integrated auto/cross-correlation level between transmitted waveforms (first term) and the integrated mainlobe level of auto-correlation (second term).
%One general criteria for evaluating the goodness of orthogonality is the range-\gls{ISLR}. 
The range-\gls{ISLR} (time-\gls{ISLR}) is the ratio of range-\gls{ISL} over the mainlobe energy, i.e.,
% \begin{comment}
% {\color{red}
% \begin{equation}\label{eq:f_tld}
% 	\tilde{f}(\bS) = \frac{\sum\limits_{m=1}^{M}\sum\limits_{l=1}^{M}\sum\limits_{k=-N+1}^{N-1}|r_{m,l}(k)|^2 - \sum\limits_{m=1}^{M}|r_{m,m}(0)|^2 }{\sum\limits_{m=1}^{M}|r_{m,m}(0)|^2},
% \end{equation}
% which can be equivalently written as,}
% \end{comment}
%It can be shown that \eqref{eq:f_tld} can be written as function of $\bS$ as follow,
\begin{equation}\label{eq:f_tld2}
	\tilde{f}(\bS) = \textstyle \frac{\sum\limits_{m,l=1}^{M_t}\sum\limits_{k=-N+1}^{N-1}\norm{\tilde{\bs}_m^H\bJ_k\tilde{\bs}_l}_2^2 - \sum\limits_{m=1}^{M_t}\norm{\tilde{\bs}_m^H\tilde{\bs}_m}_2^2 }{\sum\limits_{m=1}^{M_t}\norm{\tilde{\bs}_m^H\tilde{\bs}_m}_2^2},
\end{equation}
% \begin{equation}\label{eq:f_tld2}
% 	\tilde{f}(\bS) = \frac{\sum\limits_{m=1}^{M}\sum\limits_{l=1}^{M}\sum\limits_{k=-N+1}^{N-1}\norm{\tilde{\bs}_m^H\bJ_k\tilde{\bs}_l}_2^2 - \sum\limits_{m=1}^{M}\norm{\tilde{\bs}_m^H\tilde{\bs}_m}_2^2 }{\sum\limits_{m=1}^{M}\norm{\tilde{\bs}_m^H\tilde{\bs}_m}_2^2},
% \end{equation}
where $\bJ_k = \bJ_{-k}^T$ donates the $N \times N$ shift matrix \cite{10.5555/2422911}.
% with the following definition \cite{10.5555/2422911},
%  \begin{equation}
%  \bJ_k(u,v)  = 
%  	\begin{dcases}
%  	1, & u-v = k\\
%  	0, & u-v \neq k.\\
%  	\end{dcases}
%  \end{equation}
%As can be seen $\tilde{f}(\bS)$ is a fractional quartic function.
Note that, when the transmit set of sequences are unimodular,  $\sum_{m=1}^{M_t}\norm{\tilde{\bs}_m^H\tilde{\bs}_m}_2^2 = M_tN^2$, and $\tilde{f}(\bS)$ is a scaled version of the range-\gls{ISLR} defined in \cite{8706639}. As can be seen $\tilde{f}(\bS)$ is a fractional quartic function. 

%\section{Problem Formulation and Design Metrics}\label{sec:Problem Formulation and Design Issues}
\subsection{Problem Formulation}
We aim to design sets of sequences that simultaneously possess good properties in terms of both spatial- and range-\gls{ISLR}, under limited transmit power, bounded \gls{PAR}, constant modulus and discrete phase constraints. The optimization problem can be represented as,
%are interested to design a set of waveforms to have a good transmitter beampattern shape, and good auto- and cross-correlation simultaneously. One approach is to minimize the \gls{ISLR} in both spatial and fast-time domain. 
%In this regard, the following problem can be considered,
\begin{equation}\label{eq:MOOP}
% 	\mathcal{P}
	\begin{dcases}
	\min_{\bS} 	& \bar{f}(\bS), \tilde{f}(\bS) \\
	s.t 	    & C
	\end{dcases}
\end{equation}
where $C \in \{C_1, C_2, C_3, C_4\}$, with
\begin{equation}\label{eq:constraints}
    \begin{aligned}
    C_1:& 0 < \norm{\bS}_F^2 \leqslant M_t N \\
    C_2:& 0 < \norm{\bS}_F^2 \leqslant M_t N, \ \frac{\max {|s_{m,n}|}^2}{\frac{1}{M_t N} \norm{\bS}_F^2} \leqslant \gamma_p\\
    C_3:& s_{m,n} = e^{j\phi_{m,n}}; \  \phi \in \Phi_{\infty} \\
    C_4:& s_{m,n} = e^{j\phi_{m,n}}; \ \phi \in \Phi_L\\
    \end{aligned}
\end{equation}
% \begin{equation}\label{eq:constraints}
%     \begin{aligned}
%     C_1:& 0 < \norm{\bS}_F^2 \leqslant M_t N \\
%     C_2:& 0 < \norm{\bS}_F^2 \leqslant M_t N, \ \frac{\max {|s_{m,n}|}^2}{\frac{1}{M_t N} \norm{\bS}_F^2} \leqslant \gamma_p\\
%     C_3:& s_{m,n} = e^{j\phi}; \  \phi \in \Phi_{\infty}\\
%     C_4:& s_{m,n} = e^{j\phi}; \ \phi \in \Phi_L\\
%     \end{aligned}
% \end{equation}
%In \eqref{eq:constraints}, $C_1$ represents the limited energy. $C_2$ is the \gls{PAR} with limited energy constraint, where, $m=\{1,\ldots,M\}$, $n=\{1,\dots,N\}$, and $\gamma_p$ is the maximum admissible \gls{PAR}. In $C_3$ and $C_4$ constraints, $\Phi_{\infty}$ and $\Phi_L$ denotes the continuous and discrete phase respectively. In case of continuous phase constraint we assume that, the phase can be any value in the interval $\Phi_{\infty} = [-\pi,\pi)$. In addition, we consider to design a set of \gls{MPSK} sequences under discrete phase constraint, where the phase ($\phi$) is chosen through a finite phase set of $\Phi_L = \left \{\phi_0,\phi_1,\dots,\phi_{L-1}\right \} \in \left \{0, \frac{2\pi}{L}, \dots, \frac{2\pi(L-1)}{L} \right \}$, with $L$ being the alphabet size. 
where $m=\{1,\ldots,M_t\}$, and $n=\{1,\dots,N\}$. In \eqref{eq:constraints}, 
\begin{itemize}
    \item $C_1$ represents the limited transmit power constraint.
    \item $C_2$ is the \gls{PAR} constraint with limited power, and $\gamma_p$ indicates the maximum admissible \gls{PAR}.
    \item $C_3$ is the constant modulus constraint with $\Phi_{\infty} = [-\pi,\pi)$.
    \item $C_4$ is the discrete phase constraint with $\Phi_L = \left \{\phi_0,\phi_1,\dots,\phi_{L-1}\right \} \in \left \{0, \frac{2\pi}{L}, \dots, \frac{2\pi(L-1)}{L} \right \}$, and $L$ is the alphabet size.
\end{itemize}
%In case of constant modulus constraint ($C_3$ and $C_4$), The energy of signal is, by construction, $\norm{\bS}_F^2 = MN$. Therefore, \eqref{eq:MOOP} under $C_1, \dots, C_4$ constraints are sorted from the largest to the least feasible set, i.e,
The first constraint ($C_1$) is convex while the second constraint ($C_2$) is non-convex due to the fractional inequality. Besides, the equality constraints $C_3$ and $C_4$ ($s_{m,n} = e^{j\phi} $ \footnote{For the convenience we use $\phi$ instead of $\phi_{m,n}$ in the rest of the paper.} or $ |s_{m,n}| = 1$) are not affine.
The aforementioned constraints can be sorted from the smallest to the largest feasible set as, 
\begin{equation}
	C_4 \subset C_3 \subset C_2 \subset C_1.
\end{equation}
% {\color{blue} I guess we need the constraints above in the first problem, or we should say something on that.}\\ 
Problem \eqref{eq:MOOP} is a bi-objective optimization problem in which a feasible solution that minimizes the both the objective functions may not exist \cite{8706639,deb2001multi}. Scalarization, a well known technique converts the bi-objective optimization problem to a single objective problem, by replacing a weighted sum of the objective functions. Using this technique, the following Pareto-optimization problem will be obtained,
% \begin{equation}\label{eq:sum_weighted}
% 	f_o(\bS) = \eta\bar{f}(\bS) + (1-\eta)\tilde{f}(\bS)
% \end{equation}
\begin{equation}\label{eq:sum_weighted}
	\mathcal{P}
	\begin{dcases}
	\min_{\bS} 	& f_o(\bS) \triangleq \eta\bar{f}(\bS) + (1-\eta)\tilde{f}(\bS) \\
	s.t 	    & C,
	\end{dcases}
\end{equation}
% where $f_o(\bS) = \eta\bar{f}(\bS) + (1-\eta)\tilde{f}(\bS)$. 
The coefficient $\eta \in \left[0,1\right]$ is a weight factor that effects trade-off between spatial- and range-\gls{ISLR}. In \eqref{eq:sum_weighted}, $\bar{f}(\bS)$ is a fractional quadratic function of $\bar{\bs}_n$, and $\tilde{f}(\bS)$ is fractional quartic function of $\tilde{\bs}_m$. Hence, the objective is a non-convex and multi-variable function.  Therefore, we encounter a non-convex, multi-variable and NP-hard optimization problem \cite{8706639,7967829}. 
%In the following, we propose an efficient framework to solve the problem under $C_1, \dots, C_4$ constraints.

\section{Proposed Waveform Design}\label{sec:Proposed method}
%In this section, we deal with the joint optimization of spatial-\gls{ISLR} and range-\gls{ISLR} by solving the bi-objective optimization problem $\mathcal{P}$. In order to find a Pareto solution for problem $\mathcal{P}$ we propose an iterative method based on \gls{CD} framework. 
To tackle the fractional optimization problems, several approaches including expanded \gls{SDR} \cite{7126203, 6805184}, Dinkelbach \cite{Dinkel, GFP}, polynomial optimization \cite{sedighi2020localization, doi:10.1137/S1052623400366802} and Grab-n-Pull \cite{GHARANJIK20191, 7472288} can be used. In this paper, to solve \eqref{eq:sum_weighted} directly, we propose \gls{CD} framework, which is applicable for both fractional quadratic and quartic problems under four different constraints, i.e., $C_1$, $C_2$, $C_3$, and $C_4$. Under this framework, the multi variable problem is solved as a sequence of single variable problems. Further this single variable problems admit a global solution.
\subsection{\gls{CD} based framework}
The methodologies based on \gls{CD}, generally start with a feasible matrix $\bS=\bS^{(0)}$ as the initial waveform set. Then, in each iteration, the waveform set is updated entry by entry several times \cite{8706639, 7967829, 9054519, 9078952, 9052442, 9082109, 9093027, 9103620, doi:10.1137/120891009, boyd2004convex, wright2015coordinate}. In particular, an entry of $\bS$ is considered as the only variable while others are held fixed and then the objective function is optimized with respect to this identified variable. Let us assume that $s_{t,d}$ ($t \in \{1, \dots, M_t$ and $d \in \{1, \dots, N\}$) is the only variable. There are several rules to update the matrix $\bS$: (a) randomized i.e., the entry ($s_{t,d}$) is chosen uniformly randomly at each single variable update, (b) cyclic i.e., iterate over all different $s_{t,d}$ entries and (c) \gls{MBI} (greedy) i.e., optimizing the problem for each entry separately and choosing the best one. Note that in case of large number of variables, the use of \gls{MBI} rule naturally increases the convergence time drastically. 
In this paper, we consider cyclic rule to update the waveform. In this case, the fixed are stored in the matrix $\bS_{-(t,d)}^{(i)}$ as the following,
% This procedure will be performed for all the entries in one iteration.
% optimizing each entry of $\bS$ sequentially instead of the entire matrix $\bS$ \cite{8706639, 7967829, 9054519, 9078952, 9052442, 9082109, 9054143,boyd2004convex,wright2015coordinate}. 
% In particular, the objective is optimized with respect to one identified variable with others held fixed and the procedure is repeated for other variables. 
% Such a methodology is efficient when the problem of updating every entry can be written in a simplified form with respect to that variable. 

% Let us assume that $s_{t,d}$ is the only variable and the other entries are included in matrix $\bS^{(i)}_{-(t,d)}$ as the following,
\begin{equation*}\label{eq:S_(t,d)}
	\bS^{(i)}_{-(t,d)} \triangleq 
		\begin{bmatrix}
		s_{1,1}^{(i)} & \dots  & \dots & \dots & \dots & \dots  & s_{1,N}^{(i)} \\
		\vdots & \vdots & \vdots & \vdots & \vdots & \vdots & \vdots \\
		s_{t,1}^{(i)} & \dots  & s_{t,d-1}^{(i)} & 0 & s_{t,d+1}^{(i-1)} & \dots  & s_{t,N}^{(i-1)}  \\
		\vdots & \vdots & \vdots & \vdots & \vdots & \vdots & \vdots \\
		s_{M_t,1}^{(i-1)} & \dots  & \dots & \dots & \dots & \dots  & s_{M_t,N}^{(i-1)}\\
		\end{bmatrix},
\end{equation*}
where, the superscripts $(i)$ and $(i-1)$ show the updated and non-updated entries at iteration $i$. This methodology is efficient when the problem in \eqref{eq:sum_weighted} is written in a simplified form with respect to that variable. In this regards, the optimization problem with respect to variable $s_{t,d}$ can be written as follows (see Appendix \ref{app:2} for details), 
\begin{equation}\label{eq:P_s_td}
	\mathcal{P}_{s_{t,d}}
	\begin{dcases}
	\min_{s_{t,d}} 	&  f_o(s_{t,d}, \bS^{(i)}_{-(t,d)})\\
	s.t 	    & C\\
	\end{dcases}
\end{equation}
where, $f_o(s_{t,d}, \bS^{(i)}_{-(t,d)})$ and the constraints are given by,
\begin{equation*}
	f_o(s_{t,d}, \bS^{(i)}_{-(t,d)}) \triangleq \eta\bar{f}(s_{t,d}, \bS^{(i)}_{-(t,d)}) + (1-\eta)\tilde{f}(s_{t,d}, \bS^{(i)}_{-(t,d)}),
\end{equation*}
\begin{equation}\label{eq:f_bar_std}
	\bar{f}(s_{t,d}, \bS^{(i)}_{-(t,d)}) \triangleq \frac{a_0s_{t,d}+a_1+a_2s_{t,d}^*+a_3|s_{t,d}|^2}{b_0s_{t,d}+b_1+b_2s_{t,d}^*+b_3|s_{t,d}|^2},
\end{equation}
\begin{equation}\label{eq:f_tld_std}
\begin{aligned}
	\tilde{f}&(s_{t,d}, \bS^{(i)}_{-(t,d)}) \triangleq \\ &\frac{c_0s_{t,d}^2+c_1s_{t,d}+c_2+c_3s_{t,d}^*+c_4{s_{t,d}^*}^2+c_5|s_{t,d}|^2}{|s_{t,d}|^4+d_1|s_{t,d}|^2+d_2},
	\end{aligned}
\end{equation}
%and the constraints are modified as,
\begin{equation}\label{eq:constraints2}
    \begin{aligned}
    C_1:& |s_{t,d}|^2 \leqslant \gamma_e,\\
	C_2:& |s_{t,d}|^2 \leqslant \gamma_e, \quad \gamma_l \leqslant |s_{t,d}|^2 \leqslant \gamma_u,\\
	C_3:& s_{t,d} = e^{j\phi}; \quad \phi \in \Phi_{\infty},\\
	C_4:& s_{t,d} = e^{j\phi}; \quad \phi \in \Phi_L,\\
    \end{aligned}
\end{equation}
Note that, in \eqref{eq:f_bar_std}, \eqref{eq:f_tld_std} and \eqref{eq:constraints2} the coefficients $a_v$, $b_v$, ($v\in\{0,\dots,3\}$), $c_w$ ($w\in\{0,\dots,5\}$) and boundaries $\gamma_l$, $\gamma_u$ and $\gamma_e$, depend on $\bS^{(i)}_{-(t,d)}$ all of which are defined in Appendix \ref{app:2}.

At $i^{th}$ iteration, for $t = 1, \ldots, M_t$, and $d = 1, \ldots, N$, the $(t,d)^{th}$ entry of $\bS$ will be updated by solving \eqref{eq:P_s_td}. After updating all the entries, a new iteration will be started, provided that the stopping criteria is not met. This procedure will continue until the objective function converges to an optimal value. A summary of the proposed method is reported (like a pseudo-code) in \textbf{Algorithm \ref{alg:waveform_design}}. 
% {\color{red}
% In $i^{th}$ iteration the $(t,d)^{th}$ entry of $\bS$ is updated by solving \eqref{eq:P_s_td} and the matrix update procedure is continued until the objective function converges to the optimum value. 
% }

To optimize the code entries, notice that the optimization variable is a complex number and can be expressed as $s_{t,d} = re^{j\phi}$, where $r \geqslant 0$ and $\phi \in [-\pi,\pi)$ are the amplitude and phase of $s_{t,d}$, respectively. By substituting $s_{t,d}$ with $re^{j\phi}$ and performing standard mathematical manipulations, the problem $\mathcal{P}_{s_{t,d}}$ can be rewritten with respect to $r$ and $\phi$ as follows,
\begin{equation}\label{eq:P_r_phi}
	\mathcal{P}_{r,\phi}
	\begin{dcases}
	\min_{r,\phi} 	& f_o\left(r, \phi \right)\\
	s.t 	    & C\\
\end{dcases}
\end{equation}
with $f_o\left(r, \phi \right) \triangleq \eta\bar{f}\left(r, \phi \right) + (1-\eta)\tilde{f}\left(r, \phi \right)$, where,
\begin{equation}\label{eq:f_bar_rphi}
	\bar{f}\left(r, \phi \right) \triangleq \frac{a_0re^{j\phi}+a_1+a_2re^{-j\phi}+a_3r^2}{b_0re^{j\phi}+b_1+b_2re^{-j\phi}+b_3r^2},
\end{equation}
\begin{equation}\label{eq:f_tld_rphi}
\begin{aligned}
	&\tilde{f}\left(r, \phi \right) \triangleq \\ &\frac{c_0r^2e^{j2\phi}+c_1re^{j\phi}+c_2+c_3re^{-j\phi}+c_4r^2e^{-j2\phi}+c_5r^2}{r^4 + d_1r^2 + d_2}.
	\end{aligned}
\end{equation}
%The constraints also will be modified as,
\begin{equation}\label{eq:constraints3}
    \begin{aligned}
    C_1:& 0 \leqslant r \leqslant \sqrt{\gamma_e},\\
	C_2:& 0 \leqslant r \leqslant \sqrt{\gamma_e}, \sqrt{\gamma_l} \leqslant r \leqslant \sqrt{\gamma_u},\\
	C_3:& r = 1; \quad \phi \in \Phi_{\infty},\\
	C_4:& r = 1; \quad \phi \in \Phi_L.\\
    \end{aligned}
\end{equation}
%As can be seen the problem $\mathcal{P}_{r,\phi}$ under $C_3$ and $C_4$ constraints are the special case of $C_3$ and $C_4$ with $r=1$. In this case they depend only on $\phi$.
% {\color{red}
% Let $s_{t,d}^{(i)} = r^{\star}e^{j\phi^{\star}}$ be the optimum solution of $\mathcal{P}_{r,\phi}$ at $i^{th}$ iteration. We propose algorithm \ref{alg:waveform_design} based on \gls{CD} to optimize all $MN$ entries of $\bS$ in an iterative procedure. This algorithm considers a feasible set of sequences as the initial waveforms. Then in each iteration, it selects $s_{t,d}^{(i-1)}$ as variable and updates that with optimum entry ${s_{t,d}^{(i)}}$. This procedure repeats for other entries and is continued until all $MN$ pulses are optimized at least once.} 
Let $s_{t,d}^{\star} = r^{\star}e^{j\phi^{\star}}$ be the optimized solution of Problem $\mathcal{P}_{r,\phi}$.
%We propose \textbf{Algorithm \ref{alg:waveform_design}} to optimize all $MN$ entries of $\bS$ in an iterative procedure. 
Towards obtaining this solution, \textbf{Algorithm \ref{alg:waveform_design}} considers a feasible set of sequences as the initial waveforms. Then, at each single variable update, it selects $s_{t,d}^{(i-1)}$ as the variable and updates it with the optimized ${s_{t,d}^{(i)}}$, denoted by ${s_{t,d}^{\star}}$. This procedure is repeated for other entries and is undertaken until all the entries are optimized at least once.
After optimizing the $M_t N^{th}$ entry, the algorithm examines the convergence metric for the objective function. If the stopping criteria is not met the algorithm repeats the aforementioned steps. We consider $(f_o(\bS^{(i)}) - f_o(\bS^{(i-1)}) \leq \zeta$, ($\zeta$ is the stopping threshold, $\zeta > 0$) as the stopping criterion of the proposed method.
\begin{algorithm}[t]
	\caption{: Pseudo-code for transmit waveform design}
	\label{alg:waveform_design}
	\textbf{Input:} Initial set of feasible sequences, $\bS^{(0)}$.\\
% 	\textbf{Output:} Optimized set of sequences, $\bS^\star$.\\
	\textbf{Initialization:} $i := 0$.\\ 
	\textbf{Optimization:} 
	\begin{enumerate}
		\item {\bf while} $(f_o(\bS^{(i-1)}) - f_o(\bS^{(i)}) ) > \zeta$ {\bf do}
		\item \hspace{5mm} $i := i+1$;
		\item \hspace{5mm} {\bf for} $t=1,\dots,M_t$ {\bf do}
		\item \hspace{10mm} {\bf for} $d=1,\dots,N$ {\bf do}
		\item \hspace{15mm} Optimize  $s_{t,d}^{(i-1)}$  and obtain $s^{\star}_{t,d}$;  %Obtain $s_{t,d}^{(i)}$ by solving \eqref{eq:P_r_phi};
		\item \hspace{15mm} Update $s^{(i)}_{t,d} = s^{\star}_{t,d}$;
		\item \hspace{15mm} $\bS^{(i)} = \bS^{(i)}_{-(t,d)} |_{s_{t,d}=s^{(i)}_{t,d}}$;%$\bS^{(i)} = \bS^{(i)}_{-(t,d)} |_{s_{t,d}=s_{t,d}^{(i)}}$;
		\item \hspace{10mm} {\bf end for}
		\item \hspace{5mm} {\bf end for}
		\item {\bf end while}
		\end{enumerate}
	\textbf{Output:} $\bS^{\star} = \bS^{(i)}$.
\end{algorithm}
% \begin{algorithm}
% 	\caption{: \gls{CD} framework for waveform design}
% 	\label{alg:waveform_design}
% 	\textbf{Input:} Initial set of feasible sequences, $\bS^{(0)}$\\
% 	\textbf{Output:} Optimized set of sequences, $\bS^\star$ 
% 	\begin{enumerate}
% 		\item {\bf Initialization}.
% 		\begin{itemize}
% 			\item Set $i,t,d := 1$; 
% 		\end{itemize}
% 		\item {\bf Optimization}.{\color{red}  I guess this part is not correct(iterations)}
% 		\begin{itemize}
% 			\item Obtain $s_{t,d}^{(i)}$ by solving \eqref{eq:P_r_phi};
% 			\item $\bS^{(i)} = \bS^{(i)}_{-(t,d)} |_{s_{t,d}=s_{t,d}^{(i)}}$;
% 			\item If $t=M$ then $t:=1$; otherwise $t:=t+1$;
% 			\item If $d=N$ then $d:=1$; otherwise $d:=d+1$; 
% 		\end{itemize}
% 		\item {\bf Stopping criterion}.
% 		\begin{itemize}	
% 			\item If $(f_o(\bS^{(i)}) - f_o(\bS^{(i-1)})) < \zeta$, go to (4); otherwise $i:=i+1$ and go to (2); 
% 		\end{itemize}
% 		\item {\bf Output}.
% 		\begin{itemize}
% 			\item $\bS^{\star} = \bS^{(i)}$ 
% 		\end{itemize}
% 	\end{enumerate}
% \end{algorithm}
With the defined methodology, it now remains to solve $\mathcal{P}_{r,\phi}$ for the different constraints. This is considered next.
\subsection{Solution for limited power constraint}\label{subsec:energy_constraint}
Problem $\mathcal{P}_{r,\phi}$ under $C_1$ constraint can be written as follows (see Appendix \ref{app:3} for details),
\begin{equation}\label{eq:P_e}
	\mathcal{P}_e
	\begin{dcases}
	\min_{r,\phi} 	& f_o\left(r, \phi \right)\\
	s.t 	    & C_1: 0 \leqslant r \leqslant \sqrt{\gamma_e}.\\
	\end{dcases}
\end{equation}
where $f_o\left(r, \phi \right) = \eta\bar{f}\left(r, \phi \right) + (1-\eta)\tilde{f}\left(r, \phi \right)$ and,
\begin{equation}\label{eq:f_bar_rphi2}
	\bar{f}\left(r, \phi \right) = \frac{a_3r^2 + 2(a_{0r}\cos{\phi}-a_{0i}\sin{\phi})r + a_1}{b_3r^2 + 2(b_{0r}\cos{\phi}-b_{0i}\sin{\phi})r + b_1},
\end{equation}
% \begin{equation}\label{eq:f_tld_rphi2}
% \begin{aligned}
% 	&\tilde{f}\left(r, \phi \right) = \\ &\frac{(2c_{0r}\cos{2\phi}-2c_{0i}\sin{2\phi}+c_5)r^2 + 2(c_{1r}\cos{\phi}-c_{1i}\sin{\phi})r + c_2}{r^4 + d_1r^2 + d_2}.
% 	\end{aligned}
% \end{equation}
\begin{equation}\label{eq:f_tld_rphi2}
\begin{aligned}
	&\tilde{f}\left(r, \phi \right) = [(2c_{0r}\cos{2\phi}-2c_{0i}\sin{2\phi}+c_5)r^2\\
	&+ 2(c_{1r}\cos{\phi}-c_{1i}\sin{\phi})r + c_2]\frac{1}{r^4 + d_1r^2 + d_2}.
	\end{aligned}
\end{equation}
The solution to $\mathcal{P}_e$ will be obtained by finding the critical points of the objective function and selecting the one that minimizes the objective. As $f_o(r,\phi)$ is a differentiable function, the critical points of $\mathcal{P}_e$ contain the solutions to $\nabla f_o(r,\phi) = 0$ and the boundaries ($0, \sqrt{\gamma_e}$), which satisfy the constraint ($ 0 \leqslant r \leqslant \sqrt{\gamma_e}$). 
To solve this problem, we use alternating optimization, where we first optimize for $r$ keeping $\phi$ fixed and vice-versa. 
% While this procedure can be carried out iteratively, we short after one iteration since the algorithm repeats the procedure for other entries.

\subsubsection{Optimization with respect to $r$}
Let us assume that the phase of the code entry $s_{t,d}^{(i-1)}$ is $\phi_0 = \tan^{-1}\left({\frac{\Im(s_{t,d}^{(i-1)})}{\Re(s_{t,d}^{(i-1)})}}\right)$. By substituting $\phi_0$ in $\frac{\partial f_o(r,\phi)}{\partial r}$, it can be shown that the solution to the condition $\frac{\partial f_o(r,\phi_0)}{\partial r}=0$ 
% the roots of partial derivative $\frac{\partial f_o(r,\phi_0)}{\partial r}=0$ 
can be obtained by finding the roots of the following degree $10$ real polynomial (see Appendix \ref{app:4} for details),
\begin{equation}\label{eq:dfr_root}
 	\textstyle \sum_{k=0}^{10} p_kr^k = 0.
\end{equation}
% The \eqref{eq:dfr_root} and $r$ are real polynomial function and real variable respectively. Hence only the real roots indicate the extrema points. 
Further, since $r$ is real, we seek only the real extrema points. Let us assume that the roots are $r_v$, $v=\{1,\dots, 10\}$; therefore the critical points of problem $\mathcal{P}_e$ with respect to $r$ can be expressed as,
\begin{equation}\label{eq:r_critical_points}
	R_e = \left\{r\in\{ 0, \sqrt{\gamma_e}, r_1, \dots, r_{10} \} | \Im(r)=0,  0 \leqslant r \leqslant \sqrt{\gamma_e} \right\}.
\end{equation}
Thus, the optimum solution for $r$ will be obtained by,
\begin{equation}\label{eq:optimum_r}
 	r_e^{\star} = \arg\min_{r} \left \{f_o(r, \phi_0) | r \in R_e \right \}.
\end{equation}

\subsubsection{Optimization with respect to $\phi$}
Let us keep $r$ fixed and optimize the problem with respect to $\phi$. Considering $\cos(\phi) = {(1-\tan^2(\frac{\phi}{2}))}/{(1+\tan^2(\frac{\phi}{2}))}$, $\sin(\phi) = {2\tan(\frac{\phi}{2})}/{(1+\tan^2(\frac{\phi}{2}))}$ and using the change of variable $z \triangleq \tan(\frac{\phi}{2})$, it can be shown that finding the roots of $\frac{\partial f_o(r_e^{\star},\phi)}{\partial \phi}$ is equivalent finding the roots of the following $8$ degree real polynomial (see Appendix \ref{app:5} for details),
\begin{equation}\label{eq:dfz_root}
	\textstyle \sum_{k=0}^{8} q_kz^k.
\end{equation}
Similar to \eqref{eq:dfr_root}, we only admit real roots. Let us assume that $z_v$, $v=\{1,\dots,8\}$ are the roots of \eqref{eq:dfz_root}. Hence, the critical points of $\mathcal{P}_e$ with respect to $\phi$ can be expressed as, 
\begin{equation}\label{eq:phi_critical_points}
	\Phi = \left\{2\arctan{(z_v)} | \Im(z_v)=0 \right\}.
\end{equation}
Therefore, the optimum solution for $\phi$ is, 
\begin{equation}\label{eq:optimum_phi_e}
	\phi_e^{\star} = \arg\min_{\phi} \left \{f_o(r_e^{\star}, \phi) | \phi \in \Phi \right \}.
\end{equation}
Subsequently the optimum solution for $s_{t,d}$ is, $s_{t,d}^{(i)} = r_e^{\star} e^{j\phi_e^{\star}}$.
% \begin{equation}\label{eq:Pe_optimum_s_td}
% 	s_{t,d}^{(i)} = r_e^{\star} e^{j\phi^{\star}}.
% \end{equation}

\begin{remark}
Since, $0$ and $\sqrt{\gamma_e}$ are members of $R_e$, two critical points always exist, and $R_e$ is never a null set. On the other hand, as $f_o(r_0,\phi)$ is function of $\cos{\phi}$ and $\sin{\phi}$, it is periodic, real and differentiable. Therefore, it has at least two extrema and hence its derivative has at least two real roots; thus $\Phi_e$ never becomes a null set. As a result in each single variable update, the problem has a solution and never becomes infeasible.
\end{remark}
% \textit{\underline{Note}}: As, $0$ and $\sqrt{\gamma_e}$ are the members of $R_e$, hence at least always there are two critical points and $R_e$ never becomes a null set. On the other hand as $f_o(r_0,\phi)$ is a real, differentiable and periodic function, it has at least two extrema. hence, its derivative has at least two real roots and $\Phi_e$ never becomes a null set. Therefore in each iteration the problem has a solution and never becomes infeasible.

\subsection{Solution for \gls{PAR} constraint} \label{subsec:PAR_constraint}
Problem $\mathcal{P}_{r,\phi}$ under $C_2$ constraint is a special case of $C_1$ and the procedures in subsection \ref{subsec:energy_constraint} are valid for limited power and \gls{PAR} constraint. The only difference lies in the boundaries and critical points with respect to $r$. Considering the $C_2$ constraint, the critical points can be expressed as the following,
\begin{equation}\label{eq:Pp_r_critical_points}
	\begin{aligned}
	R_p = &\{r\in\{ \max\{0, \sqrt{\gamma_l}\}, \min\{\sqrt{\gamma_u}, \sqrt{\gamma_e}\}, r_1, \dots, r_{10} \} |\\
	&\Im(r)=0,  \max\{0, \sqrt{\gamma_l}\} \leqslant r \leqslant \min\{\sqrt{\gamma_u}, \sqrt{\gamma_e}\}\}.
	\end{aligned}
\end{equation}
Therefore, the optimum solution for $r$ and $\phi$ is,
\begin{equation}\label{eq:optimum_r_phi_p}
\begin{aligned}
    r_p^{\star} &= \arg\min_{r} \left \{f_o(r, \phi_0) | r \in R_p \right \},\\
    \phi_p^{\star} &= \arg\min_{\phi} \left \{f_o(r_p^{\star}, \phi) | \phi \in \Phi \right \},
\end{aligned}
\end{equation}
and, the optimum entry can be obtained by, $s_{t,d}^{(i)} = r_p^{\star} e^{j\phi_p^{\star}}$.
% \begin{equation}\label{eq:Pp_optimum_s_td}
% 	s_{t,d}^{(i)} = r_p^{\star} e^{j\phi^{\star}}.
% \end{equation}

\subsection{Solution for Continuous Phase}\label{subsec:Continuous Phase}
The continuous phase constraint ($C_3$) is a special case of limited power ($C_1$) constraint. In this case $r=1$, and the optimum solution for $\phi$ is,
\begin{equation}\label{eq:optimum_phi_c}
    \phi_c^{\star} = \arg\min_{\phi} \left \{f_o(r, \phi) | \phi \in \Phi, r=1 \right \}.
\end{equation}
The optimum entry can be obtained by  $s_{t,d}^{(i)} =e^{j\phi_c^{\star}}$.
% \begin{equation}\label{eq:Pc_optimum_s_td}
% 	s_{t,d}^{(i)} =e^{j\phi^{\star}},
% \end{equation}
\subsection{Solution for discrete phase }\label{subsec:Discrete Phase}
We consider the design of a set of \gls{MPSK} sequences for the discrete phase problem. In this case, $\mathcal{P}_{r,\phi}$ can be written as follows (see Appendix \ref{app:6} for details),
\begin{equation}\label{eq:P_d}
\mathcal{P}_d
	\begin{dcases}
	\min_{\phi} 	& f_d(\phi) = \frac{e^{j3\phi}\sum_{k=0}^{6} g_ke^{-jk\phi}}{e^{j\phi}\sum_{k=0}^{2} h_k^{-jk\phi}} \\
	s.t 	    &  C_4:\phi \in \Phi_L.\\
	\end{dcases}
\end{equation}
As the problem under $C_4$ constraint is discrete, the optimization procedure is different compared with other constraints. In this case all the discrete points lie on the boundary of the optimization problem; hence, all of them are critical points for the problem. 
% On the other word it is not possible to use gradient or derivative to find the critical points. 
Therefore, one approach for solving this problem is to obtain all the possibilities of the objective function $f_o(\phi)$ over the set $\Phi_L = \left \{\phi_0,\phi_1,\dots,\phi_{L-1}\right \} \in \left \{0, \frac{2\pi}{L}, \dots, \frac{2\pi(L-1)}{L} \right \}$ and choose the phase which minimizes the objective function. It immediately occurs that such an evaluation could be cumbersome; however, for \gls{MPSK} alphabet, an elegant solution can be obtained as detailed below.

The objective function can be formulated with respect to the indices of $\Phi_L$ as follows,
% \begin{equation}
% 	\begin{aligned}  
% 	f_d(\phi_0) &= \frac{\sum_{k=0}^{6} g_k}{\sum_{k=0}^{2} h_k},\\
% 	f_d(\phi_1) &= \frac{e^{j3\frac{2\pi}{L}}\sum_{k=0}^{6} g_ke^{-jk\frac{2\pi}{L}}}{e^{j\frac{2\pi}{L}}\sum_{k=0}^{2} h_k^{-jk\frac{2\pi}{L}}}, \\ 
% 	&\vdots \\
% 	f_d(\phi_l) &= \frac{e^{j3\frac{2\pi l}{L}}\sum_{k=0}^{6} g_ke^{-jk\frac{2\pi l}{L}}}{e^{j\frac{2\pi l}{L}}\sum_{k=0}^{2} h_k^{-jk\frac{2\pi l}{L}}},\\
% 	&\vdots \\
% 	f_d(\phi_{L-1}) &= \frac{e^{j3\frac{2\pi(L-1)}{L}}\sum_{k=0}^{6} g_ke^{-jk\frac{2\pi(L-1)}{L}}}{e^{j\frac{2\pi(L-1)}{L}}\sum_{k=0}^{2} h_k^{-jk\frac{2\pi(L-1)}{L}}},
% 	\end{aligned}
% \end{equation}
% As can be seen $f_d(\phi_l)$ can be written as follow,
\begin{equation}\label{eq:f_l}
	f_d(\phi_l) = f_d(l) = \frac{e^{j3\frac{2\pi l}{L}}\sum_{k=0}^{6} g_ke^{-jk\frac{2\pi l}{L}}}{e^{j\frac{2\pi l}{L}}\sum_{k=0}^{2} h_ke^{-jk\frac{2\pi l}{L}}},
\end{equation}
where $l=\{0,\dots,L-1\}$, and the summation terms on numerator and denominator exactly follow the definition of $L$-points \gls{DFT} of sequences $\{g_0,\dots,g_6\}$ and $\{h_0,h_1,h_2\}$ respectively. Therefore, the problem $\mathcal{P}_d$ can be written as,
\begin{equation}\label{eq:P_l}
	\mathcal{P}_l
	\begin{dcases}
	\min_{l} 	& f_d(\phi_l)=\frac{\bw_{L,3} \odot \mathcal{F}_L\{g_0,g_1,g_2,g_3,g_4,g_5,g_6\}}{\bw_{L,1} \odot \mathcal{F}_L\{h_0,h_1,h_2\}} 
	\end{dcases},
\end{equation} 
where, $\bw_{L,\nu} = [1, e^{-j\nu\frac{2\pi}{L}}, \ldots, e^{-j\nu\frac{2\pi(L-1)}{L}}]^T \in \mathbb{C}^{L}$ and $\mathcal{F}_L$ is $L$ point \gls{DFT} operator. Due to aliasing phenomena, when $L < 7$, the objective function would be changed. Let $N_{f_d}$ and $D_{f_d}$ be the summation terms in nominator and denominator of $f_d(\phi_l)$ respectively, it can be shown that,
% The recent problem is valid when $L\geqslant7$ and for other values it is different. Let $N_{f_d}$ and $D_{f_d}$ be the summation terms in nominator and denominator of $f_d(\phi_l)$ respectively. For $L=6$, $N_{f_d} = \sum_{k=0}^{5} g_ke^{-jk\frac{2\pi l}{6}} + g_6e^{-j6\frac{2\pi l}{6}}$. 
% \begin{equation}
% 	f_d(\phi_l) = \frac{e^{j3\frac{2\pi l}{L}}\left(\sum_{k=0}^{5} g_ke^{-jk\frac{2\pi l}{6}} + g_6e^{-j6\frac{2\pi l}{6}}\right)}{e^{j\frac{2\pi l}{L}}\sum_{k=0}^{2} h_k^{-jk\frac{2\pi l}{6}}}\\ 
% \end{equation}
% As, $g_6e^{-j6\frac{2\pi}{6}} = g_6$, hence, $N_f=\mathcal{F}_L\{g_0+g_6,g_1,g_2,g_3,g_4,g_5\}$.
% \begin{equation*}\label{eq:f_l}
% 	\mathcal{F}_L\{g_0+g_6,g_1,g_2,g_3,g_4,g_5\}.
% \end{equation*}
% \begin{equation}\label{eq:f_l}
% 	f_d(\phi_l) = \frac{\bw_{L,3} \odot \mathcal{F}_L\{g_0+g_6,g_1,g_2,g_3,g_4,g_5\}}{\bw_{L,1} \odot \mathcal{F}_L\{h_0,h_1,h_2\}}.
% \end{equation}
% Likewise for others values $L=5, 4, 3$, respectively it can be shown that,
\begin{equation*}
	\begin{aligned} 
	L = 6 & \Rightarrow N_{f_d} = \mathcal{F}_L\{g_0+g_6,g_1,g_2,g_3,g_4,g_5\}\\
	L = 5 & \Rightarrow N_{f_d} = \mathcal{F}_L\{g_0+g_5,g_1+g_6,g_2,g_3,g_4\}\\ 
	L = 4 & \Rightarrow N_{f_d} = \mathcal{F}_L\{g_0+g_4,g_1+g_5,g_2+g_6,g_3\}\\
	L = 3 & \Rightarrow N_{f_d} = \mathcal{F}_L\{g_0+g_3+g_6,g_1+g_4,g_2+g_5\},
	\end{aligned}
\end{equation*}
and for $L=2$, $N_{f_d} = \mathcal{F}_L\{g_0+g_2+g_4+g_6,g_1+g_3+g_5\}$ and $D_{f_d} = \mathcal{F}_L\{h_0+h_2,h_1\}$.

% \begin{equation*}
% 	\begin{aligned} 
% 	f_d(\phi_l) &=\frac{\bw_{L,3} \odot \mathcal{F}_L\{g_0+g_5,g_1+g_6,g_2,g_3,g_4\}}{\bw_{L,1} \odot \mathcal{F}_L\{h_0,h_1,h_0\}}\\ 
% 	f_d(\phi_l) &= \frac{\bw_{L,3} \odot \mathcal{F}_L\{g_0+g_4,g_1+g_5,g_2+g_6,g_3\}}{\bw_{L,1} \odot \mathcal{F}_L\{h_0,h_1,h_2\}}\\
% 	f_d(\phi_l) &= \frac{\bw_{L,3} \odot \mathcal{F}_L\{g_0+g_3+g_6,g_1+g_4,g_2+g_5\}}{\bw_{L,1} \odot \mathcal{F}_L\{h_0,h_1,h_0\}}\\
% 	f_d(\phi_l) &= \frac{\bw_{L,3} \odot \mathcal{F}_L\{g_0+g_2+g_4+g_6,g_1+g_3+g_5\}}{\bw_{L,1} \odot \mathcal{F}_L\{h_0+h_2,h_1\}}
% 	\end{aligned}
% \end{equation*}
According to aforementioned discussion the optimum solution of \eqref{eq:P_l} is,
\begin{equation}\label{eq:l_star}
    l^{\star} = \arg\displaystyle{\min_{l=1,\dots,L}} \left\{f_d(\phi_l)\right\}.
\end{equation} 
Hence, $\phi_d^{\star} = \frac{2\pi(l^{\star}-1)}{L}$ and the optimum entry is $s_{t,d}^{(i)} = e^{j\phi_d^{\star}}$.

\subsection{Convergence}\label{subsec:Convergence}
% As $A(\theta) = a(\theta)a^H(\theta)$, $A(\theta)$ is a positive definite ($A(\theta) \succ 0$), and $\bs_n^H\sum_{k=1}^K\bA(\theta_k)\bs_n > 0$, $\forall \theta_k \in [-\pi,\pi)$, hence,
% \begin{equation*}
% 	\bar{f}(\bS) > 0, \quad \forall \bS \neq 0
% \end{equation*}
% For all $\bS \neq 0$, $|r_{m,l}(k)|^2 \geqslant 0$ and the cross- and auto-correlation sidelobes are positive, hence,
% \begin{equation*}
% 	\tilde{f}(\bS) > 0, \quad \forall \bS \neq 0
% \end{equation*}
The convergence of proposed method can be discussed in two aspects, the convergence of objective function and the convergence of the waveform set $\bS$.
With regard to objective function, as $\bar{f}(\bS) > 0$ and $\tilde{f}(\bS) > 0$, therefore, $f_o(\bS) > 0$, $\forall \bS \neq 0$, and this expression is also valid for the optimum solution of \textbf{Algorithm \ref{alg:waveform_design}} ($f_o(\bS^{\star}) > 0$).

On the other hand, the \textbf{Algorithm \ref{alg:waveform_design}} minimizes the objective function in each step leading to a monotonic decrease of the function value. Since the function value is lower bounded, it can be argued that the algorithm converges to a specific value. Particularly, if the algorithm starts with feasible $\bS^{(0)}$ we have,
\begin{equation*}
	f_o(\bS^{(0)}) \geqslant \dots \geqslant f_o(\bS^{(i)}) \geqslant \dots \geqslant f_o(\bS^{\star}) > 0.
\end{equation*}

Finally, the \gls{MBI} updating rule (greedy), which evaluates the new objective value by updating each entry separately and choosing the best one, ensures the convergence of argument \cite{6563125, 8454321, chen2012maximum} to stationary point. However, the \gls{MBI} selection rule could be costly with large number of variables.
% However, we numerically observed that, applying the \gls{MBI} updating rule has similar performance compared to applying cyclic updating rule in Algorithm \ref{alg:waveform_design} {\color{red} Have you checked this? (similar performance?) please, if you didn't check, remove it or check}. 
% In case of using \gls{MBI} updating rule, the convergence time increases drastically.
In cyclic rule which is considered in this paper, there are three key assumptions in convergence of the argument: (a) separable constraints, (b) differentiable objective, and (c) unique minimizer at each step \cite{Bertsekas/99}. 
% These three conditions are satisfied under $C_1$, $C_2$ and $C_3$ constraints, hence their convergence to a stationary point is guaranteed. In case of $C_4$ constraint, we numerically observed that the problem converges.

In this paper we consider the convergence of objective function and numerically observed that the problem converges under limited energy, \gls{PAR}, continuous and discrete phase constraints.

\subsection{Computational Complexity}\label{subsec:Computational complexity}
In each single variable update, \textbf{Algorithm \ref{alg:waveform_design}} needs to perform the following steps:
\begin{itemize}
  \item \textit{Calculate the coefficient $a_v$, $b_v$ and $c_w$ in \eqref{eq:P_r_phi}}: Calculating $a_v$ and $b_v$ needs $M_t^2N$ operation, while $c_w$ need $M_t^2N\log_2(N)$ due to using fast convolution (see Appendix \ref{app:2} for details). Using a recursive relation, the computational complexity of the coefficients $a_v$ and $b_v$ can be reduced to $M_t^2$ and for $c_w$ can be reduced to $M_t N\log_2(N)$. Typically, in many practical \gls{MIMO} radar systems, $N>>M_t$. Hence, considering the fact that $a_v$ and $b_v$ can be obtained in parallel, the overall computational complexity of calculating the coefficients is ${\cal{O}}(M_t N\log_2(N))$.
  \item \textit{Solve the optimization problem \eqref{eq:P_r_phi}}: 
  %Solving the problem \eqref{eq:P_r_phi} depends on $C_1,\dots,C_4$ constraints. 
  Under $C_1$ and $C_2$ constraints, \textbf{Algorithm \ref{alg:waveform_design}} needs finding the roots of $10$ and $8$ degree polynomials\footnote{
%   Obtaining the roots of $K$ degree polynomial needs $K^3$ operation \cite{1324718}.
For finding the roots of polynomial we use ``roots'' function in MATLAB. This function is based on computing the eigenvalues of the companion matrix. Thus the computational complexity of this method is ${\cal{O}}(k^3)$, where $k$ is the degree of the polynomial \cite{1324718, GoluVanl96}} in \eqref{eq:dfr_root} and \eqref{eq:dfz_root}, which take an order of $10^3$ and $8^3$ operations respectively, while under $C_3$ the algorithm needs finding roots \eqref{eq:dfz_root} and takes an order of $8^3$ operations. In case of $C_4$ constraint we obtain \eqref{eq:f_l} using two $L$-points \gls{FFT} which each has $L\log_2(L)$ operations.
  
  %\underline{Note}: Obtaining the roots of $K$ degree polynomial needs $K^3$ operation \cite{1324718}.
  \item \textit{Optimizing all the entries of matrix $\bS$}: To this end we need to repeat the two aforementioned steps $M_t N$ times.
\end{itemize}
Let us assume that ${\cal{K}}$ iterations are required for convergence of the algorithm. Therefore, the overall computational complexity of 
\textbf{Algorithm \ref{alg:waveform_design}} 
is ${\cal{O}}({\cal{K}}M_t N(10^3 + 8^3 + M_t N\log_2(N)))$ under $C_1$ and $C_2$ constraints, while under $C_3$ is ${\cal{O}}({\cal{K}}M_t N(8^3 + M_t N\log_2(N)))$. In case of $C_4$ the computational complexity is ${\cal{O}}({\cal{K}}M_t N(L\log_2(L) + M_t N\log_2(N)))$.
%  {\color{red} Can we conclude the overall computational complexity is ${\cal{O}}(N^3 M^2)$ and remove your last paragraph? Also, after defining accurately the ``iteration'', we should say considering $\cal{K}$ iterations required for convergence of the algorithm, the overall computational complexity is ${\cal{O}}({\cal{K}}N^3 M^2)$. }

\section{Numerical Results}\label{sec:Numerical Results}
In this section, we provide some representative numerical examples to illustrate the effectiveness of the proposed algorithm. Towards this end, unless otherwise explicitly stated, we consider the following assumptions. For transmit parameters we consider \gls{ULA} configuration with $M_t=8$ transmitters and the antenna distance is set as $d_t = \frac{\lambda}{2}$. We also consider a \gls{ULA} configuration at the receive side with $M_r=8$ antennas. We select the desired and undesired angular regions to be $\Theta_d = [-55^o,-35^o]$ and $\Theta_u = [-90^o,-60^o] \cup [-30^o,90^o]$ respectively. For purpose of simulation, we consider an uniform sampling of these regions  with a  grid size of $5^o$. The stopping condition for \textbf{Algorithm~\ref{alg:waveform_design}} is set at $\zeta=10^{-6}$. 

\subsection{Convergence}\label{subsec:Convergence_Num}
\begin{figure*}
    \centering
    \begin{subfigure}{.24\textwidth}
        \centering
        \includegraphics[width=1\linewidth]{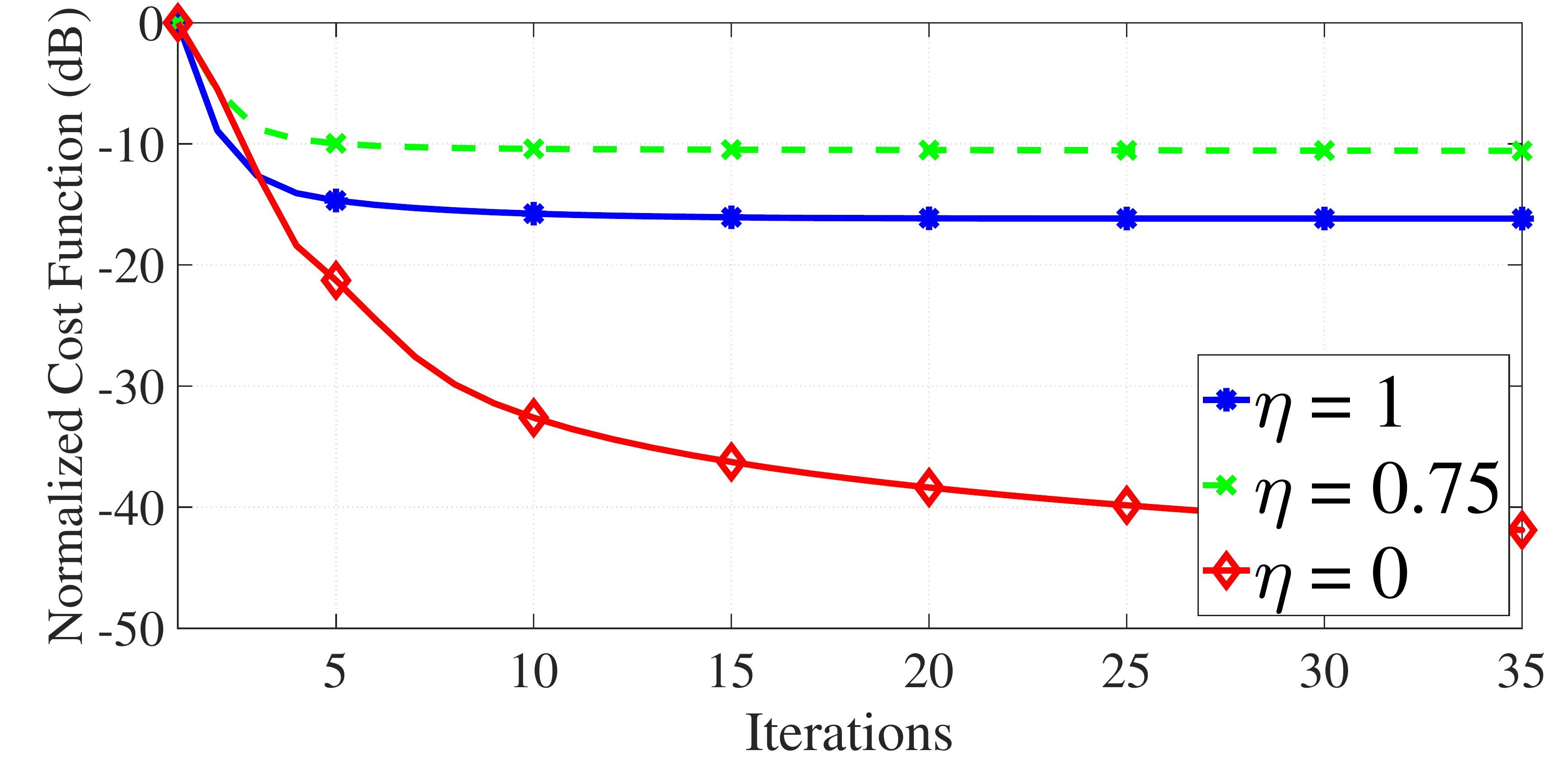}
		\caption[]{$C_1$.}\label{fig:Convergence_C1}
	\end{subfigure}
    \begin{subfigure}{.24\textwidth}
        \centering
        \includegraphics[width=1\linewidth]{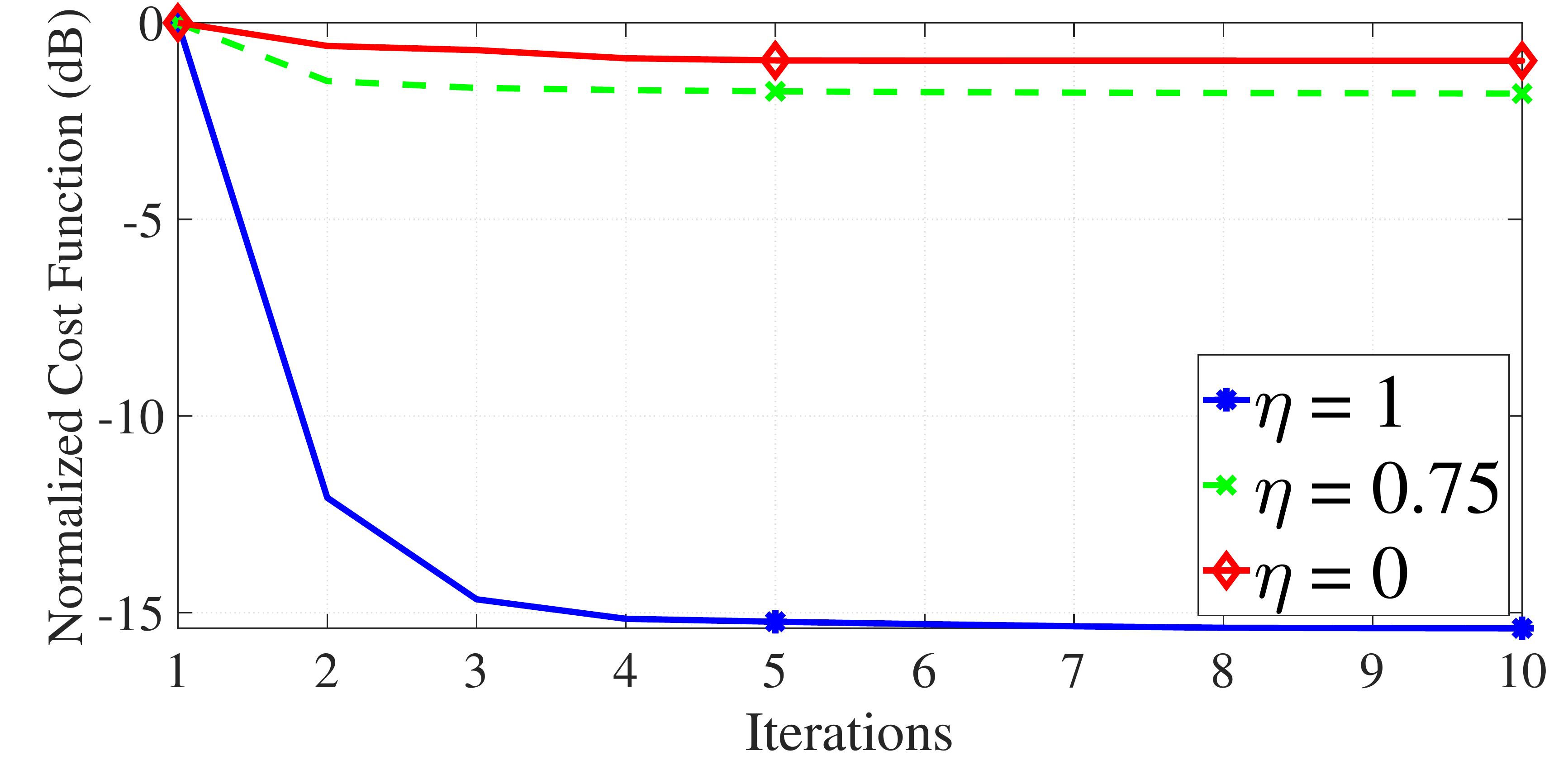}
		\caption[]{$C_2$, $\gamma_p = 1.5 dB$.}\label{fig:Convergence_C2}
    \end{subfigure}
    \begin{subfigure}{.24\textwidth}
        \centering
        \includegraphics[width=1\linewidth]{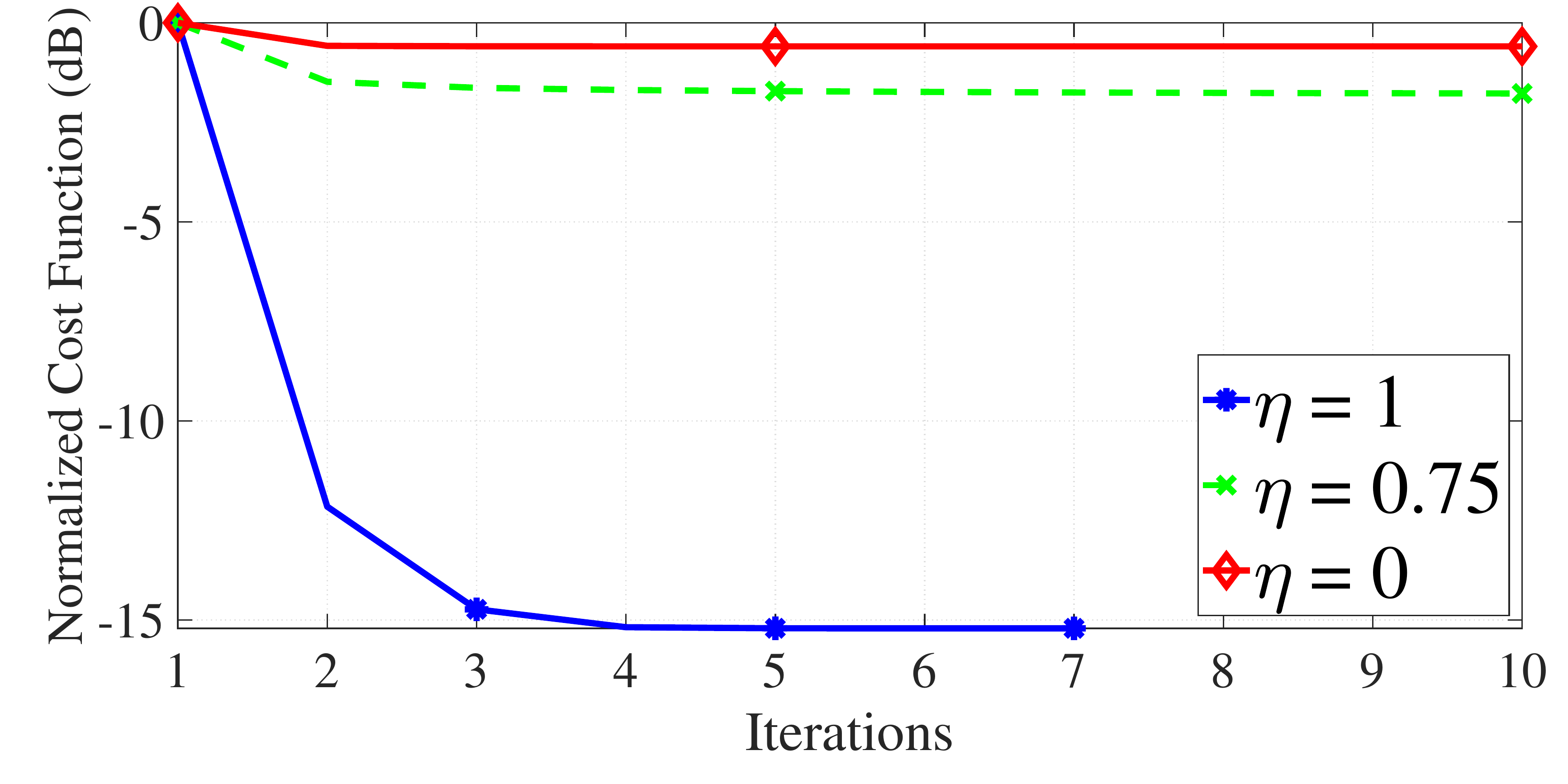}
		\caption[]{$C_3$.}\label{fig:Convergence_C3}
    \end{subfigure}
    \begin{subfigure}{.24\textwidth}
        \centering
        \includegraphics[width=1\linewidth]{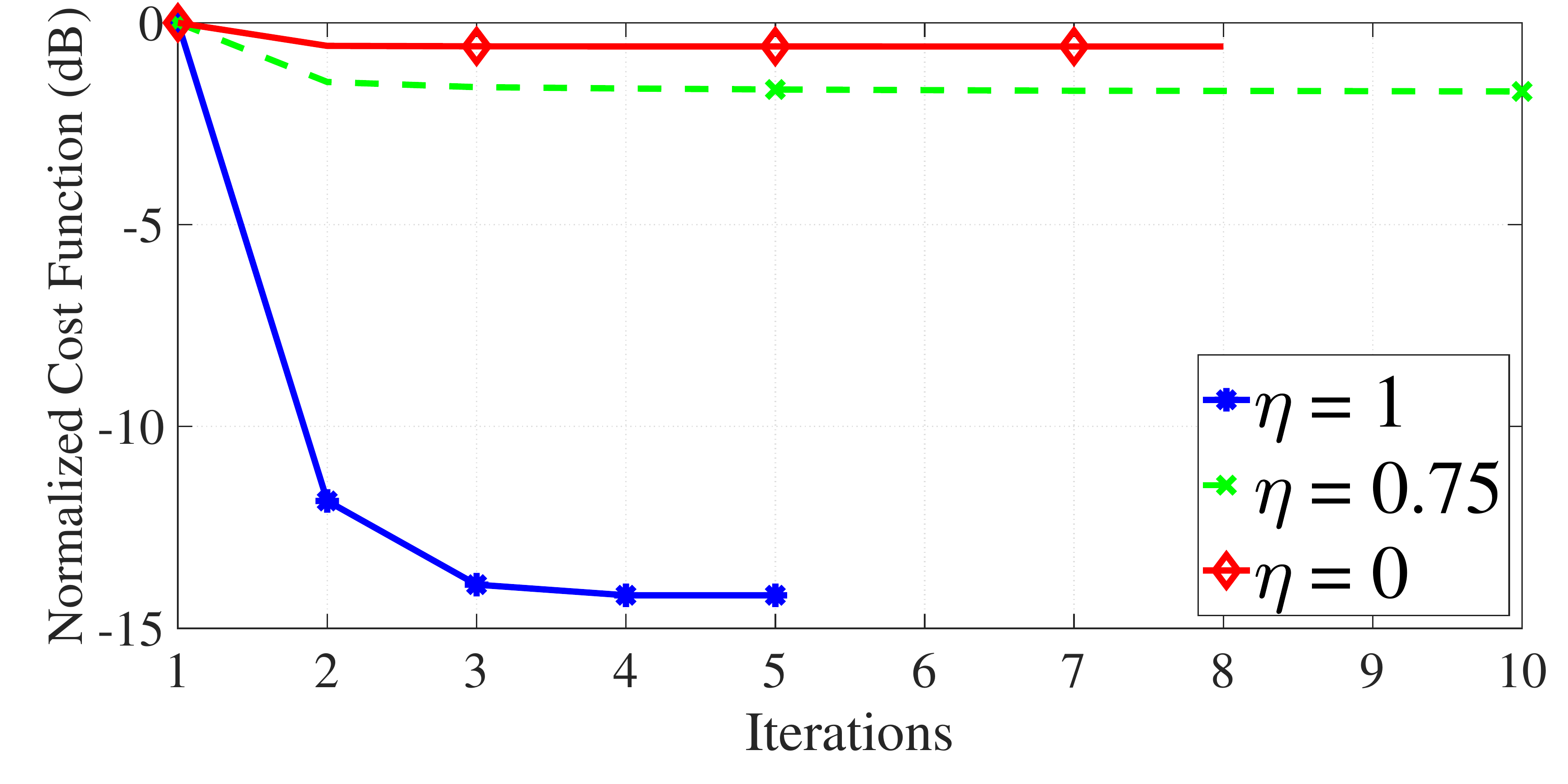}
		\caption[]{$C_4$, $L = 8$.}\label{fig:Convergence_C4}
    \end{subfigure}
    \caption[]{Convergence behavior of the proposed algorithm for different constraint and values of $\eta$} ($M_t=8$, $N=64$).\label{fig:Convergence_C1C2C3C4}
\end{figure*}

\figurename{~\ref{fig:Convergence_C1C2C3C4}} depicts the convergence behavior of the proposed algorithm under $C_1$, $C_2$, $C_3$, and $C_4$ constraints under different scalarization coefficients $\eta$. 
% {\color{red}Is it correct? Is it a random-phase sequence? I guess you should define a $\phi_i$ then say that every $\phi_i$ is a random variable that is identically independent and can be chosen from a set which every entry of that set is adopted from the following discrete phase set (something like this). right now, the definition is incorrect!\\
% In this figure, the initial waveform $\bS_0 \in \mathbb{C}^{M \times N}$ is a set of random \gls{MPSK} sequences with alphabet size $L=8$. Here, every code entry is given by,
% \begin{equation}\label{eq:S_0}
% 	s_{m,n}^{(0)} = e^{j\frac{2\pi(l-1)}{L}},
% \end{equation} 
% with  $l \in \{1,\dots,L\}$. } 
Since MPSK sequences are feasible for the all constraints, we consider a set of random \gls{MPSK} sequences ($\bS_0 \in \mathbb{C}^{M_t \times N}$) with alphabet size $L=8$ as an initial waveform. Here, every code entry is given by,
\begin{equation}\label{eq:S_0}
	s_{m,n}^{(0)} = e^{j\frac{2\pi(l-1)}{L}},
\end{equation} 
where $l$ is the random integer variable uniformly distributed in $[1,L]$.
According to \figurename{~\ref{fig:Convergence_C1C2C3C4}}, the objective function decreases monotonically for all values of $\eta$ and for all the constraints. Furthermore, for any $\eta$, the performance ordering of limited power, \gls{PAR}, continuous and discrete phase can be predicted from the relation $	C_4 \subset C_3 \subset C_2 \subset C_1$.
% Furthermore, \figurename{~\ref{fig:Convergence_different_eta}} shows the convergence behavior of proposed method with a fixed $\eta = 1$. As can be seen among $C_1, \dots, C_4$ constraints, limited energy, \gls{PAR}, continuous and discrete phase respectively has the best performance in terms of the objective function. This behavior was predictable, because the problem under $C_1, \dots, C_4$ constraints are sorted from the largest to the smallest feasible set.
% \begin{figure}
% 	\centering
% 	\includegraphics[width=1\linewidth]{Nrm_Convergence_eta_1.eps}
% 	\caption[]{Convergence behavior of proposed method ($\eta = 1$).}\label{fig:Convergence_different_eta}
% \end{figure}

\subsection{Trade-off between spatial- and range-\gls{ISLR}}\label{subsec:Trade-off}
In this part we first assess the contradiction in waveform design for beampattern shaping and orthogonality; subsequently, we show the importance of making a trade-off between spatial- and range-\gls{ISLR} to obtain a better performance. 
\subsubsection{Relation between Beampattern Shaping and Orthogonality}
\begin{figure*}
    \centering
    \begin{subfigure}{.24\textwidth}
        \centering
        \includegraphics[width=1\linewidth]{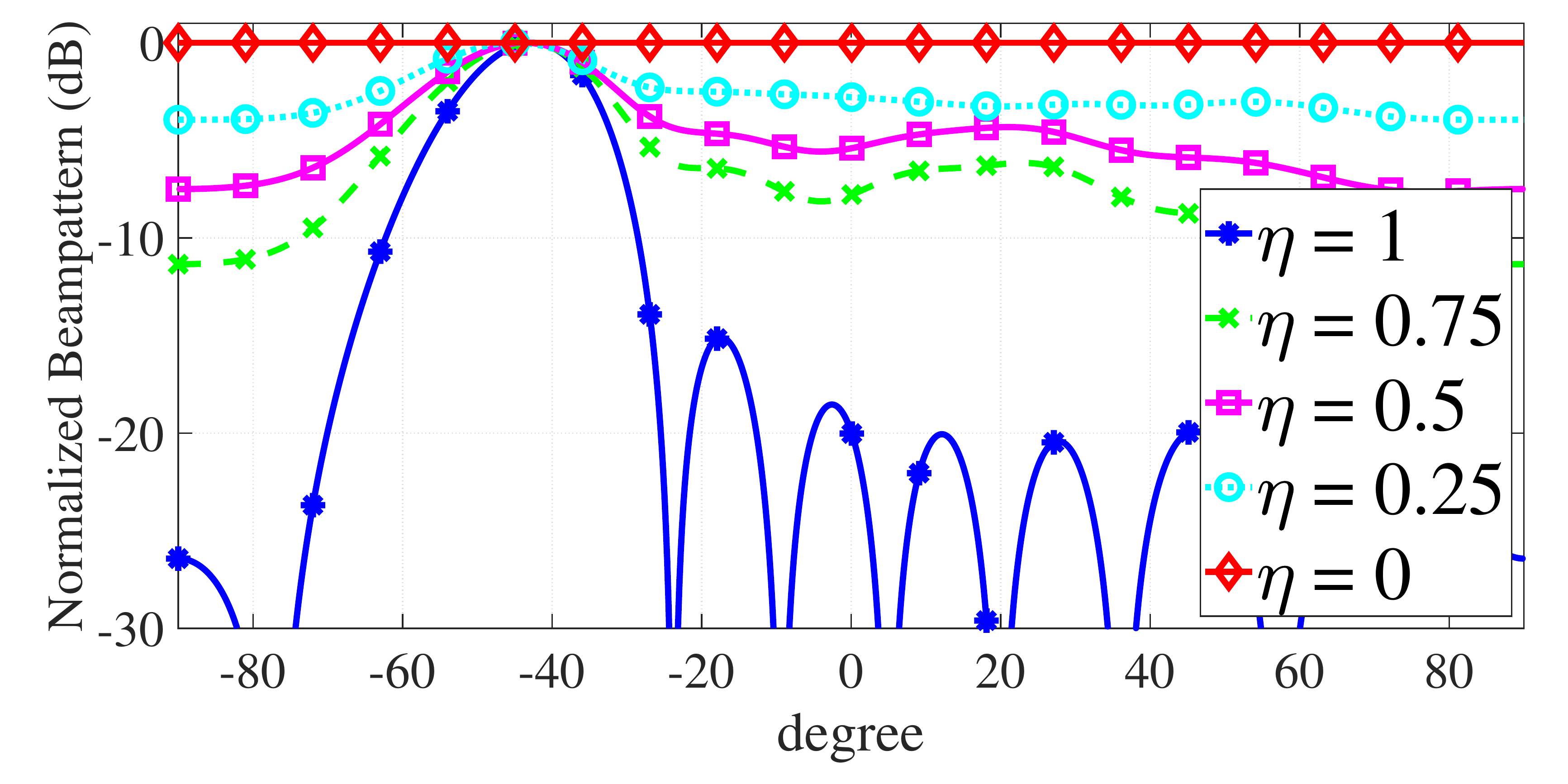}
		\caption[]{$C_1$.}\label{fig:BeamPatternE_vs_eta}
    \end{subfigure}
    \begin{subfigure}{.24\textwidth}
        \centering
        \includegraphics[width=1\linewidth]{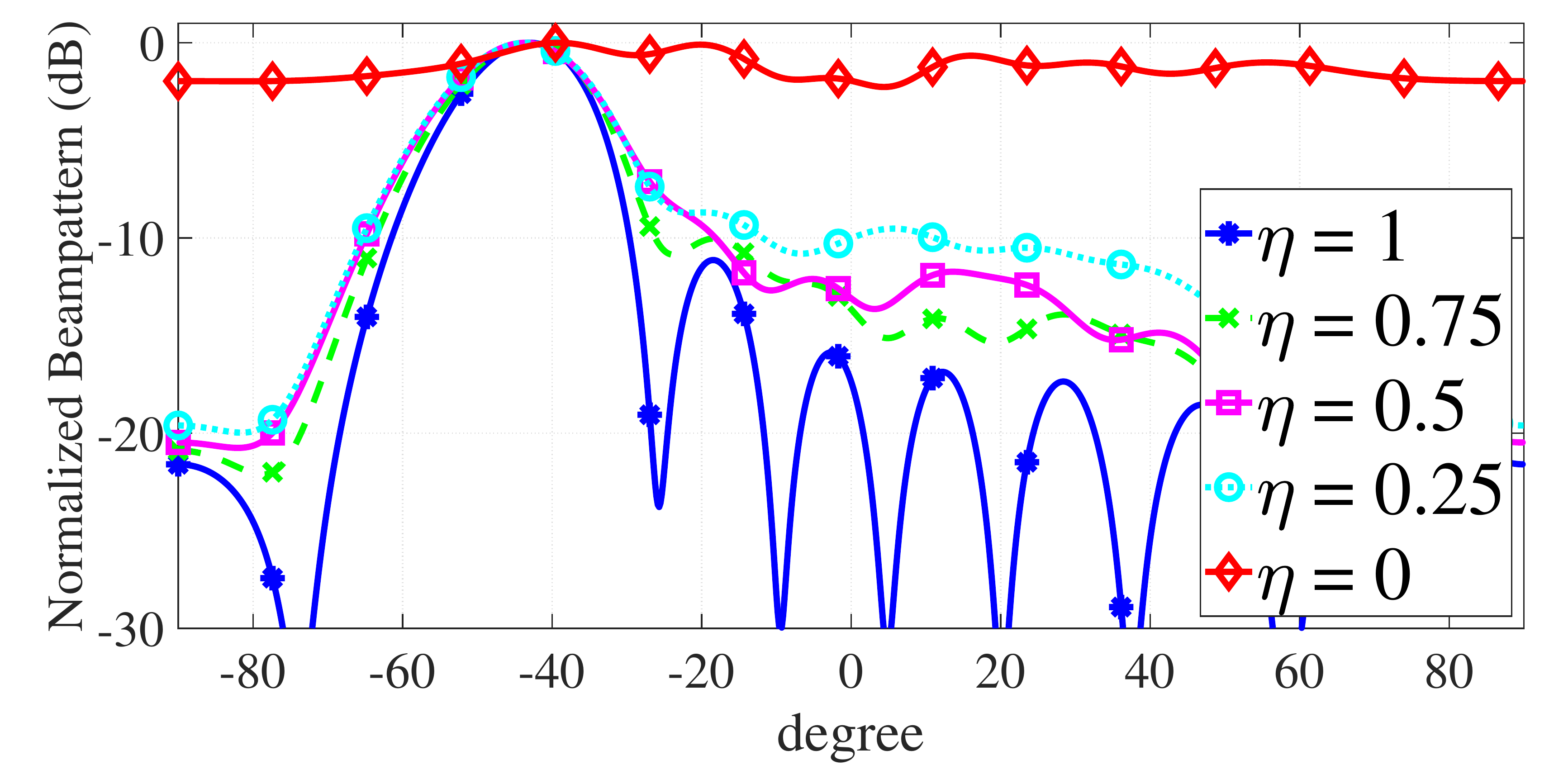}
		\caption[]{$C_2$, $\gamma_p = 1.5 dB$.}\label{fig:BeamPatternP_vs_eta}
    \end{subfigure}
    \begin{subfigure}{.24\textwidth}
        \centering
        \includegraphics[width=1\linewidth]{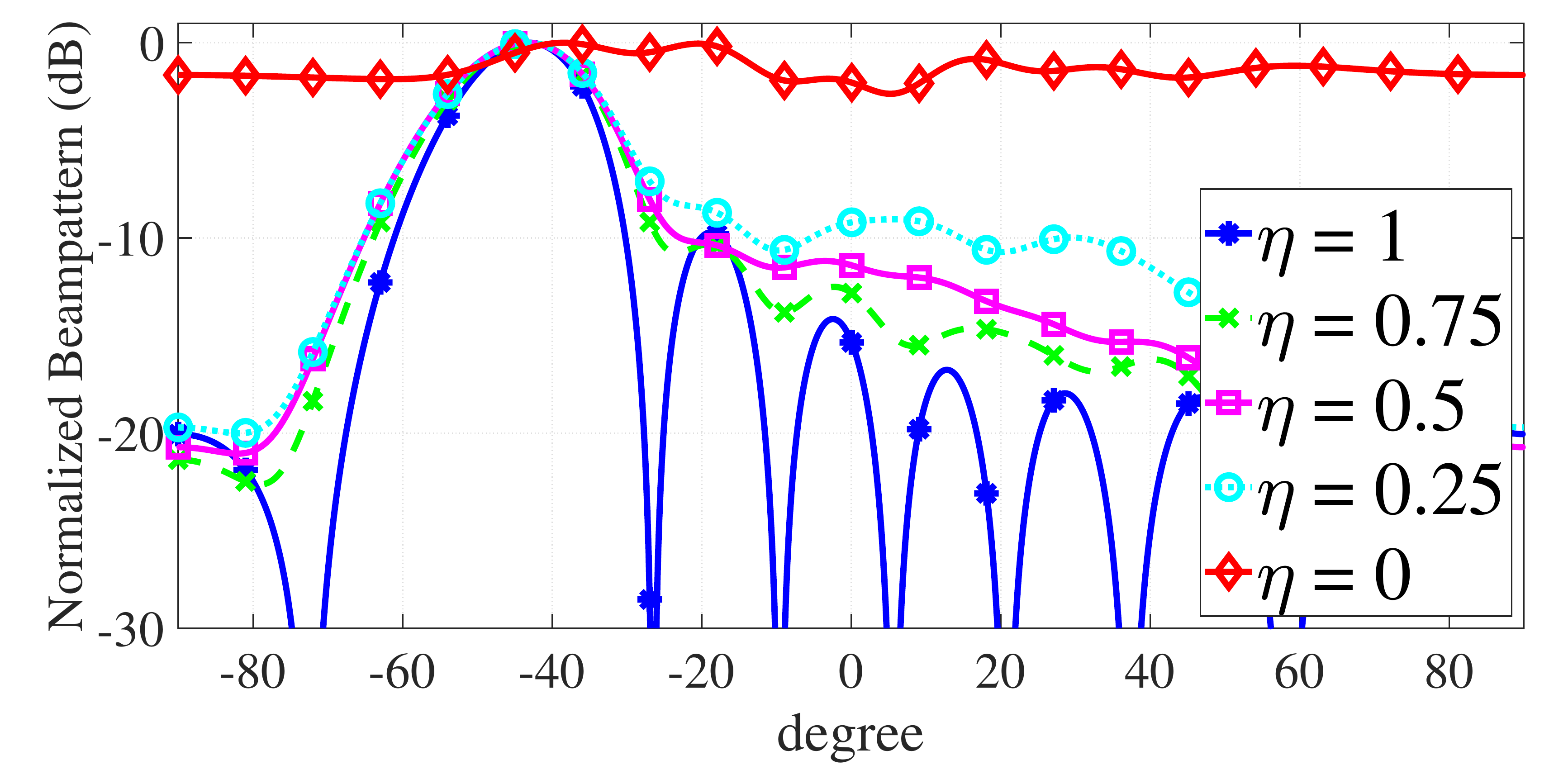}
		\caption[]{$C_3$.}\label{fig:BeamPatternC_vs_eta}
    \end{subfigure}
    \begin{subfigure}{.24\textwidth}
        \centering
        \includegraphics[width=1\linewidth]{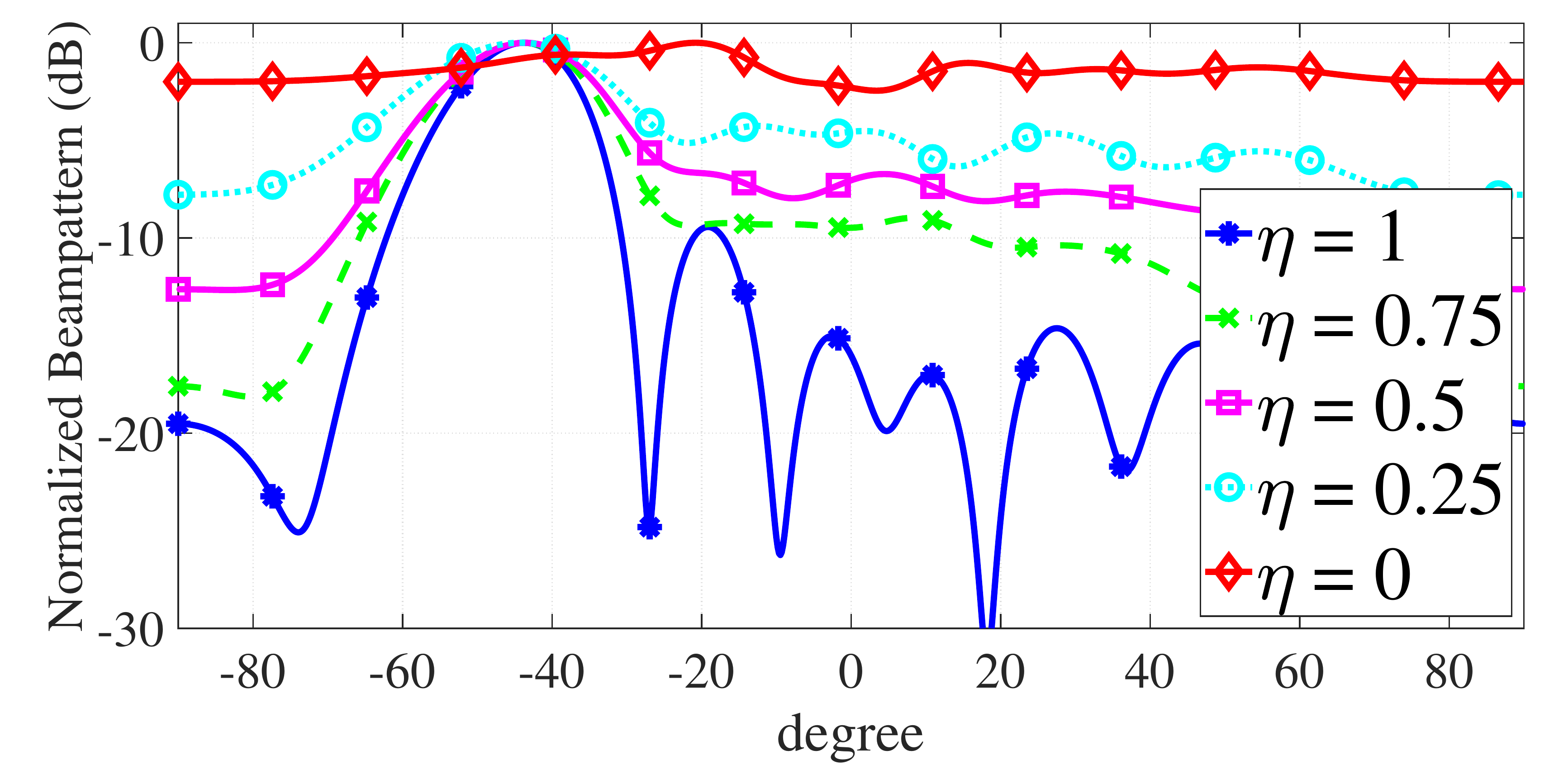}
		\caption[]{$C_4$, $L = 8$.}\label{fig:BeamPatternD_vs_eta}
    \end{subfigure}
    % \caption[]{Transmit beampattern of the optimized set of sequences ($M_t=8$, $N=64$, $\Theta_d = [-55^o,-35^o]$ and $\Theta_u = [-90^o,-60^o] \cup [-30^o,90^o]$).}\label{fig:BeamPattern_vs_eta}
    \caption[]{Transmit beampattern under different constraint and value of $\eta$ ($M_t=8$, $N=64$, $\Theta_d = [-55^o,-35^o]$ and $\Theta_u = [-90^o,-60^o] \cup [-30^o,90^o]$).}\label{fig:BeamPattern_vs_eta}
\end{figure*}

\figurename{~\ref{fig:BeamPattern_vs_eta}} shows the beampattern of the proposed algorithm under $C_1, \dots, C_4$ constraints with different values of $\eta$. Setting $\eta = 0$  results in an almost omni directional beam. By increasing $\eta$, radiation pattern takes the shape of a beam with $\eta=1$ offering the optimized pattern. 
%the $\eta$ value, the beampattern gradually be formed and $\eta = 1$ obtains the best beampattern.  
%
% \begin{comment}
% through the following equation,
% \begin{equation}\label{eq:S_me}
% 	\tilde{\bs}_{me} \triangleq \arg\max_{\tilde{\bs}_m} \{ \norm{\tilde{\bs}_m} | m=\{1,\dots,M\}\}.
% \end{equation} 
% Since the energy of signals in constant modulus are equal, thus to obtain $\tilde{\bs}_{me}$ under $C_3$ and $C_4$ we can choose randomly one through the sequences.

% Henceforth we assume that $\tilde{\bs}_{me}$ is the $4^{th}$ sequence of the transmitted waveform.
% \end{comment}
%

% On the other hand, \tablename{~\ref{tab:CC_vs_eta}} shows a three-dimensional representation of the amplitude of correlation of a particular sequence, $\tilde{\bs}_{me}$, with the other waveforms in the optimized set $\bS^{\star}$. Here $\tilde{\bs}_{me}$ denotes the sequence in the optimized set $\bS^{\star}$ having the maximum power\footnote {Note that, $\tilde{\bs}_{me}$ can be randomly chosen under $C_3$, and $C_4$ constraints, but will be obtained by 
% $
% 	\tilde{\bs}_{me} \triangleq \arg\max_{\tilde{\bs}_m} \{ \norm{\tilde{\bs}_m} | m=\{1,\dots,M_t\}\}
% $,
% under $C_1$ or $C_2$. In addition we assume that $\tilde{\bs}_{me}$ is the $4^{th}$ sequence of $\bS^{\star}$},
% %
% The $4^{th}$ sequence shows the auto-correlation of the waveform. While $\eta = 1$ (first row in \tablename{~\ref{tab:CC_vs_eta}}), yields an optimized beam pattern, the correlation between $\tilde{\bs}_{me}$ with other sequences is rather large in all cases. 

On the other hand, \tablename{~\ref{tab:CC_vs_eta}} shows a three-dimensional representation of the amplitude of correlation of a particular sequence with the other waveforms in the optimized set $\bS^{\star}$\footnote { In order to plot the auto- and cross-correlation, we first sort the optimized waveforms based on their energy, then we move the waveform which has the maximum energy to the middle of the waveform set (at $[\frac{M_t}{2}]$). By this rearrangement, the peak of auto-correlation will always be located at the middle.}.
The $4^{th}$ sequence shows the auto-correlation of that particular waveform. With $\eta = 1$ (first row in \tablename{~\ref{tab:CC_vs_eta}}), yields an optimized beampattern, the cross-correlation with other sequences is rather large in all cases. 

This shows the transmission of scaled waveforms (phase-shifted) from all antennas, similar to traditional phased array. In this case, it would  not be possible to separate the transmit signals at the receiver (by matched filter) and the \gls{MIMO} virtual array will not be formed, thereby losing in the angular resolution. When $\eta = 0$ (last row in \tablename{~\ref{tab:CC_vs_eta}}), an orthogonal set of sequences is obtained as their cross-terms (auto- and cross-correlation lags) are small under different design constraints. The resulting omnidirectional beampattern (see \figurename{~\ref{fig:BeamPattern_vs_eta}}), however, prevents steering of the transmit power towards the desired angles, while a strong signal from the undesired directions may saturate the radar receiver. 
%Further, the energy of the radar transmitter maybe emitted on the directions which are unnecessary to be searched. 
The middle row in \tablename{~\ref{tab:CC_vs_eta}}, depicts $\eta=0.5$, a case when partially orthogonal waveforms are adopted, while some degree of transmit beampattern shaping can still be obtained (see \figurename{~\ref{fig:BeamPattern_vs_eta}}). 

\figurename{~\ref{fig:BeamPattern_vs_eta}} and \tablename{~\ref{tab:CC_vs_eta}} show that, having simultaneous beampattern shaping and orthogonality are contradictory, and the choice of $\eta$ effects a trade-off between the two and enhance the performance of radar system. This is explored next.
% 
%the auto- and cross-correlations of $\tilde{\bs}_{me}$ under $C_1, \dots, C_4$ constraints with different value of $\eta$. As can be seen, when $\eta = 0$, the auto- and cross-correlation are ideal and by increasing $\eta$ the auto- and cross-correlation gradually become worse and with $\eta = 1$ the waveform is fully correlated. 

\begin{table*}
	\centering
% 	\caption[]{Three-dimensional representation of the correlation of $\tilde{\bs}_{me}$ with other sequences in the optimized set ($M_t=8$, $N=64$).}\label{fig:CC_vs_eta}
\caption[]{Three-dimensional representation of the auto- and cross-correlation of proposed method ($M_t=8$,  $N=1024$).}\label{fig:CC_vs_eta}
	\begin{tabular}{c|c|c|c|c}	
		\hline
		\hline
		$\eta$ & $C_1$ & $C_2$, $\gamma_p = 1.5 dB$ & $C_3$ & $C_4$ ($L=8$) \\
		\hline
		1 & 
		\begin{minipage}{.215\textwidth}
            \includegraphics[width=\linewidth]{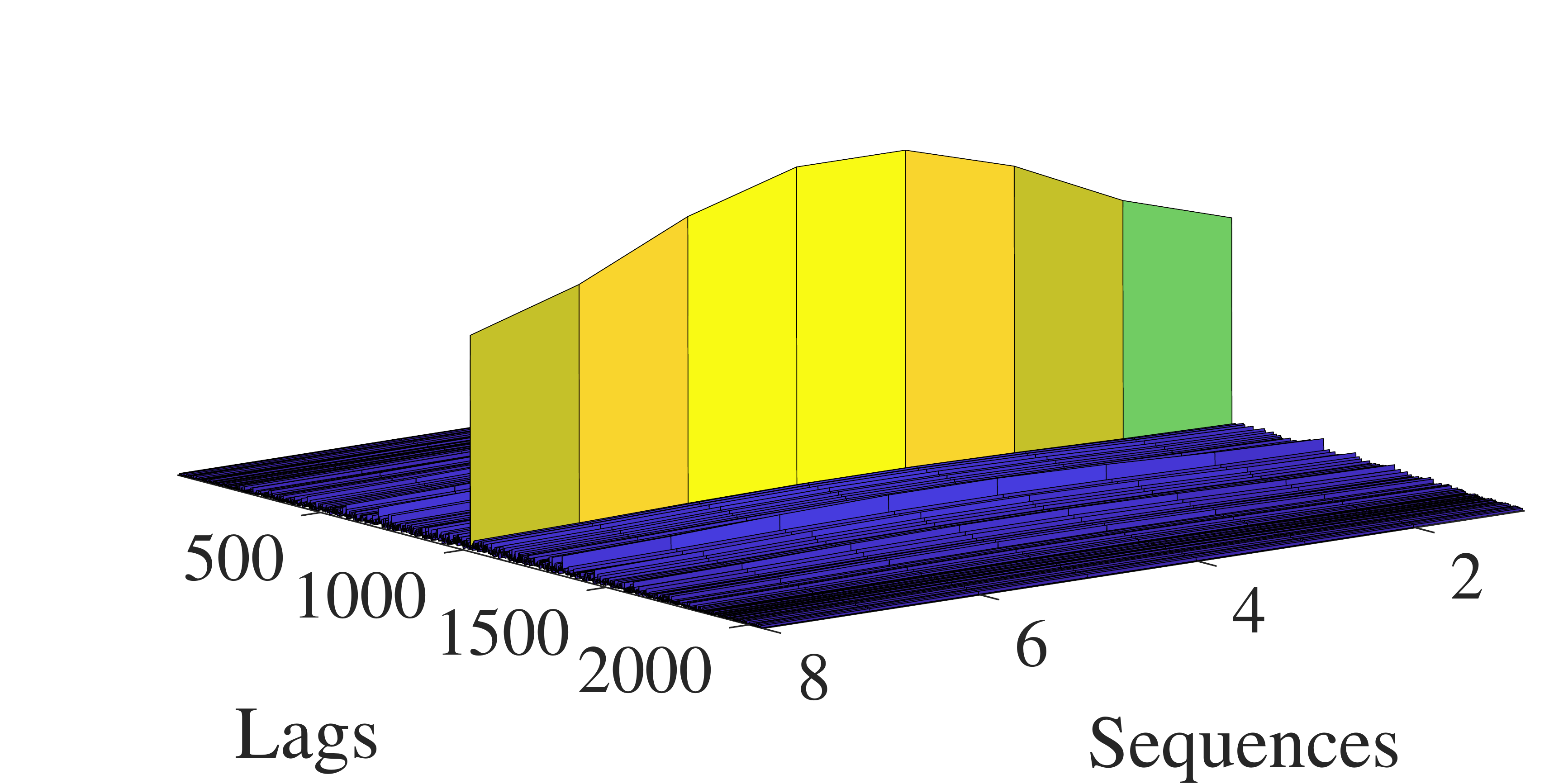}
        \end{minipage}
        & 
        \begin{minipage}{.215\textwidth}
            \includegraphics[width=\linewidth]{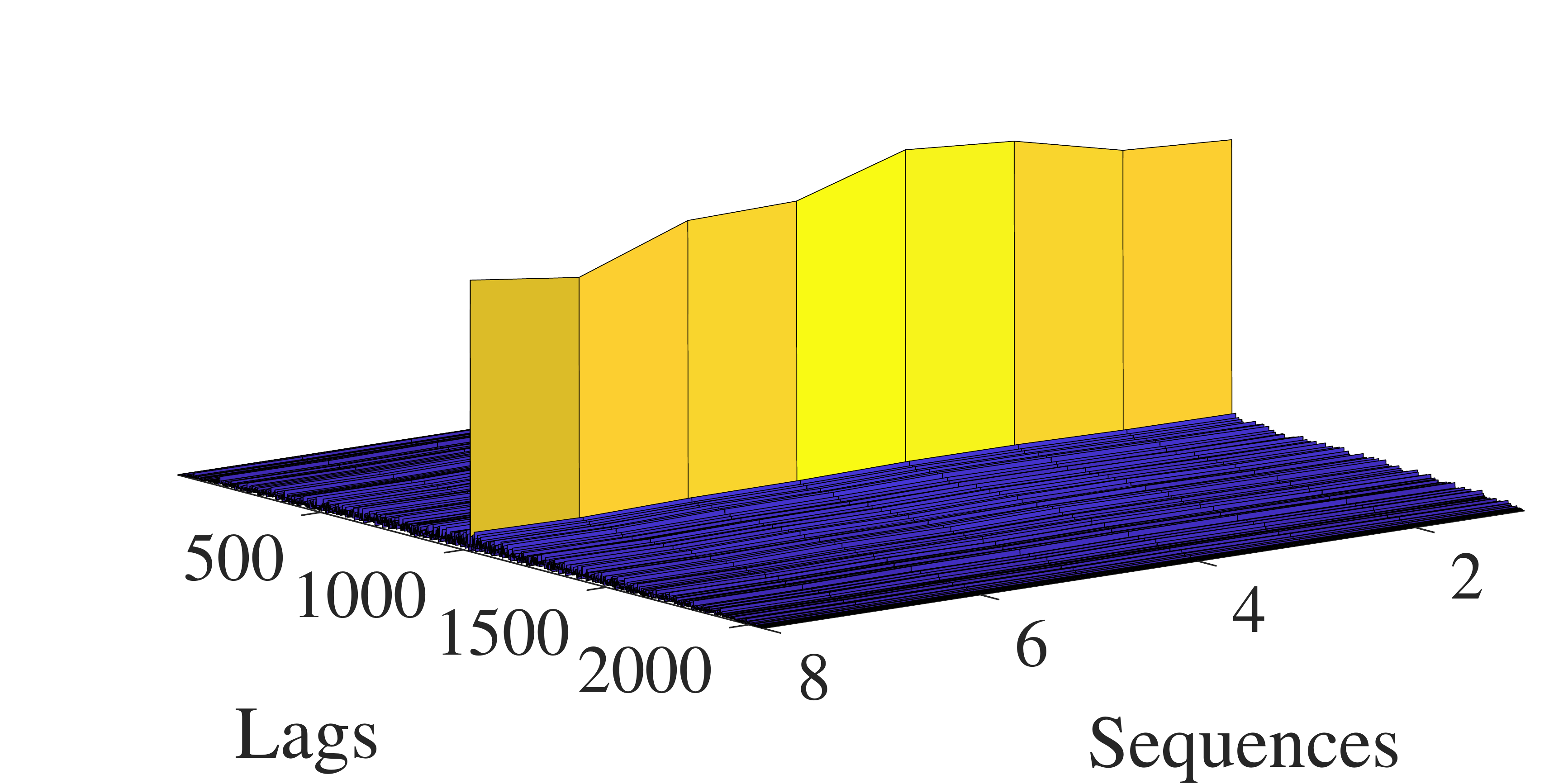}
        \end{minipage}
        & 
        \begin{minipage}{.215\textwidth}
            \includegraphics[width=\linewidth]{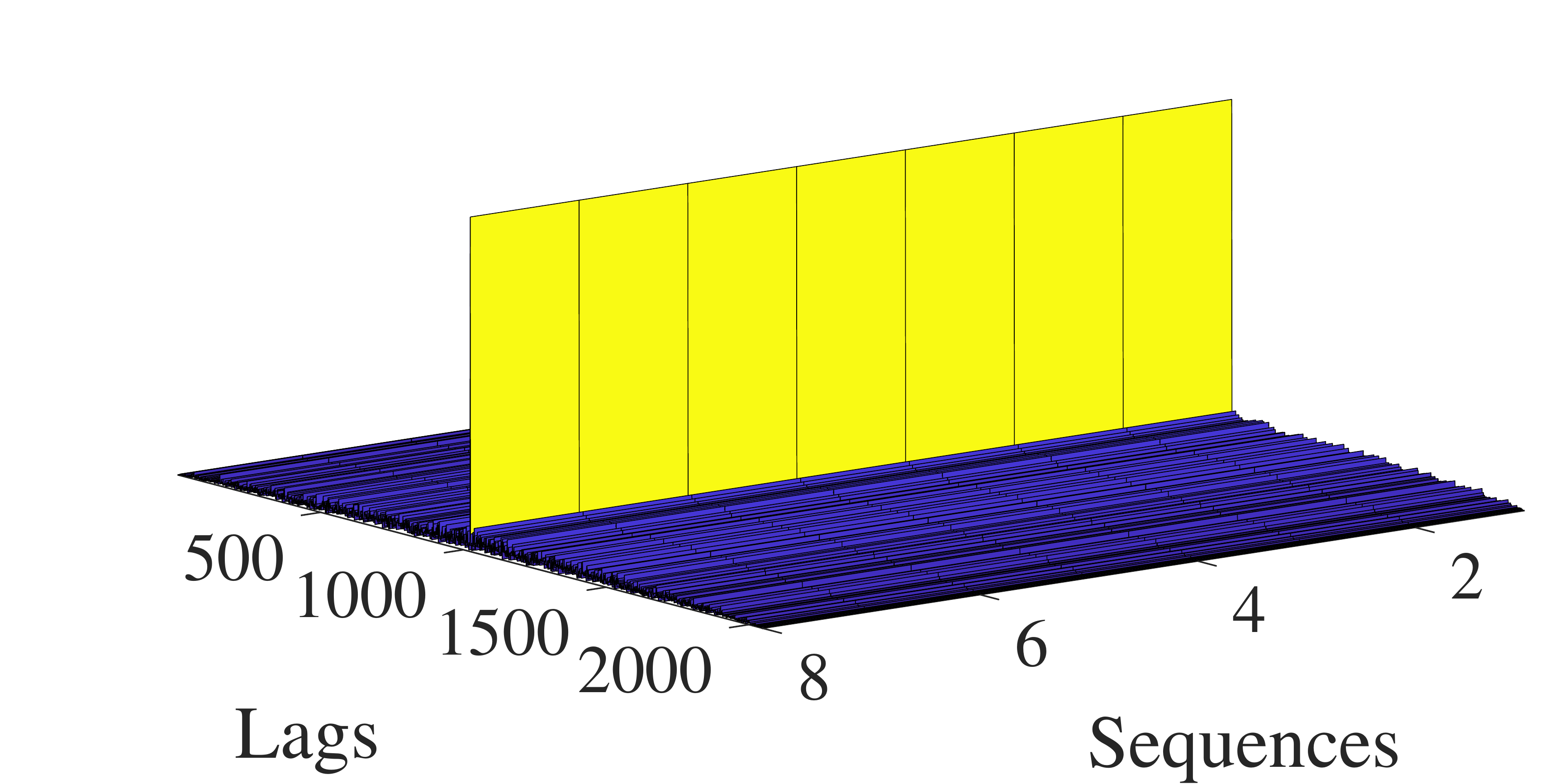}
        \end{minipage}
        & 
        \begin{minipage}{.215\textwidth}
            \includegraphics[width=\linewidth]{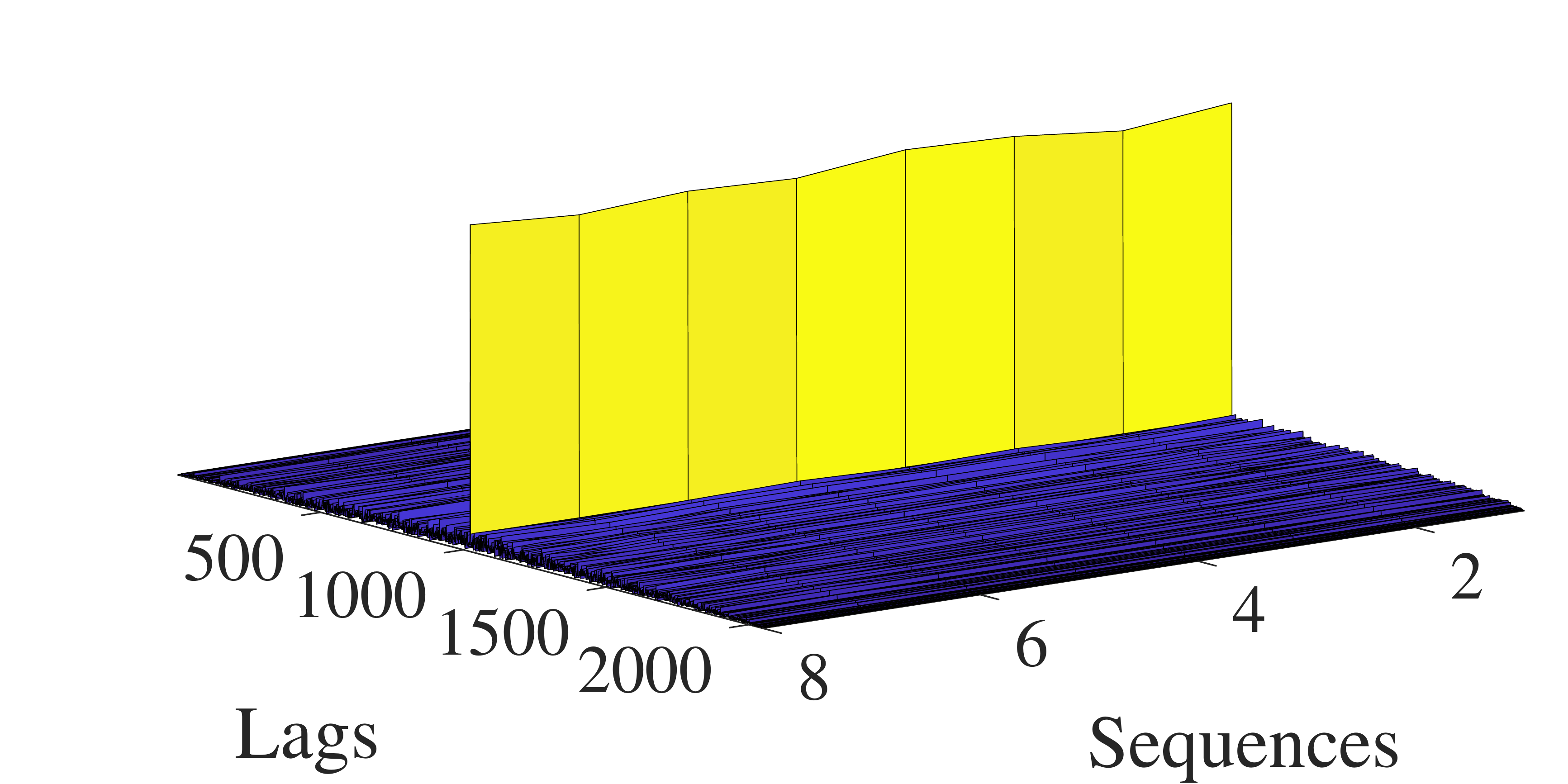}
        \end{minipage}
        \\
		0.5 
		& 
        \begin{minipage}{.215\textwidth}
            \includegraphics[width=1\linewidth]{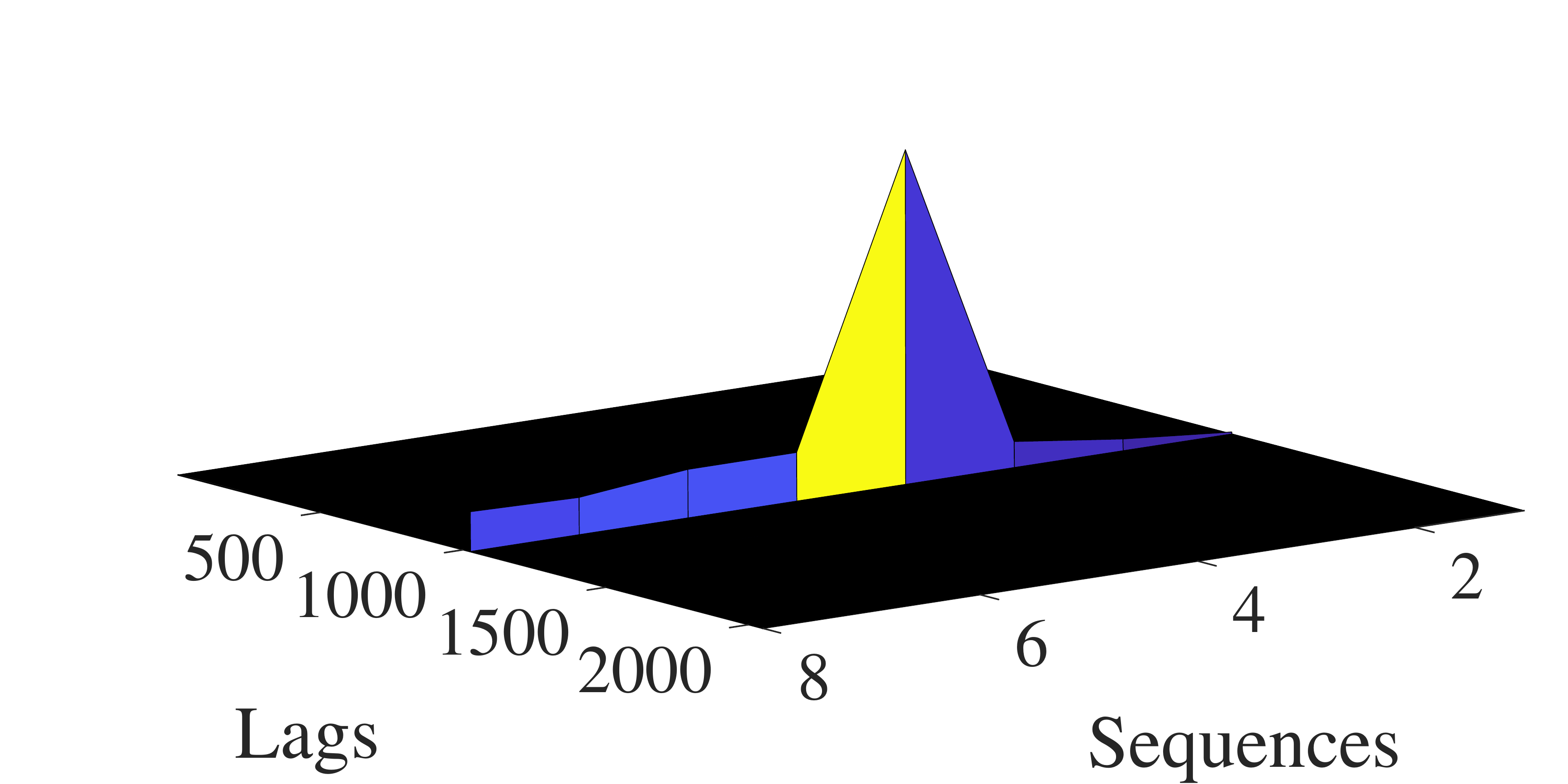}
        \end{minipage}
        &  
        \begin{minipage}{.215\textwidth}
            \includegraphics[width=1\linewidth]{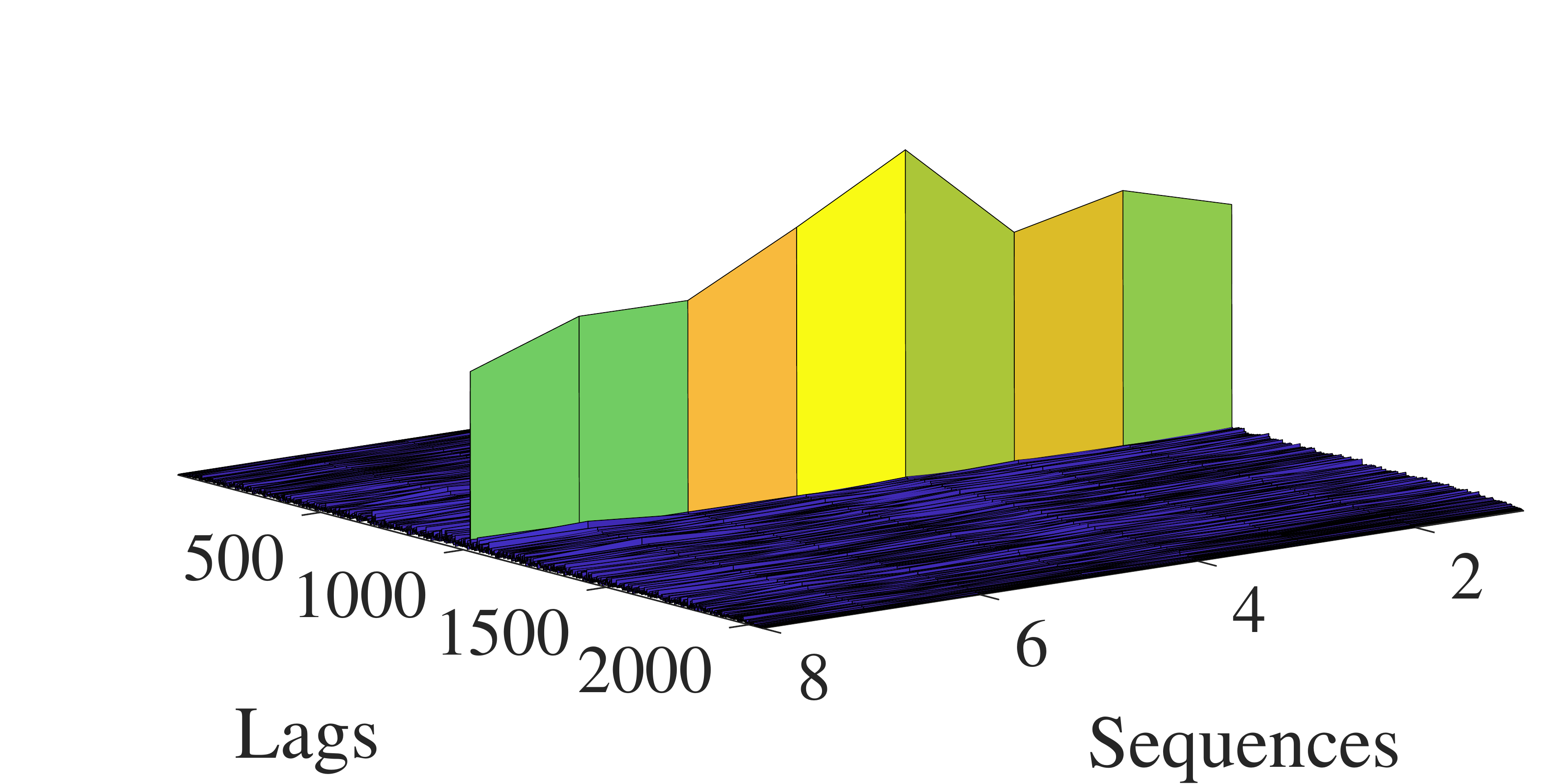}
        \end{minipage}
        & 
        \begin{minipage}{.215\textwidth}
            \includegraphics[width=1\linewidth]{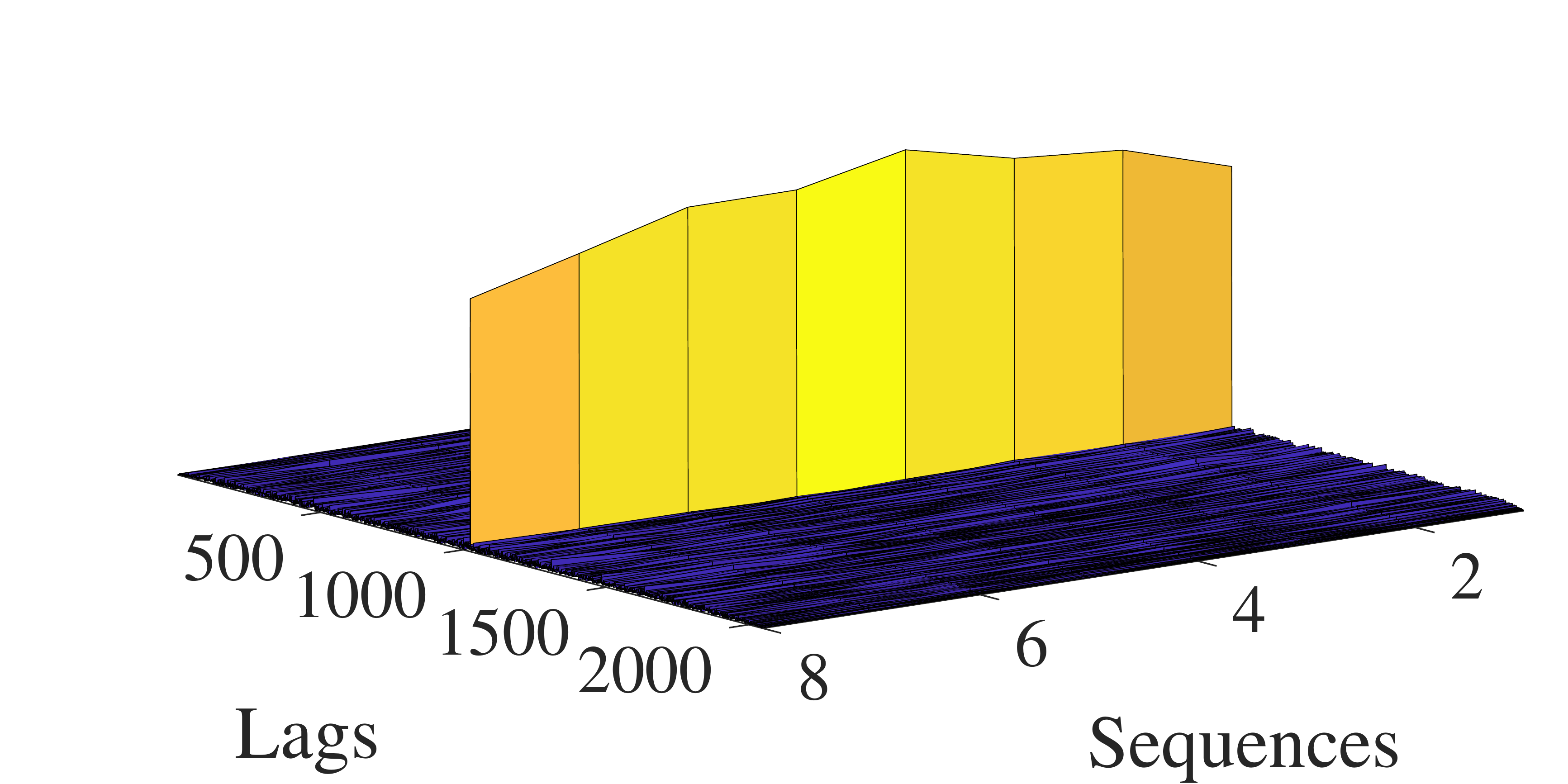}
        \end{minipage}
        & 
        \begin{minipage}{.215\textwidth}
            \includegraphics[width=1\linewidth]{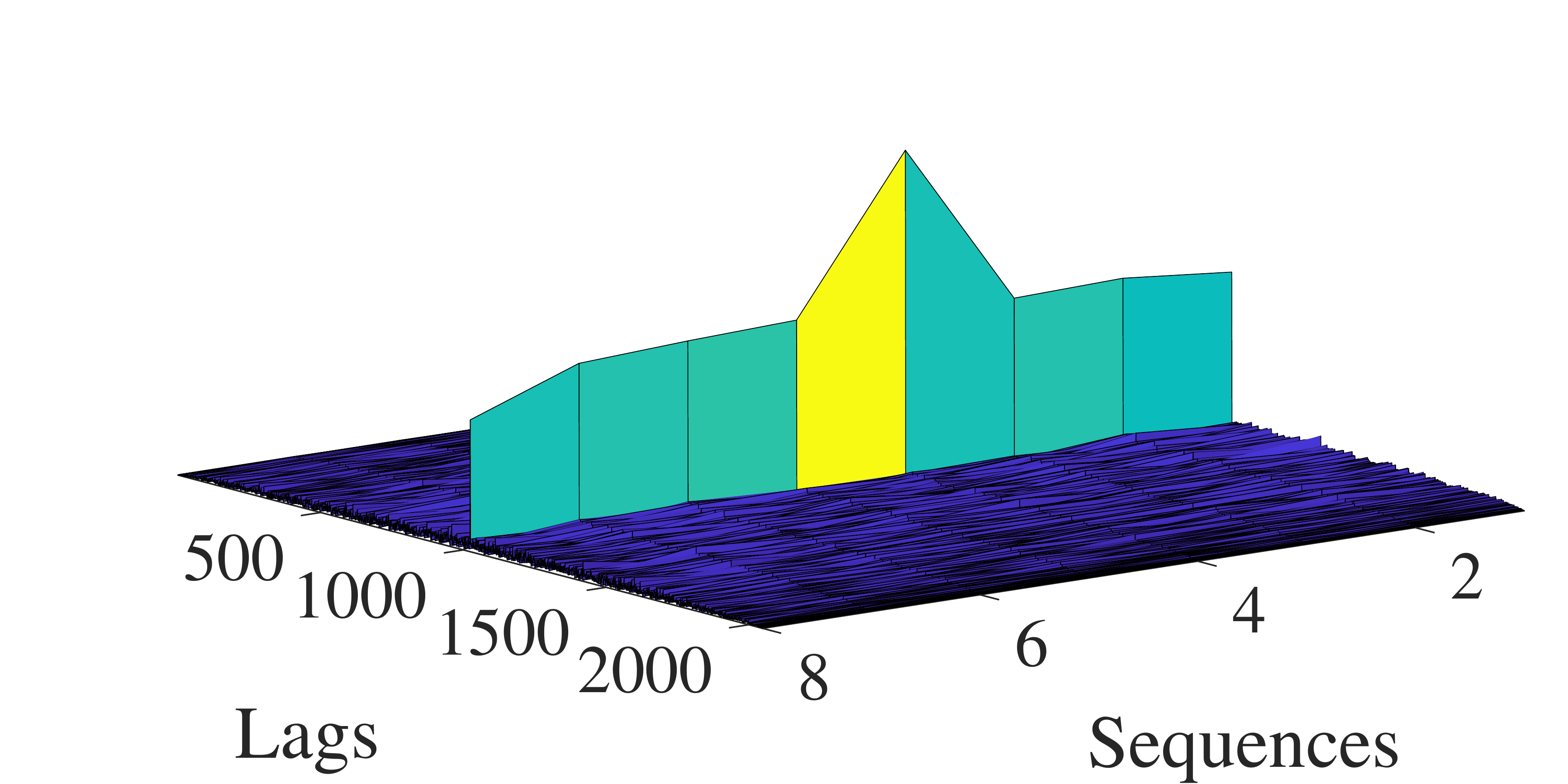}
        \end{minipage}
        \\
		0 
		& 
        \begin{minipage}{.215\textwidth}
            \includegraphics[width=1\linewidth]{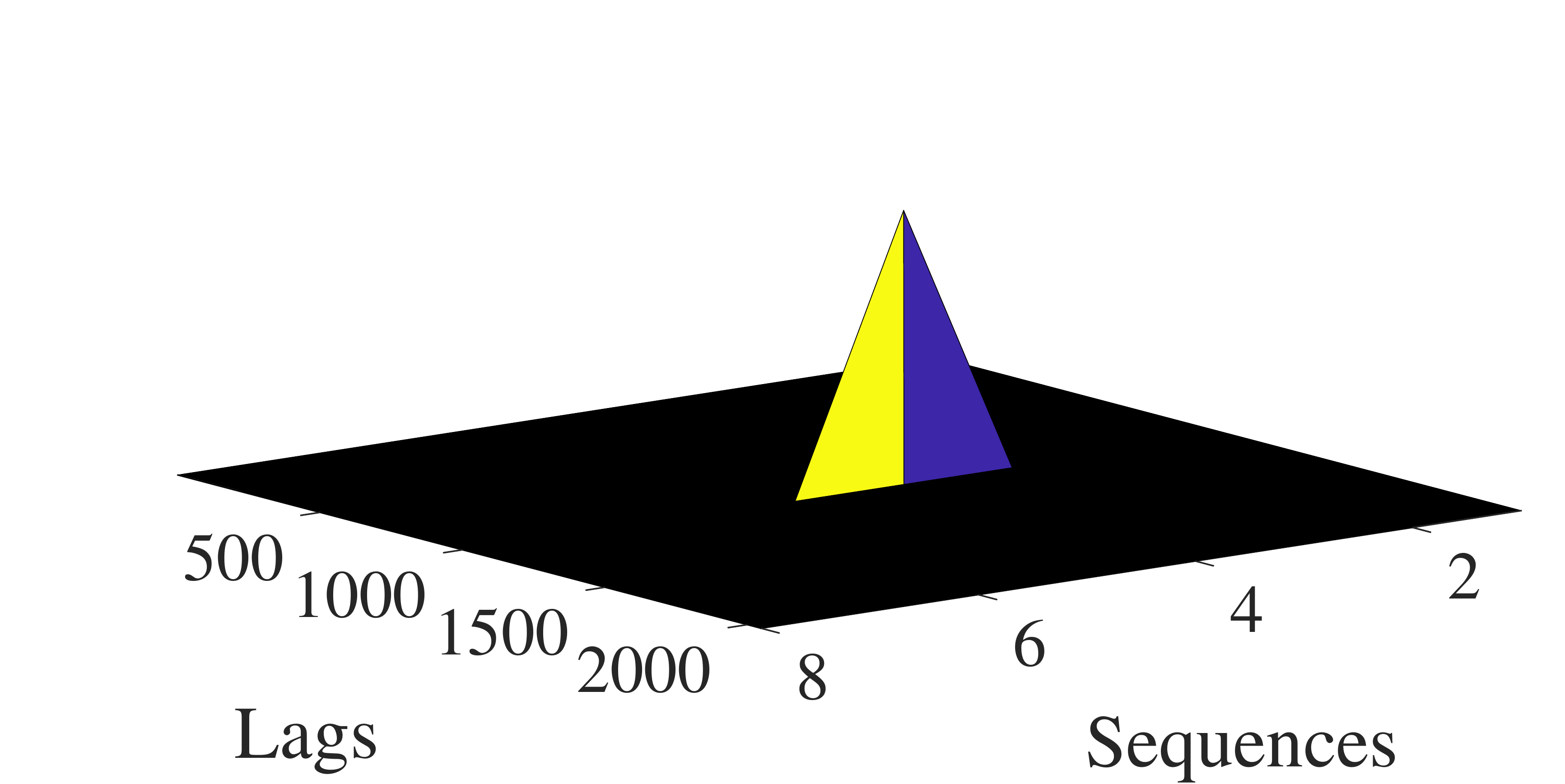}
        \end{minipage}
        & 
        \begin{minipage}{.215\textwidth}
            \includegraphics[width=1\linewidth]{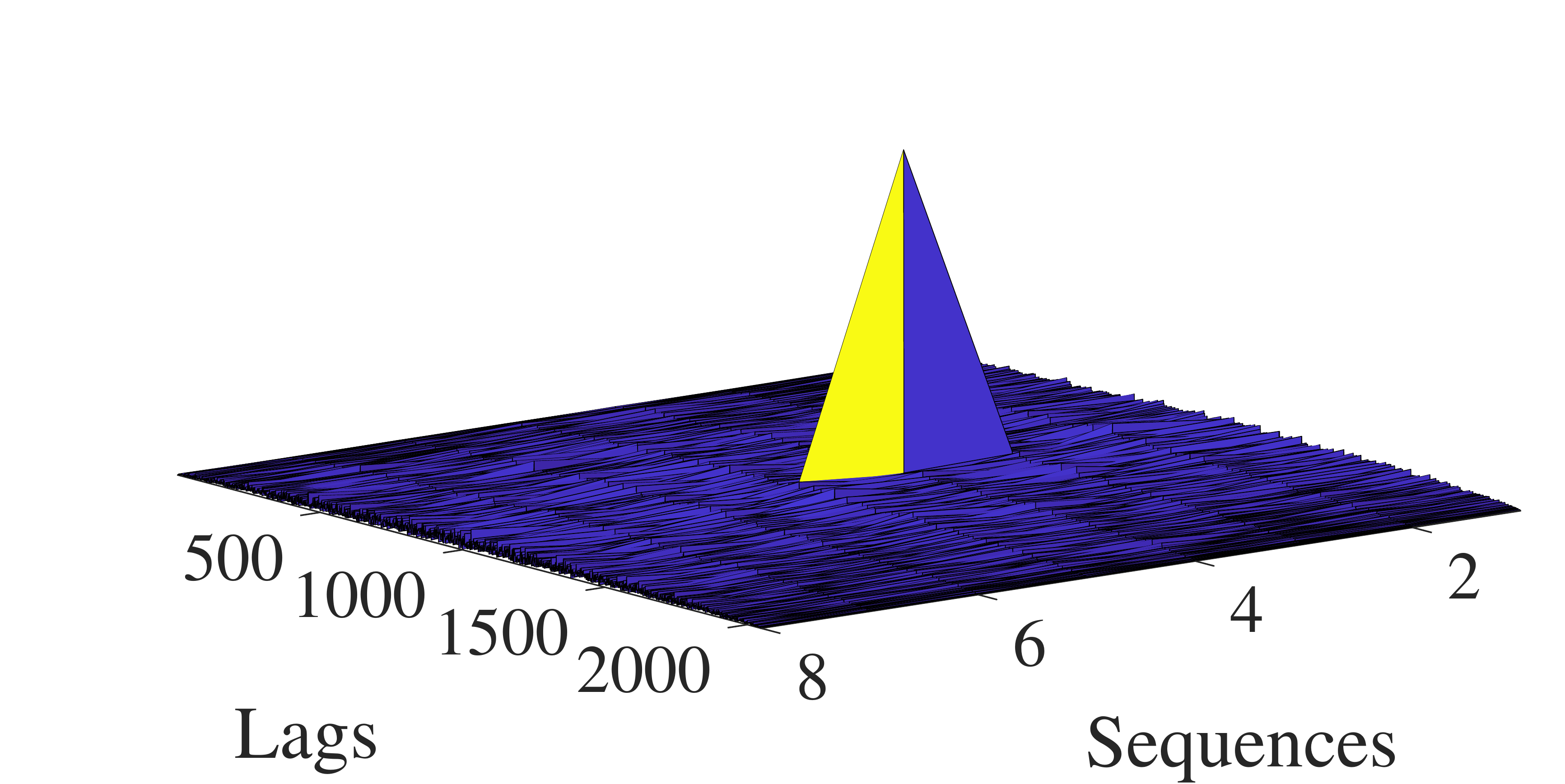}
        \end{minipage}
        & 
        \begin{minipage}{.215\textwidth}
            \includegraphics[width=1\linewidth]{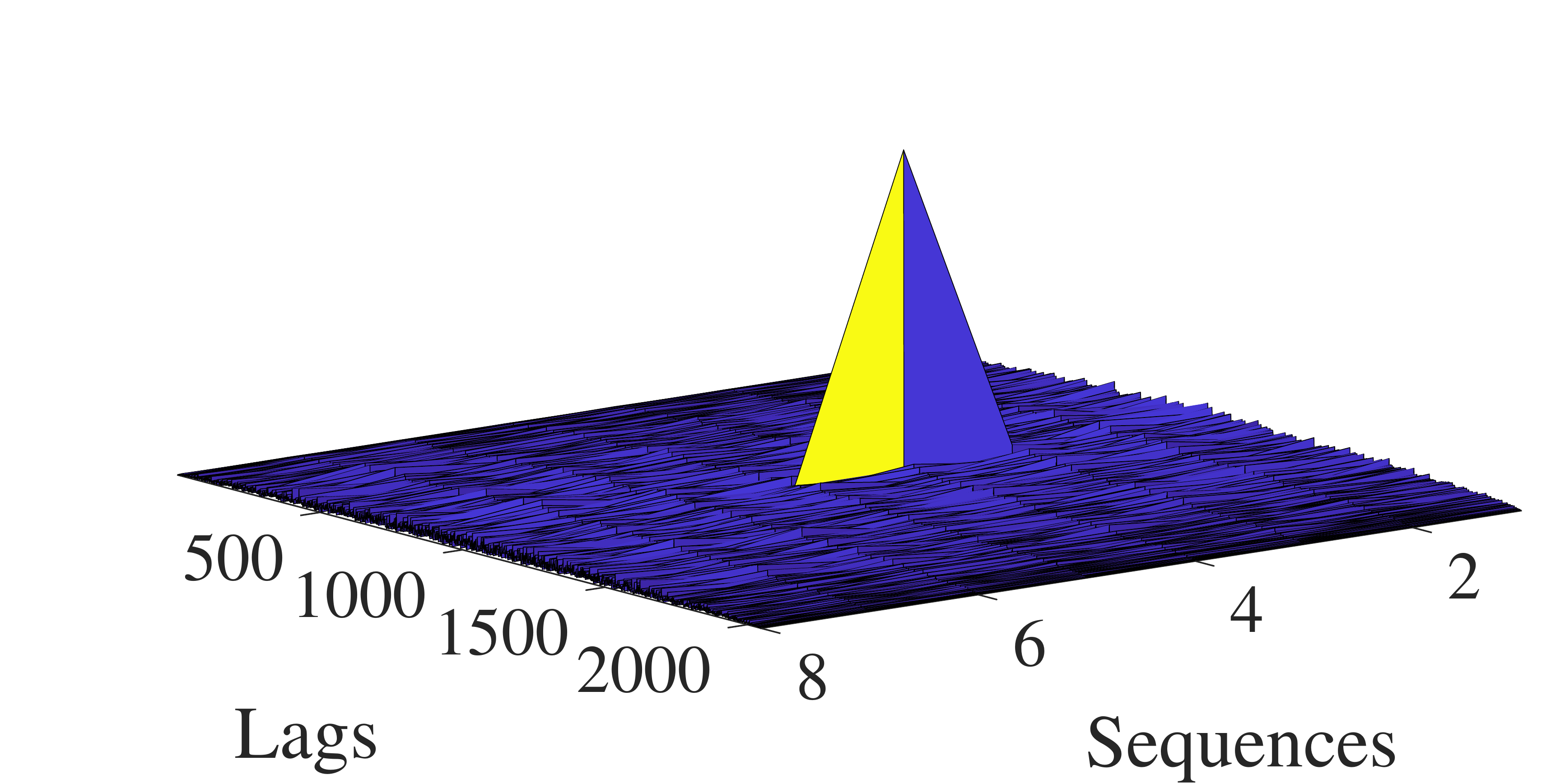}
        \end{minipage}
        & 
        \begin{minipage}{.215\textwidth}
            \includegraphics[width=1\linewidth]{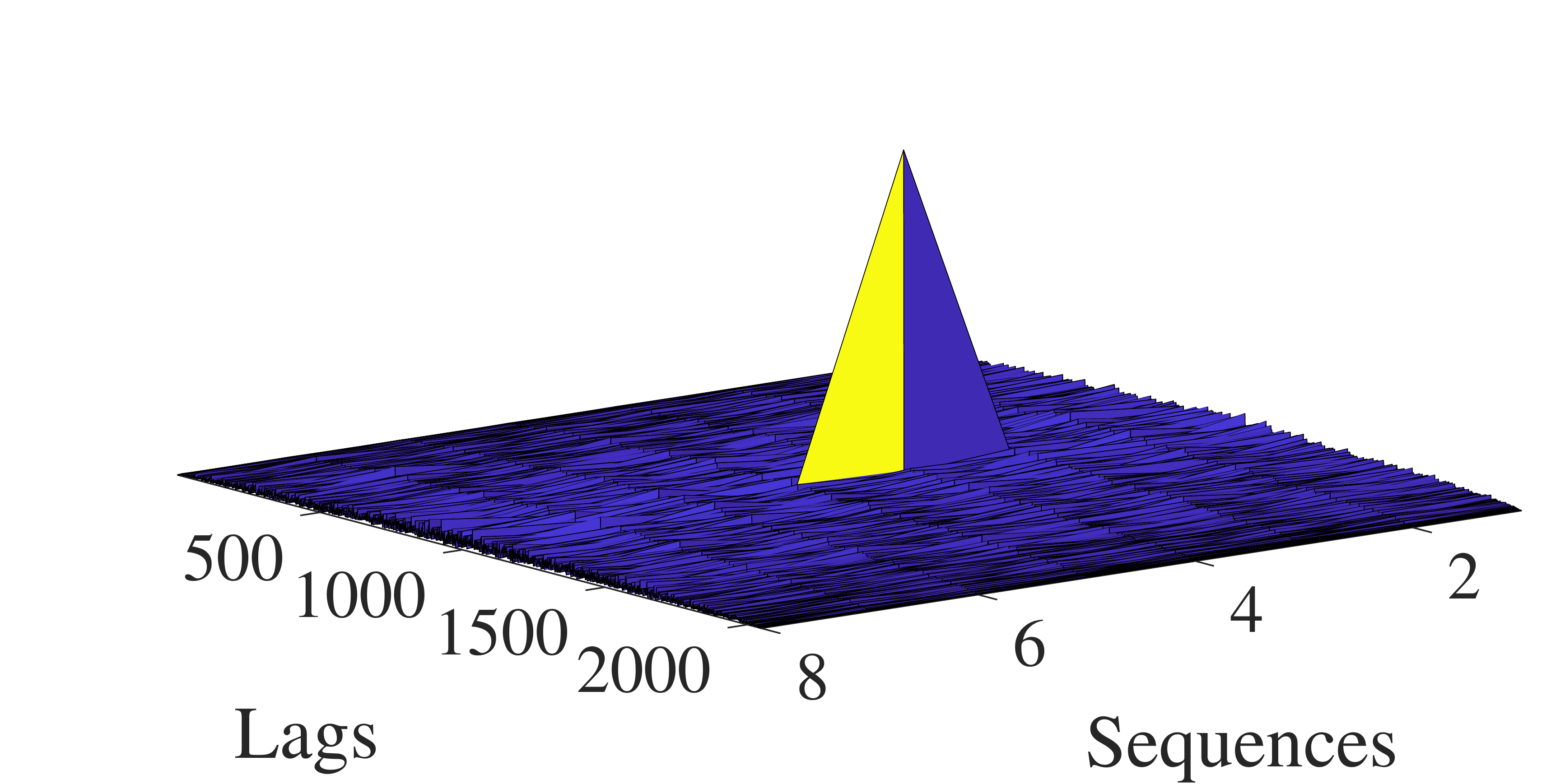}
        \end{minipage}
        \\        
		\hline
		\hline
	\end{tabular}
	\label{tab:CC_vs_eta}
\end{table*}

\subsubsection{Beampattern nulling and target discrimination} \label{subsubsec:Discrimination} 
%Although choosing $\eta = 1$ leads a good beampattern response, the optimized sequences will be fully correlated.
%In this case, we lose the waveform diversity and the conventional \gls{MIMO} processing cannot separate the waveform in receive side, hence the virtual array is not formed. 
%On the other hand by choosing $\eta = 0$ we obtain a good orthogonality and diversity through the sequences, whereas the transmit beampattern is almost omni directional. In this case, the transmit power cannot be steered on the desired angles, and the strong signal from the undesired directions may saturate the radar receiver. Further, the energy of the radar transmitter maybe emitted on the directions which are unnecessary to be searched. 
% furthermore on undesired angles, the transmitters might make an interference or the receiver saturate by strong reflections. 

To illustrate the effectiveness of choosing $ 0 < \eta < 1$, we consider a scenario where two desired targets ($T_1$ and $T_2$) with similar reflectivity, speed, and range are located in $\theta_{T_1} = -40^o$ and $\theta_{T_2} = -50^o$.
% {\color{red} please change $\btheta_t=[-40^o, -50^o]^T$ to $\theta_{T_1} = -40^o$, and $\theta_{T_2} = -50^o$. This is not interval, or vector. So, you should not write like the previous form. Please consider this also later in the papers and here for the interference.} .
% {\color{red} please write the angles, I guess something around $-50^o$}. 
The reason for selecting similar speed and range is to consider a worse case scenario where targets cannot be extracted from the range and Doppler processing. Also, we assume that three more strong targets denoted as $B_1$, $B_2$ and $B_3$ (potentially can be clutter), are located in identical speed and range, but with different angles,  $\theta_{B_1} = -9.5^o$, $\theta_{B_2} = 18.5^o$ and $\theta_{B_3} = 37^o$,
% {\color{red} please write the angles}, 
%called $B_1$, $B_2$ and $B_3$ respectively.  
We aim to design a set of transmit sequences to be able to discriminate the two desired targets, but avoiding interference from the undesired directions. 
% {\color{red} why you used bold to show direction/speed/... of targets? In my opinion, you don't need to say their range and speed, since you said they are similar and that is enough. For the angles, you can say $\theta_{T_1}$ and so on the red parts that I left. Please update. Also I removed basestation since BS is a signal independent interference. Here, we are talking about signal dependent interference.}
%Let $\btheta = [-40,-50,-9.5,18.5,37]^T$ ($deg$), $\br = [50,50,50,50,50]^T$ ($m$), $\bsigma = [30, 30, 100, 100, 100]^T$ ($m^2$) and $\bv = [30,30,0,0,0]^T$ ($m/sec$) be the azimuth angle, range, \gls{RCS} and velocity vectors of the two targets and three base stations respectively. 

\begin{figure*}
    \centering
    \begin{subfigure}{.32\textwidth}
        \centering
        \includegraphics[width=1\linewidth]{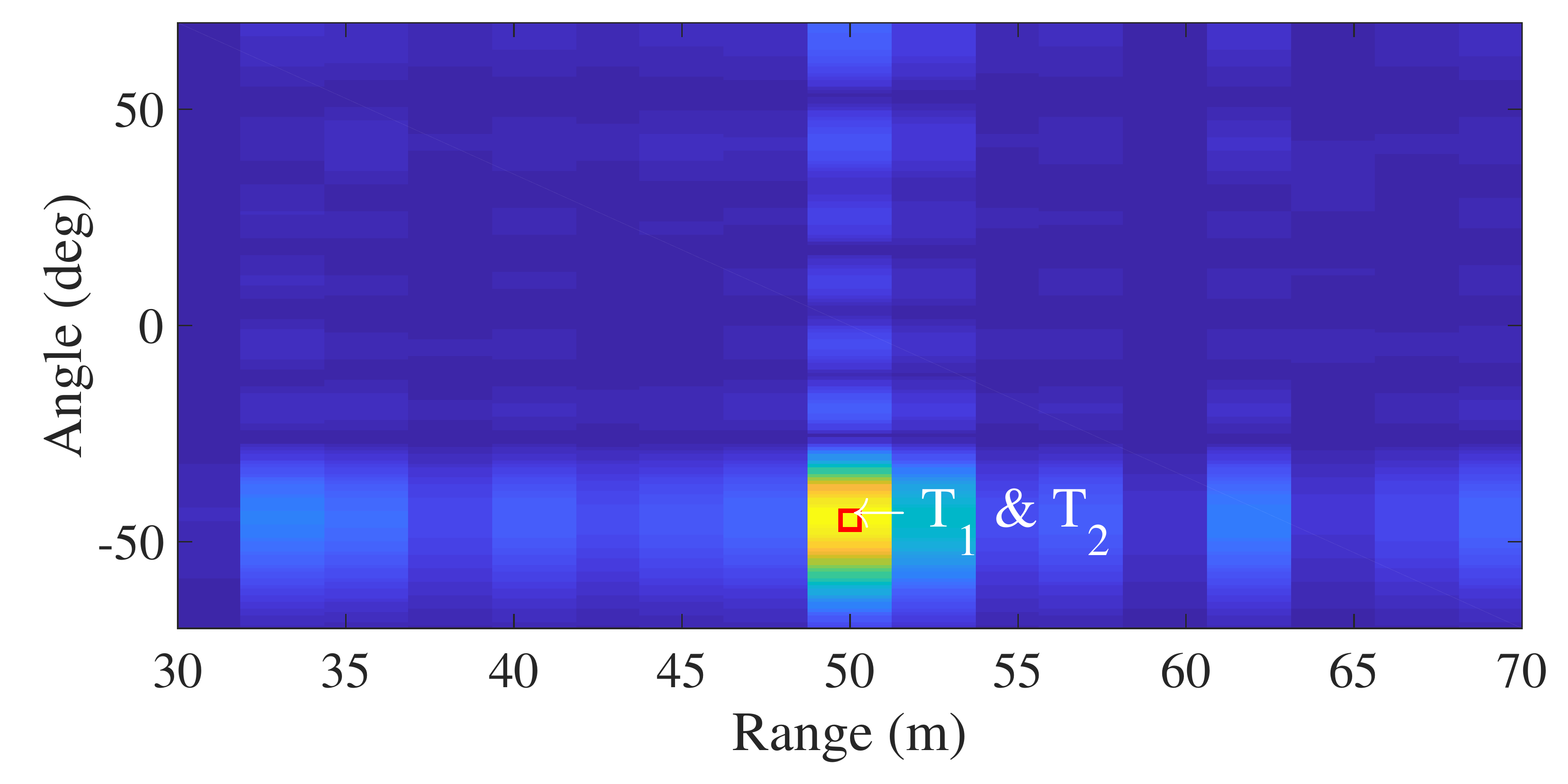}
		\caption[]{Phased array processing $\eta=1$.}\label{fig:RangeAngleArray_eta1}
    \end{subfigure}
    \begin{subfigure}{.32\textwidth}
        \centering
        \includegraphics[width=1\linewidth]{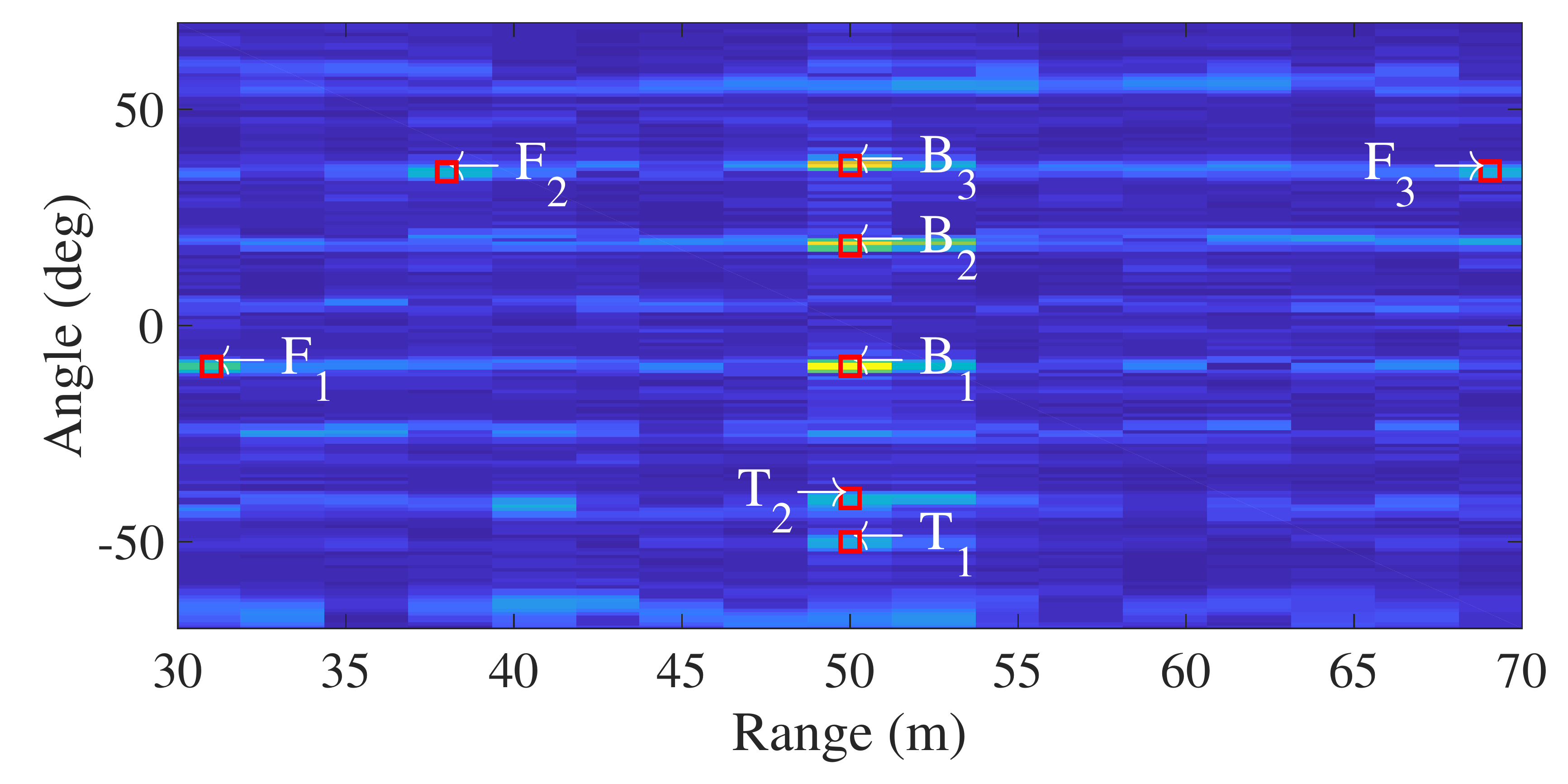}
		\caption[]{\gls{MIMO} processing $\eta=0$.}\label{fig:RangeAngleMIMO_eta0}
    \end{subfigure}
    \begin{subfigure}{.32\textwidth}
        \centering
        \includegraphics[width=1\linewidth]{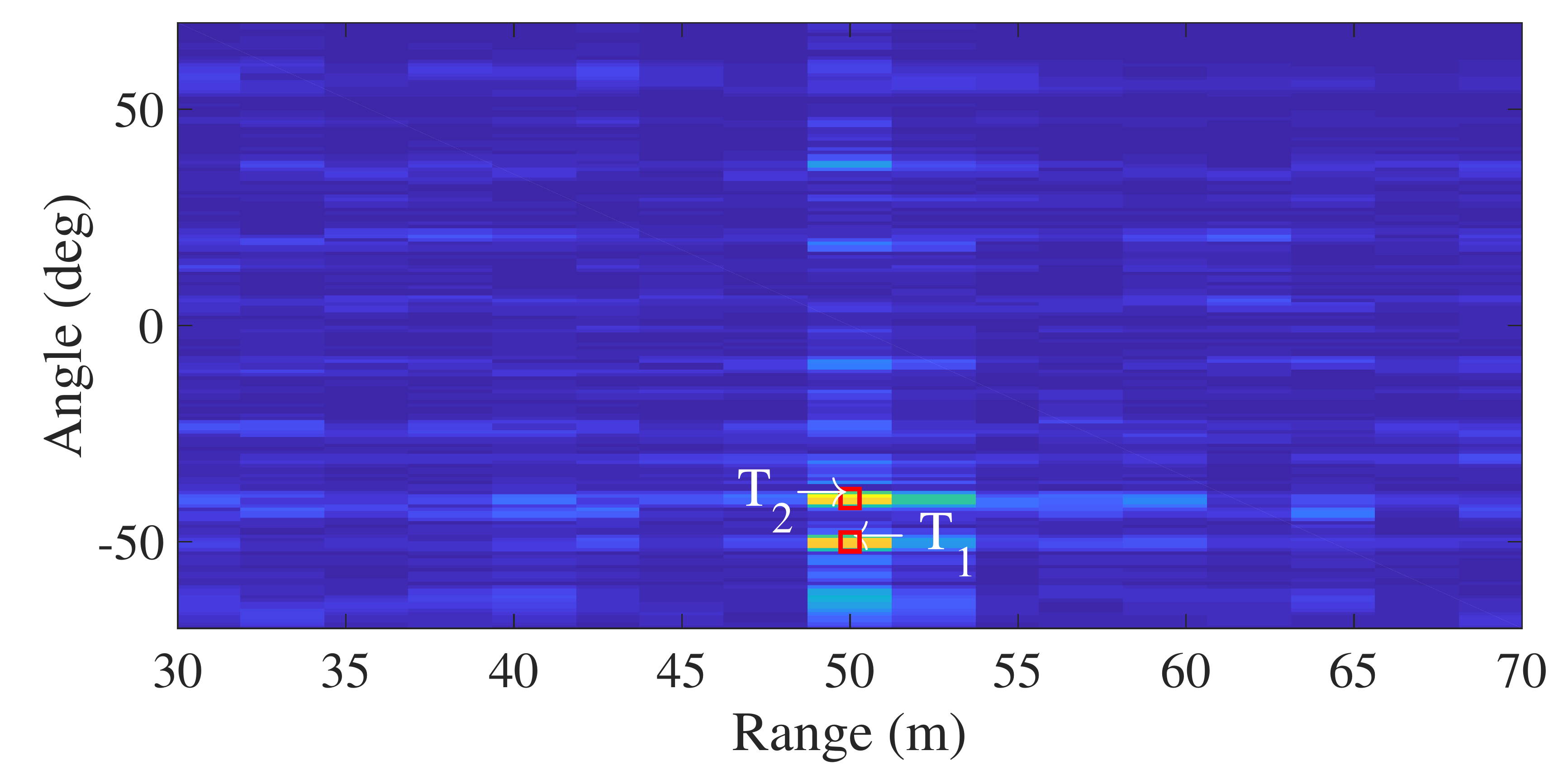}
		\caption[]{\gls{MIMO} processing $\eta=0.5$.}\label{fig:RangeAngleMIMO_eta05}
    \end{subfigure}
    \caption[]{ Illustration of the centrality of $\eta$ ($C_4$ constraint, $M_t=M_r=8$, $N=64$, $L=8$, $\theta_{T_1} = -50^o$, $\theta_{T_2} = -40^o$, $\theta_{B_1} = -9.5^o$, $\theta_{B_2} = 18.5^o$ and $\theta_{B_3} = 37^o$).}\label{fig:RangeAngle}
\end{figure*}

\figurename{~\ref{fig:RangeAngle}} shows the range-angle profile of the above scenario under the representative $C_4$ constraint with $L=8$.
% {\color{red} In this figure, the waveforms were designed under which constraint?}
When $\eta=1$, we consider the conventional phased array receiver processing for \figurename{~\ref{fig:RangeAngleArray_eta1}} and use one matched filter to extract the range-angle profile. To this end we assume $\lambda/2$ spacing for transmit and receive antenna elements, i.e., $d_t = d_r = \frac{\lambda}{2}$. Observe that, despite the mitigation of undesired targets, the two targets are not discriminated and are merged into a single target. The same scenario has been repeated in \figurename{~\ref{fig:RangeAngleMIMO_eta0}} when $\eta = 0$. Since the optimized waveforms are orthogonal in this case, we consider \gls{MIMO} processing to exploit the virtual array and improve the discrimination/identifiability. In this case, we use $M_t$ matched filters in every receive chain, each corresponding to one of the $M_t$ transmit sequences. The receive antennas have a sparse configuration with $d_r = M_t\frac{\lambda}{2}$ but the transmit antennas are a filled \gls{ULA} with $d_t = \frac{\lambda}{2}$; this forms a \gls{MIMO} virtual array with a maximum length. In this case, the optimized set of transmit sequences is able to  discriminate the two targets, but it is contaminated by the strong reflections of the undesired targets. Also, some false targets ($F_1$, $F_2$ and $F_3$) have appeared due to the high side-lobe levels of the strong reflectors. By choosing $\eta = 0.5$, we are able to discriminate the two targets and mitigate the signal of the undesired reflections in a same time. This fact is shown in \figurename{~\ref{fig:RangeAngleMIMO_eta05}}.

\tablename{~\ref{tab:RangeAngleResponse}} shows the amplitude of the desired targets and undesired reflections in the scene (after the detection chain) at different Pareto-weights ($\eta$). As can be seen from \tablename{~\ref{tab:RangeAngleResponse}}, the performance of target enhancement and interference mitigation reduces from $\eta = 1$ to $\eta=0$. Nevertheless by choosing $\eta = 0.5$ the waveform achieves a trade-off between spatial- and range-\gls{ISLR}, it can discriminate the two targets and mitigate the interference from the undesired locations.

\begin{table}
	\centering
	\caption{Amplitude of the desired and undesired targets }
	\begin{tabular}{c|c|c|c|c|c}	
		\hline
		\hline
		$\eta$ & $T_1$ & $T_2$ & $B_1$ & $B_2$ & $B_3$\\
		\hline
		$ 1$ & 9.54 dB & 9.79 dB  & -13.24 dB  & -19.93 dB & -9.4 dB \\
		$ 0.5$ & 8.78 dB & 9.71 dB & -3.5 dB & -3.51 dB & -0.6 dB \\         
		$ 0$ & -2.39 dB & -2.44 dB & 3.68 dB & 2.95 dB & 2.87 dB \\
		\hline
		\hline
	\end{tabular}
	\label{tab:RangeAngleResponse}
\end{table}

\subsubsection{Pareto-front} \label{subsubsec:Pareto}
% Pareto-front or non-dominated solutions, is a curve which gives a set of optimal solutions and helps the radar designers to choose the best solution for radar system according to the environment conditions, priorities and risks. \figurename{~\ref{fig:Pareto}} shows non-dominated (optimal) solutions of the problem under $C_1, \dots, C_4$ constraints; as an illustrative example, it also depicts the dominated solutions for $C_4$. As expected, the non-dominated solutions have better performance in terms of spatial- and range-\gls{ISLR} in compare to dominated solutions. In addition \figurename{~\ref{fig:Pareto}} also depicts the performance of the solution corresponding to $\eta=0.5$; this solution is used to generate \figurename{~\ref{fig:RangeAngleMIMO_eta05}}. It can be observed that the solution lies on the Pareto front. 

Pareto-front or non-dominated solutions, is a curve which gives a set of optimal solutions and helps the radar designers to choose the best solution for the radar system according to the environment conditions, priorities and risks. Based of our best knowledge there is no technique in literature trading off the two spatial- and range-\gls{ISLR} functions considered in the paper. In this regards, we consider to compare the performance of the proposed method with \gls{NSGA}-\RN{2}, a multi objective evolutionary algorithm \cite{10.5555/1215640}. We assume the following setup for \gls{NSGA}-\RN{2}, the number of population $n_p = 50$, crossover percentage $c_r = 70\%$, mutation percentage $m_p = 40\%$ and mutation rate $m_r = 0.05$.

\begin{figure}
	\centering
	\includegraphics[width=1\linewidth]{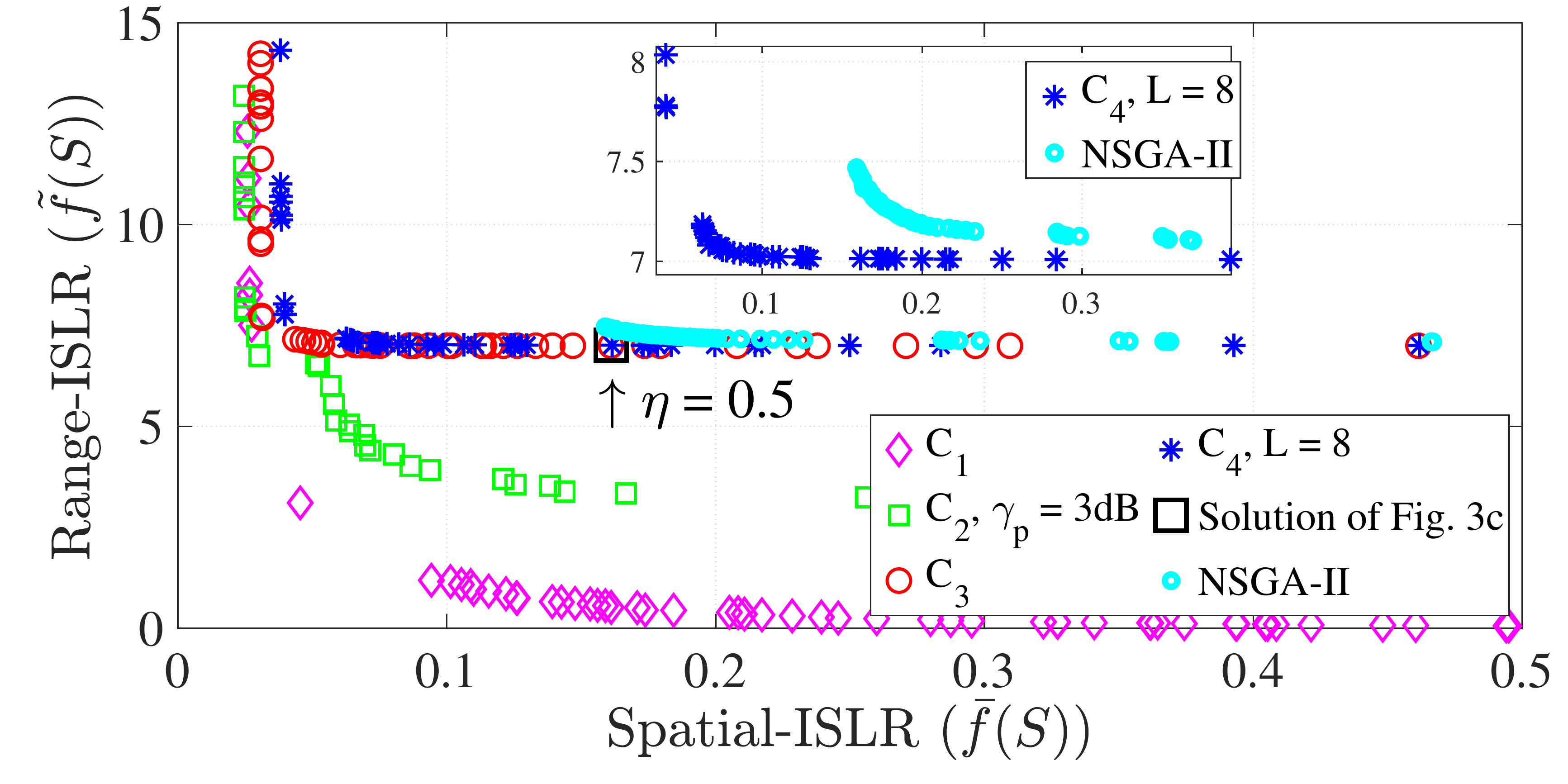}
	\caption[]{Solution for Pareto front obtained from \gls{NSGA}-\RN{2} and the proposed method ($N=64$, $M_t=8$).}\label{fig:Pareto}
\end{figure}

\figurename{~\ref{fig:Pareto}} shows the non-dominated (optimal) solutions of the proposed method under $C_1, \dots, C_4$ constraints and \gls{NSGA}-\RN{2} method under discrete phase. As can be seen the solution obtained by \gls{NSGA}-\RN{2} cannot dominate the Pareto front of proposed method. Besides the proposed methods offers more diversity in compare with \gls{NSGA}-\RN{2}. In addition \figurename{~\ref{fig:Pareto}} also depicts the performance of the solution corresponding to $\eta=0.5$; this solution is used to generate \figurename{~\ref{fig:RangeAngleMIMO_eta05}}. It can be observed that the solution lies on the Pareto front. 

% \begin{figure}
% \centering
%     \includegraphics[width=0.8\linewidth]{R4-5_CD_vs_NSGA2.eps}
% 	\caption[]{Pareto front of proposed method and \gls{NSGA}-\RN{2} ($L=8$, $M=8$ and $N=64$).}\label{fig:CD_vs_NSGA2}
% \end{figure}

The correct choice of $\eta$ is essential to achieve the objectives and that, such a choice of $\eta$ depends on the scenario. In the following, we provide an plausible method for selecting an appropriate $\eta$.
\begin{itemize}
\item {\bf Training Step}: This is an offline procedure which contains the following steps,
\begin{enumerate}
% \item We consider different scenarios and obtain the Pareto front by solving the optimization problem with different $\eta$ values. Then, we store the results in a database. 
% \item We design an artificial neural networks and train it with the different scenarios (different Pareto curves) which are stored on database, to offer the optimum solution based on the environment parameter.
\item We consider different scenarios and obtain the optimized waveform set and its $\eta$ value by using the Pareto front. Then, we store the corresponding results for every scenario in a database. 
\item We design an artificial neural networks and train it with the different scenarios and  optimized waveforms corresponding to the best $\eta$ value that are stored on the database, to offer the optimum solution based on the scenario.
\end{enumerate}

\item {\bf Functional Step}: After training successfully the artificial neural network we can consider the following procedure,
\begin{enumerate}
\item In order to form the virtual array at the receiver, at the first the \gls{MIMO} radar system transmits an orthogonal set of sequences ($\eta=0$). In this case, the \gls{MIMO} radar system is able to estimate the angles of targets and interference with high discrimination (other parameters such as range and Doppler can be estimated as well).
\item Based on the estimated parameters, the artificial neural network offers the optimized set of sequences by using the database.
\item Using the chosen set of sequences, the environment parameters are estimated.
\item We go to step 2.
\end{enumerate}
\end{itemize}

As mentioned earlier, the training step is performed offline; similar sensor training is also typically undertaken in many commercial offerings.  Further, the functional step does not involve optimization procedures but executing a neural network which is typically fast, thereby rendering the scheme practically is applicable. We consider a detailed study on this scheme for our future research.

\subsection{Minimizing spatial-\gls{ISLR} ($\eta = 1$)}\label{subsec:minimizing spatial-ISLR}
% By choosing $\eta = 1$, we focus on minimizing the spatial-\gls{ISLR}. In \figurename{~\ref{fig:BeamPattern_eta_1}}, we illustrate the beampattern of the optimized waveforms through different constraints. In this section, we use \gls{SDR} reported in \cite{7126203} as a benchmark. In case of designing discrete phase sequences, we map the results of \gls{SDR} to the nearest \gls{MPSK} sequence and call it Quantized-\gls{SDR} (Q-\gls{SDR}).
By choosing $\eta = 1$, we focus on minimizing the spatial-\gls{ISLR}. In this subsection, we compare the performance of proposed method under different constraints. In this regards, we compare with \gls{SDR} based method \cite{7126203} for $C_1$, \gls{MIAPC} for $C_2$, \gls{MIACMC} \cite{8239836} for $C_3$ and \gls{MIACMC} \cite{8401959} $C_4$ as a benchmark respectively

% we use \gls{SDR} method reported in \cite{7126203}, \gls{MIAPC}, \gls{MIACMC} \cite{8239836} and \gls{STTC} \cite{8401959} method for limited energy ($C_1$), \gls{PAR} ($C_2$), continuous ($C_3$) and discrete phase ($C_4$) as a benchmark respectively. 

\begin{figure*}
    \centering
    \begin{subfigure}{.49\textwidth}
        \centering
		\includegraphics[width=1\linewidth]{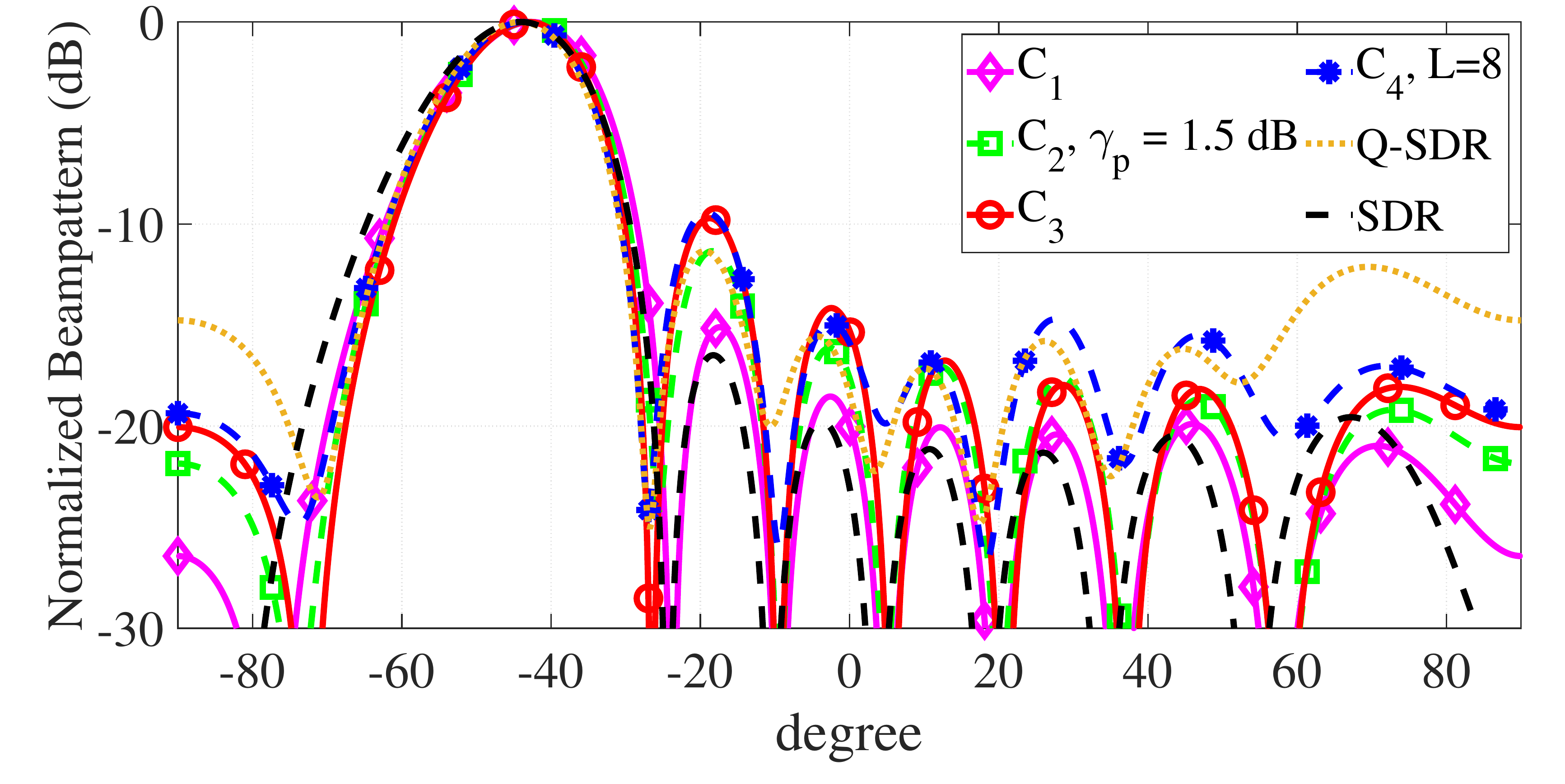}
		\caption[]{Comparison of $C_1$ and \gls{SDR}.}\label{fig:C1_vs_SDR}
    \end{subfigure}
    \begin{subfigure}{.49\textwidth}
        \centering
		\includegraphics[width=1\linewidth]{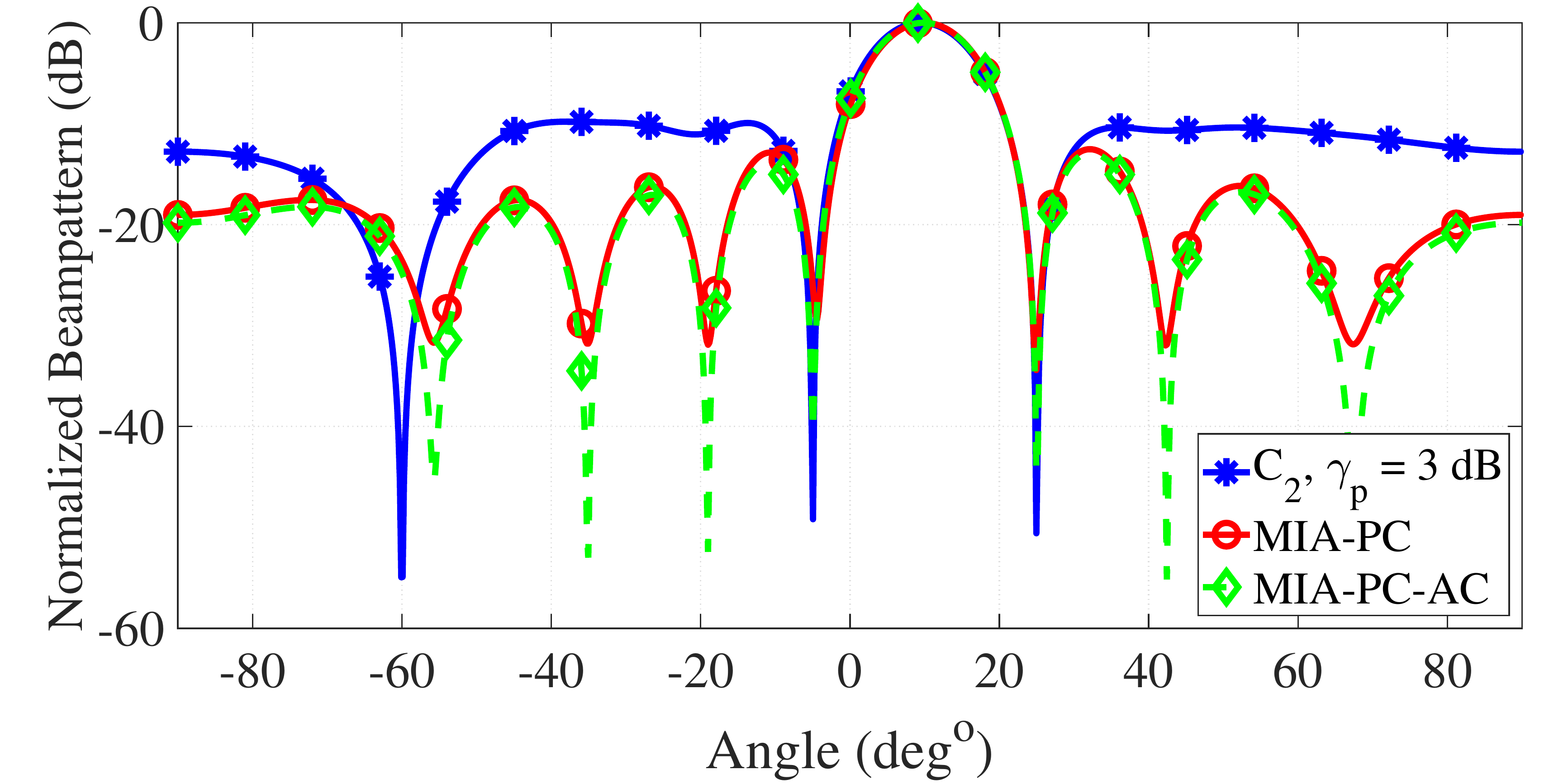}
		\caption[]{Comparison of $C_2$ and \gls{MIAPC}.}\label{fig:C2_vs_MIAPC}
    \end{subfigure}
    \begin{subfigure}{.49\textwidth}
        \centering
		\includegraphics[width=1\linewidth]{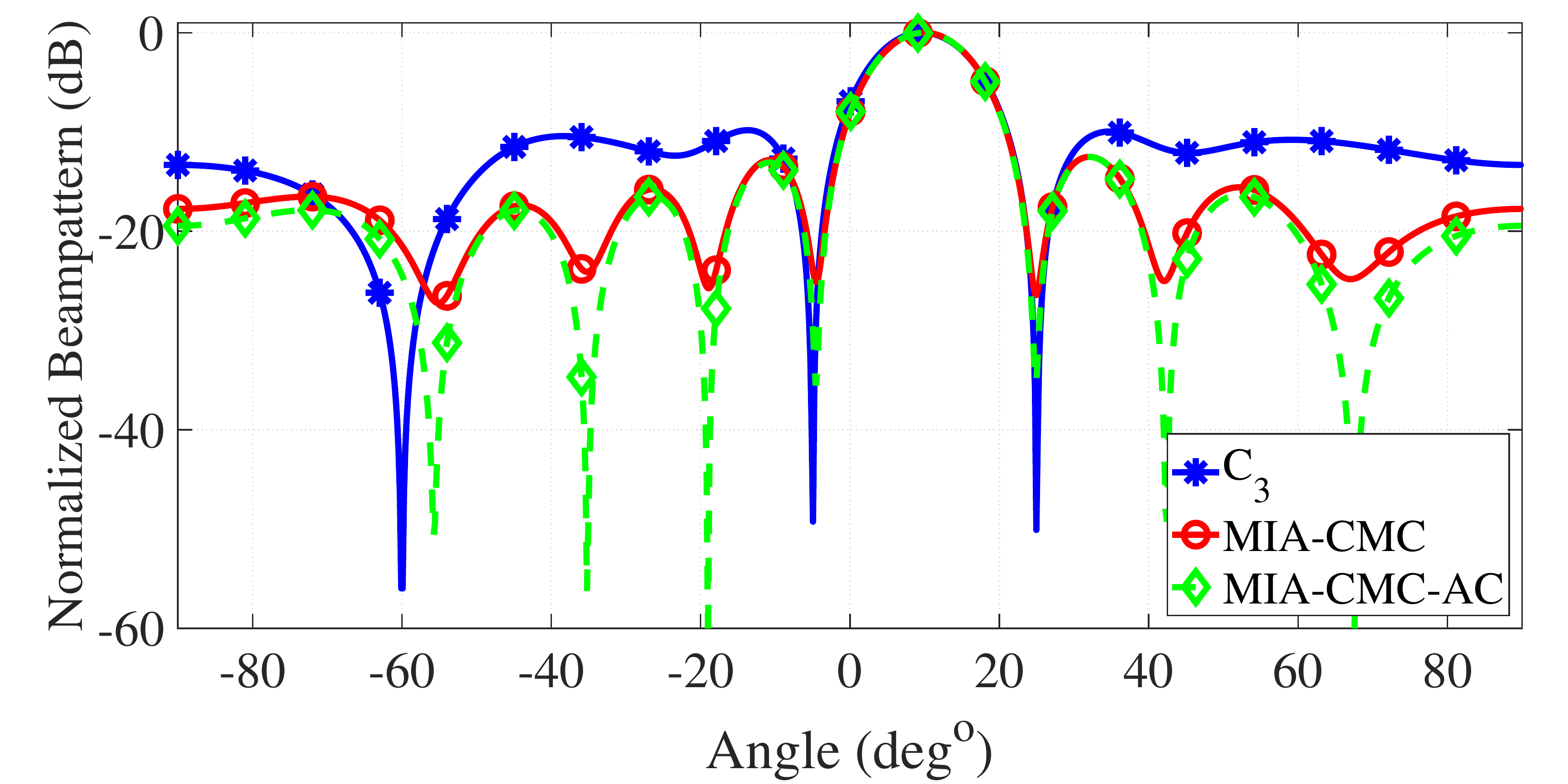}
		\caption[]{Comparison of $C_3$ and \gls{MIACMC}.}\label{fig:C3_vs_MIACMC}
    \end{subfigure}
    \begin{subfigure}{.49\textwidth}
        \centering
		\includegraphics[width=1\linewidth]{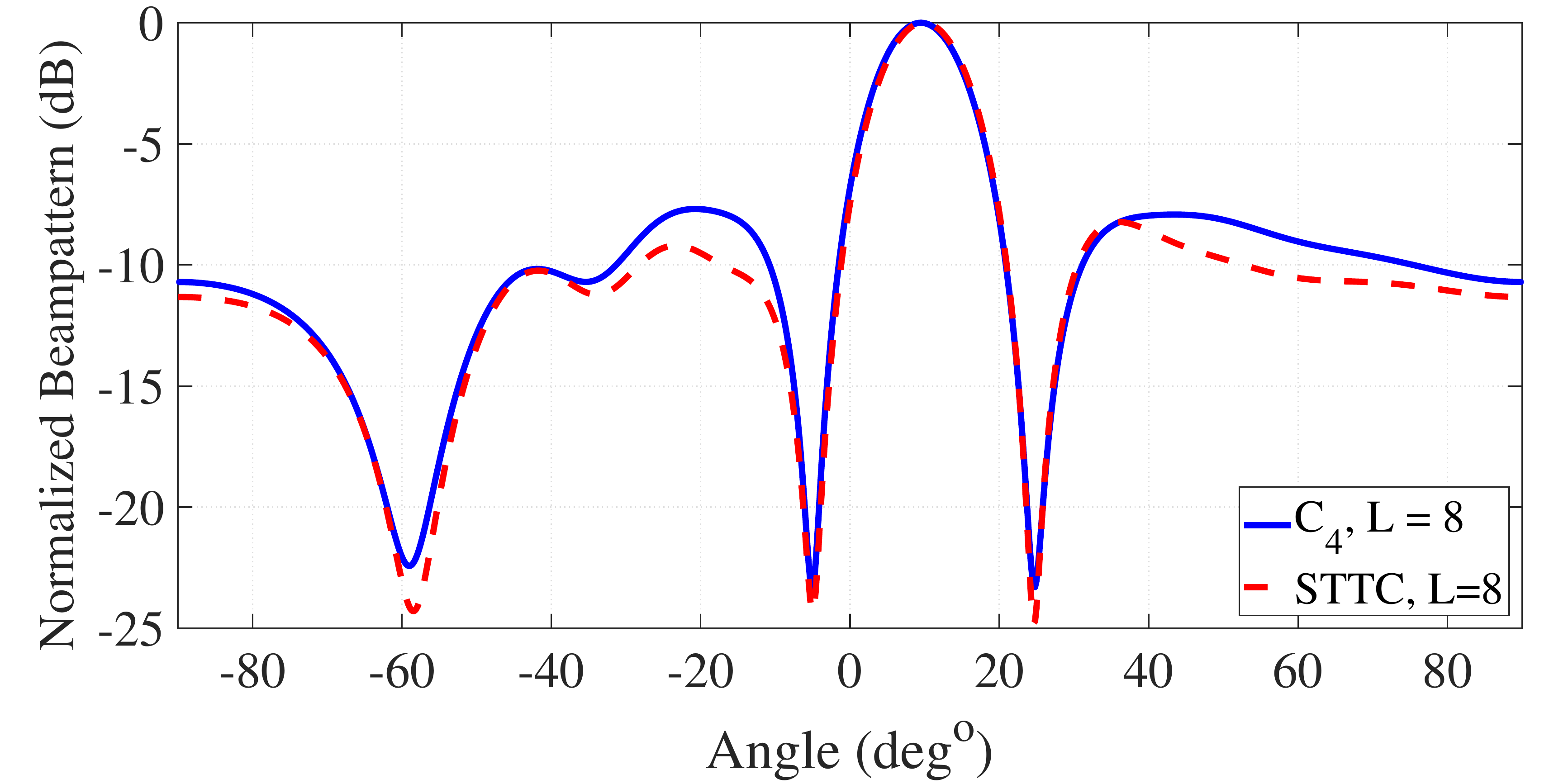}
		\caption[]{Comparison of $C_4$ and \gls{STTC}).}\label{fig:C4_vs_STTC}
    \end{subfigure}
    \caption[]{The comparison of beampattern shaping of proposed method with (a) \gls{SDR}, (b) \gls{MIAPC}, (c) \gls{MIACMC} and (d) \gls{STTC} ($M=8$ and $N=64$).}\label{fig:BeamPattern_Compare}
\end{figure*}

To compare with the \gls{SDR} method \cite{7126203}, we assume that the desired and undesired angular regions to be $\Theta_d = [-55^o,-35^o]$ and $\Theta_u = [-90^o,-60^o] \cup [-30^o,90^o]$ respectively. In \figurename{~\ref{fig:C1_vs_SDR}}, we illustrate the beampattern of the optimized waveforms through different constraints and \gls{SDR} method. In case of designing discrete phase sequences, we map the results of \gls{SDR} to the nearest \gls{MPSK} sequence and call it Quantized-\gls{SDR} (Q-\gls{SDR}).
Interestingly, the optimized waveforms through the proposed method mimics the beampattern obtained via \gls{SDR}, indicating the attractiveness of this approach in designing set of sequences with practical constraints. Notice that, there is a significant difference between the solution obtained via the proposed method under the discrete phase constraint and Q-\gls{SDR} for identical alphabet sizes. This can be justified from the fact that we consider the constraint directly in the design problem, while quantizing the waveform to the nearest \gls{MPSK} sequence does not guarantee an optimal solution.

In order to compare under \gls{PAR} continuous and discrete phase, we assume that the target and the three interferers are located at $10^o$, $-5^o$, $25^o$ and $-60^o$ respectively. We set noise power $-10$ dB, and similar values of $30$dB for target and clutter \gls{RCS}. For a fairness, we compare with \gls{MIAPC} and \gls{MIACMC} in \cite{7126203}, where the similarity constraint is not considered. Also, in \cite{8401959}, we set the similarity threshold equal to 2, the maximum admissible similarity value in \gls{STTC}. \figurename{~\ref{fig:BeamPattern_Compare}}(b),(c),(d) shows the normalized beampattern response of \gls{MIAPC}, \gls{MIACMC}, \gls{STTC} and the proposed method. Observe that the proposed method outperforms \gls{MIAPC} and \gls{MIACMC} in terms of null steering. Besides the performance of the proposed method and \gls{STTC} under discrete phase are similar.

% In order to compare under continuous and discrete phase, we assume that the target and the three interferers are located at $10^o$, $-5^o$, $25^o$ and $-60^o$ respectively. Since we do not consider the target and clutter \gls{RCS} (interferers power), noise power and similarity constraint explicitly in our paper, for fair comparison, we consider to use the following setup for \cite{8239836, 8401959}, The noise power is $-10$ dB, we set the same target and clutter \gls{RCS} equal to $30 dB$. The reason is that, we do not consider a weight for beampattern response on desired and undesired angles in the proposed method. For similarity constraint, we compare the proposed method with \gls{MIAPC} and \gls{MIACMC}. These two methods are versions of \gls{MIA} which the authors do not consider the similarity constraint. Besides, we consider the maximum admissible of similarity constraint for \gls{STTC} i.e. we set the similarity constraint equal to $2$.
    
% \figurename{~\ref{fig:BeamPattern_Compare}} shows the normalized beampattern response of \gls{MIAPC}, \gls{MIACMC}, \gls{STTC} and proposed method. Observe that the proposed method outperforms \gls{MIAPC} and \gls{MIACMC} in terms of null steering. Besides the performance of the proposed method and \gls{STTC} under discrete phase are similar.

% \begin{figure}
% 	\centering
% 	\includegraphics[width=1\linewidth]{BeamPattern_eta_1.eps}
% 	\caption[]{Beampattern of proposed method, \gls{SDR} and Q-\gls{SDR} ($\eta = 1$, $M_t=8$, $N=64$, $\Theta_d = [-55^o,-35^o]$ and $\Theta_u = [-90^o,-60^o] \cup [-30^o,90^o]$).}\label{fig:BeamPattern_eta_1}
% \end{figure}

% \begin{figure}
% \centering
% \begin{subfloat}[Non-overlapping subarrays, aperiodic cross-correlation.]
%     {\includegraphics[width=0.95\columnwidth]{figures/crossNap1.eps}}
%     \label{fig:CrossNonOver}
% \end{subfloat}
% \begin{subfloat}[Overlapping subarrays, periodic auto-correlation.]
%     {\includegraphics[width=0.95\columnwidth]{figures/autoOVp1.eps}}
%     \label{fig:AutoOver}
% \end{subfloat}
% \caption{Auto- and cross-correlations of the optimized sequences.}
% \label{fig:AutoCross}
% \end{figure}

\subsection{Minimizing range-\gls{ISLR} ($\eta = 0$)}\label{subsec:minimizing range-ISLR}
\begin{figure}
	\centering
 	\includegraphics[width=1\linewidth]{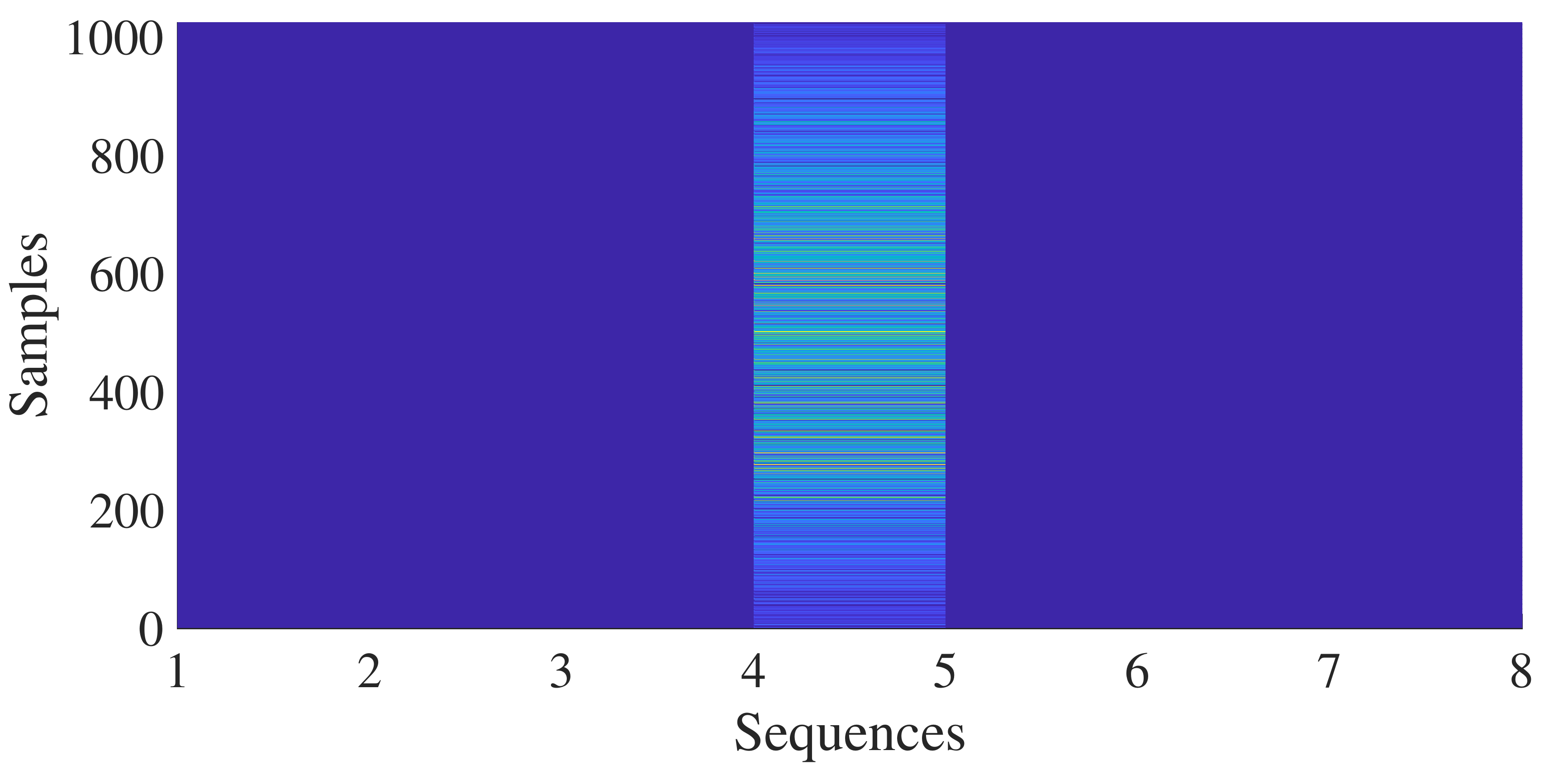}
 	\caption[]{Optimized waveform of proposed method under $C_1$ constraints ($\eta = 0$, $M_t=8$, $N=1024$).} \label{fig:OptimizedWaveform_C1}
\end{figure}

We set $\eta=0$ and evaluate the performance of the proposed method. Kindly refer to the last row of \tablename{~\ref{tab:CC_vs_eta}}, which
% {\color{red} Do you mean \tablename{~\ref{fig:CC_vs_eta}} or \figurename{~\ref{fig:OptimizedWaveform_C1}}? Please update. Also, if you mean \tablename{~\ref{fig:CC_vs_eta}}, do we need to talk again about it here? We mentioned it once.}
shows the three-dimensional representation of the auto- and cross-correlation (following the methodology in footnote 3 of section \ref{subsec:Trade-off}), under $C_1, \dots, C_4$ constraints. In this case, the proposed method designs a waveform with good orthogonality under $C_2$, $C_3$ and $C_4$ constraints, and interestingly achieves a perfect orthogonality under the $C_1$ constraint. \figurename{~\ref{fig:OptimizedWaveform_C1}} shows the absolute value of optimum sequence under $C_1$ constraint. As can be seen all the power is concentrated on one transmitter with no waveform from others. This is similar to \gls{TDM} approach for orthogonality \cite{1399141,6233300,8081604,6825710}. 
% Furthermore, as only one transmitter emits signal and others are off, the cross-correlation term in \eqref{eq:ISL} is equal to zero and we only have the auto-correlation term.

We choose Multi-CAN \cite{5072243} and MM-Corr \cite{7420715} as the benchmark and assess the range-\gls{ISLR} under $C_3$ and $C_4$ (unimodular sequences) for a fair comparison. 
% To have a fair comparison  with the benchmark (Multi-CAN \cite{5072243}), here we only assess the range-\gls{ISLR} of the  waveforms which are obtained under $C_3$ and $C_4$ constraints, i.e., unimodular sequences. 
In this case, a lower bound on the scaled range-\gls{ISLR} is $10\log(M_t - 1)$ dB \cite{7420715}. \tablename{~\ref{tab:ISLRvsM}} compares the average scaled range-\gls{ISLR} of the  proposed method with Multi-CAN, MM-Corr and the lower bound for different number of transmitters. Similar to the Multi-CAN and MM-Corr, the proposed method meets the lower bound under continuous phase constraint. Interestingly, even with discrete phase constraint where $L= 8$ and $L= 2$ (binary), the obtained set of sequences exhibits the scaled range-\gls{ISLR} values quite close to the lower bound. 

%Furthermore, with fixed transmitter number, $M_t = 8$ and different sequence length the range-\gls{ISLR} under $C_3$ and Multi-CAN meet the lower bound at $8.451$dB. 
\tablename{~\ref{tab:ISLRvsN}} shows the optimized scaled range-\gls{ISLR} values under $C_4$, for $L=8$ and $L=2$ with different sequence lengths when $M_t = 8$. As can be seen the proposed method is capable to design large sequence length without degradation. Recalling the last row of \tablename{~\ref{tab:ISLRvsN}}, we observe that the optimized sequences have range-\gls{ISLR} values quite close to the lower bound (less than $0.02$ dB difference when $M_t = 8$).

\begin{table}
	\centering
	\caption{Comparison between the average scaled range-\gls{ISLR} (dB) of the proposed method under $C_3$ and $C_4$, Multi-CAN \cite{5072243}, MM-corr \cite{7420715} and lower bound with different number of transmitters ($\eta = 0$, $N=64$).}
	\begin{tabular}[ht]
	    {c|c|c|c|c|c|c}
		\hline
		\hline
		$M_t$ & \shortstack{Lower \\ bound} & \shortstack{ Multi-\\CAN} & \shortstack{MM-\\Corr} & $C_3$ & \shortstack{$C_4$\\ ($L=8$)} & \shortstack{$C_4$ \\ ($L=2$)} \\
		\hline
		2 & 0  & 0 & 0.0003 & 0 & 0.2583 & 0.5266 \\
		3 & 3.0103 & 3.0103 & 3.0104 & 3.0103 & 3.1045 & 3.2133 \\
		4 & 4.7712 & 4.7712 & 4.7712 & 4.7712 & 4.8080 & 4.8587 \\
		5 & 6.0206 & 6.0206 & 6.0206 & 6.0206 & 6.0411 & 6.0950 \\
		6 & 6.9897 & 6.9897 & 6.9897 & 6.9897 & 7.0024 & 7.0283 \\
		7 & 7.7815 & 7.7815 & 7.7815 & 7.7815 & 7.7891 & 7.8071 \\
		8 & 8.4510 & 8.4510 & 8.4510 & 8.4510 & 8.4581 & 8.4684 \\
		\hline
		\hline
	\end{tabular}
	\label{tab:ISLRvsM}
\end{table}
% \begin{table}
% 	\centering
% 	\caption{Comparison between the scaled range-\gls{ISLR} (dB) of the proposed method under $C_3$ and $C_4$, Multi-CAN and lower bound with different number of transmitters ($\eta = 0$, $N=64$).}
% 	\begin{tabular}{c|c|c|c|c}	
% 		\hline
% 		\hline
% 		$M_t$ & Lower bound & Multi-CAN & $C_3$ & $C_4$ ($L=8$) \\
% 		\hline
% 		3 & 3.01  & 3.01  & 3.01  & 3.085 \\
% 		4 & 4.771 & 4.771 & 4.771 & 4.808 \\
% 		5 & 6.021 & 6.021 & 6.021 & 6.039 \\
% 		6 & 6.99  & 6.99  & 6.99  & 7.002 \\
% 		7 & 7.782 & 7.782 & 7.782 & 7.791 \\
% 		8 & 8.451 & 8.451 & 8.451 & 8.456 \\         
% 		\hline
% 		\hline
% 	\end{tabular}
% 	\label{tab:ISLRvsM}
% \end{table}

% \begin{table}
% 	\centering
% 	\caption{The range-\gls{ISLR} obtained by proposed method under $C_4$ constraint with different length ($\eta = 0$, $M_t=8$).}
% 	\begin{tabular}{c|c|c|c|c|c|c|c}	
% 		\hline
% 		\hline
% 		$N$ & 8 & 16 & 24 & 32 & 40 & 48 & 64 \\
% 		\hline
% 		$C_4$ & 8.456 & 8.459 & 8.458 & 8.456 & 8.458 & 8.459 & 8.457 \\         
% 		\hline
% 		\hline
% 	\end{tabular}
% 	\label{tab:ISLRvsN}
% \end{table}

\begin{table}
	\centering
	\caption{The range-\gls{ISLR} obtained by proposed method under the $C_4$ constraint with different sequence lengths ($\eta = 0$, $M_t=8$).}
	\begin{tabular}{c|c|c|c|c|c|c}	
		\hline
		\hline
		$N$ &  32 & 64 & 128  & 256 & 512  & 1024 \\
		\hline
		$L=2$ & 8.4678 & 8.4688 & 8.4687 & 8.4684 & 8.4676 & 8.4675 \\
		$L=8$ & 8.4569 & 8.458 & 8.4578 & 8.457 & 8.4568 & 8.4567 \\
		\hline
		\hline
	\end{tabular}
	\label{tab:ISLRvsN}
\end{table}

\subsection{Beampattern shaping with binary sequences}
\begin{figure}
	\centering
 	\includegraphics[width=1\linewidth]{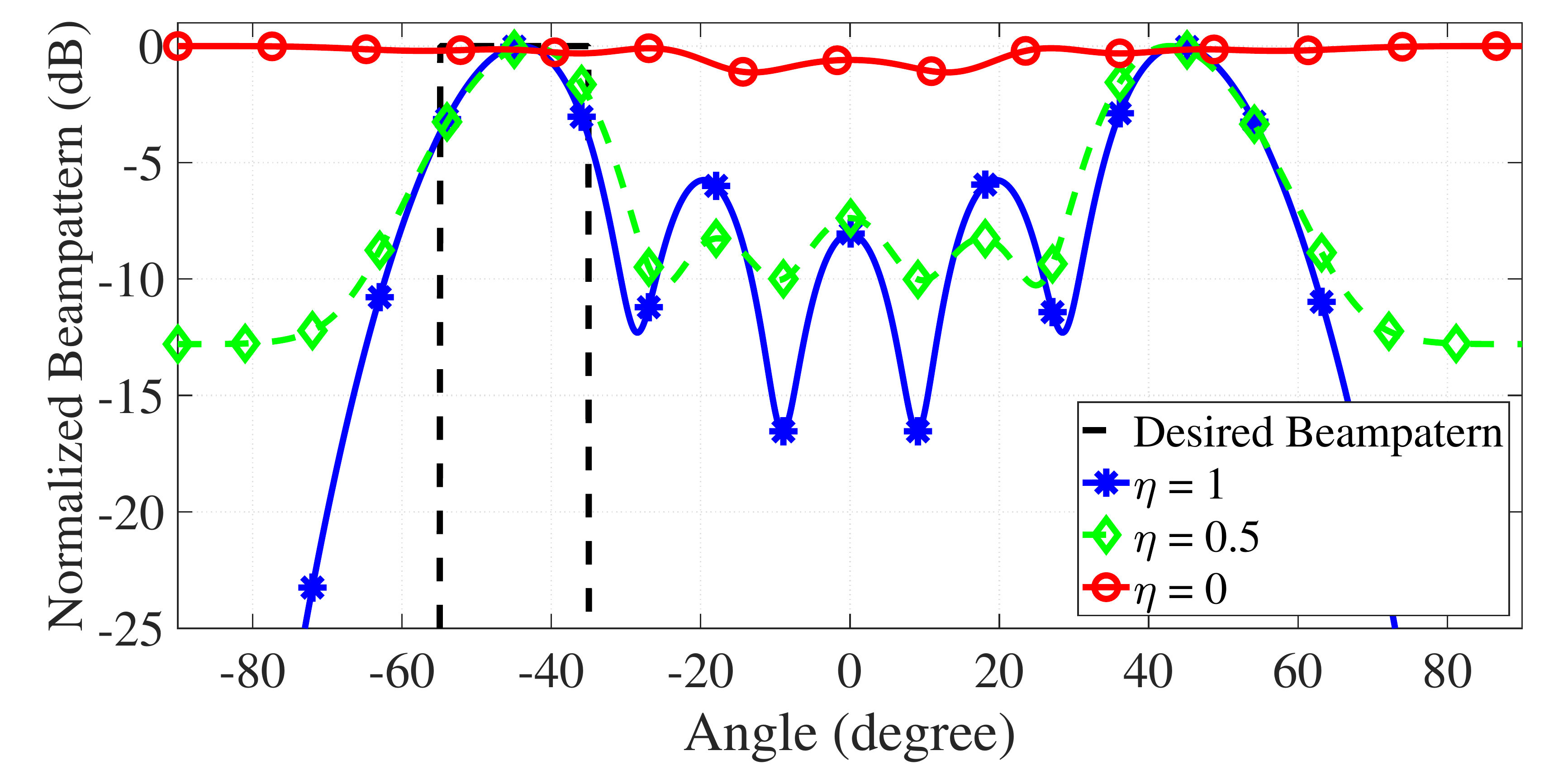}
 	\caption[]{The beampattern response in binary case ($L=2$, $M=8$, $N=1024$, $\Theta_d = [-55^o, -35^o]$ and $\Theta_u = [-90^o,-60^o] \cup [-30^o,90^o]$).}\label{fig:R1-3_bin_BP_N1024_M8}
\end{figure}

Due to the simplicity of implementing of binary sequences, these kind of waveforms are attractive for radar designers. Here we assess the beampattern performance of proposed binary waveform design. \figurename{~\ref{fig:R1-3_bin_BP_N1024_M8}} shows the beampattern response of the proposed method in binary case with different value of $\eta$, where we assume that $\Theta_d = [-55^o, -35^o]$ and $\Theta_u = [-90^o,-60^o] \cup [-30^o,90^o]$. As can be seen with $\eta=1$ we obtain the optimum beampattern response and by decreasing the $\eta$ the beampattern worsens. Besides, the beampattern response in binary case is symmetric about $0^o$. Indeed, in a case when the waveforms are real (binary sequences), the beampattern will be symmetric. 
  
%   {\it proof:} The beampattern response at $\theta$ can be written as,
%   \begin{equation}\label{eq:beampattern}
% 	P(\bS, \theta) = \frac{1}{N}\sum_{n=1}^{N} \ba^H(\theta) \bar{\bs}_n \bar{\bs}_n^H \ba(\theta)
%   \end{equation}
%   Therefore $P(\bS, -\theta)$ can be written as,
%   \begin{equation}
% 	P(\bS, -\theta) = \frac{1}{N}\sum_{n=1}^{N} \ba^H(-\theta) \bar{\bs}_n \bar{\bs}_n^H \ba(-\theta)
%   \end{equation}
%   Since $\ba^H(-\theta) = \ba^T(\theta)$ and $\ba(-\theta) = \ba^*(\theta)$, by some mathematical manipulation it can be shown that,
%   \begin{equation}
%   \begin{aligned}
% 	P(\bS, -\theta) &= \frac{1}{N}\sum_{n=1}^{N} \ba^H(\theta) \bar{\bs}_n^* \bar{\bs}_n^T \ba(\theta)
%   \end{aligned}
%   \end{equation}
%   Since in binary case $\bar{\bs}_n$ is a real vector, hence $\bar{\bs}_n^* \bar{\bs}_n^T$ is a real matrix and $\bar{\bs}_n^* \bar{\bs}_n^T = \left(\bar{\bs}_n^* \bar{\bs}_n^T\right)^* = \bar{\bs}_n \bar{\bs}_n^H$. Therefore it can be concluded that,
%   \begin{equation}
% 	P(\bS, -\theta) = P(\bS, \theta)
%   \end{equation}
  
\begin{figure}
	\centering
 	\includegraphics[width=1\linewidth]{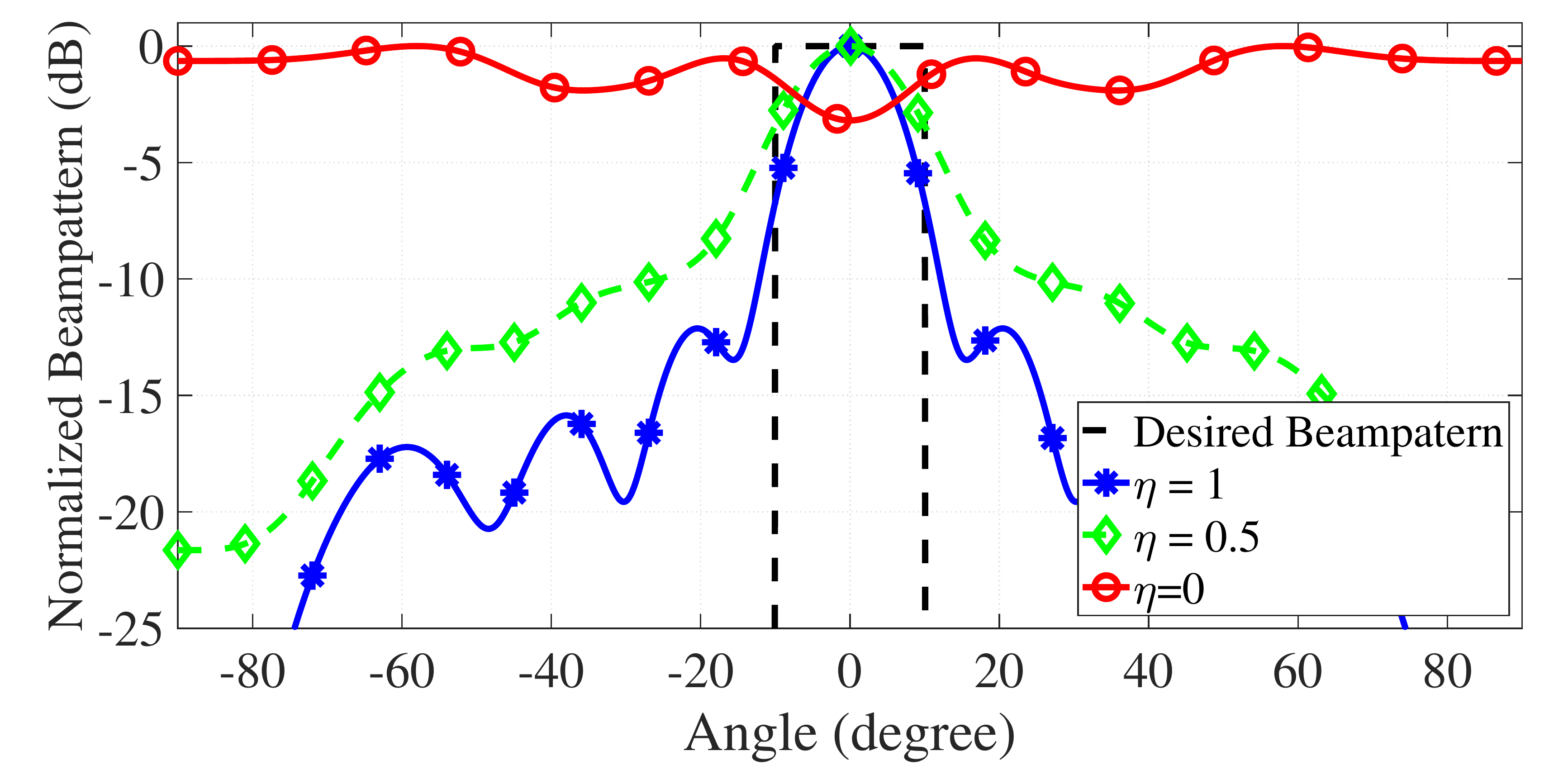}
 	\caption[]{The beampattern response in binary case ($L=2$, $M=8$, $N=1024$, $\Theta_d = [-10^o, 10^o]$ and $\Theta_u = [-90^o, -15^o] \cup [15^o, 90^o]$).}\label{fig:R1-3_bin_BP_N1024_M8_eta1_theta0}
\end{figure}

%   Therefore, in case of designing binary waveforms, the beampattern is even function (symmetric), which is related to the use of real waveforms. 
  In 4D-imaging application of automotive radar systems, the desired region for beampattern shaping can be limited to the angles around zero, where binary codes can be used.
  %which we need to steer the beam on $0^o$. 
  \figurename{~\ref{fig:R1-3_bin_BP_N1024_M8_eta1_theta0}} shows the beampattern response at $\Theta_d = [-10^o, 10^o]$ and $\Theta_u = [-90^o, -15^o] \cup [15^o, 90^o]$ for different $\eta$.

%   \begin{figure*}
%     \centering
%         \includegraphics[width=1\linewidth]{R1-3_bin_BP_N1024_M8_eta1_theta0.eps}
% 		\caption[]{The beampattern response in binary case ($\eta = 0$, $L=2$, $M=8$, $N=1024$ and $\bTheta_d = [-10^o, 10^o]$).}\label{fig:R1-3_bin_BP_N1024_M8_eta1_theta0}
%   \end{figure*}

\subsection{The impact of alphabet size and \gls{PAR}}
\begin{figure*}
    \centering
    \begin{subfigure}{.49\textwidth}
        \centering
		\includegraphics[width=1\linewidth]{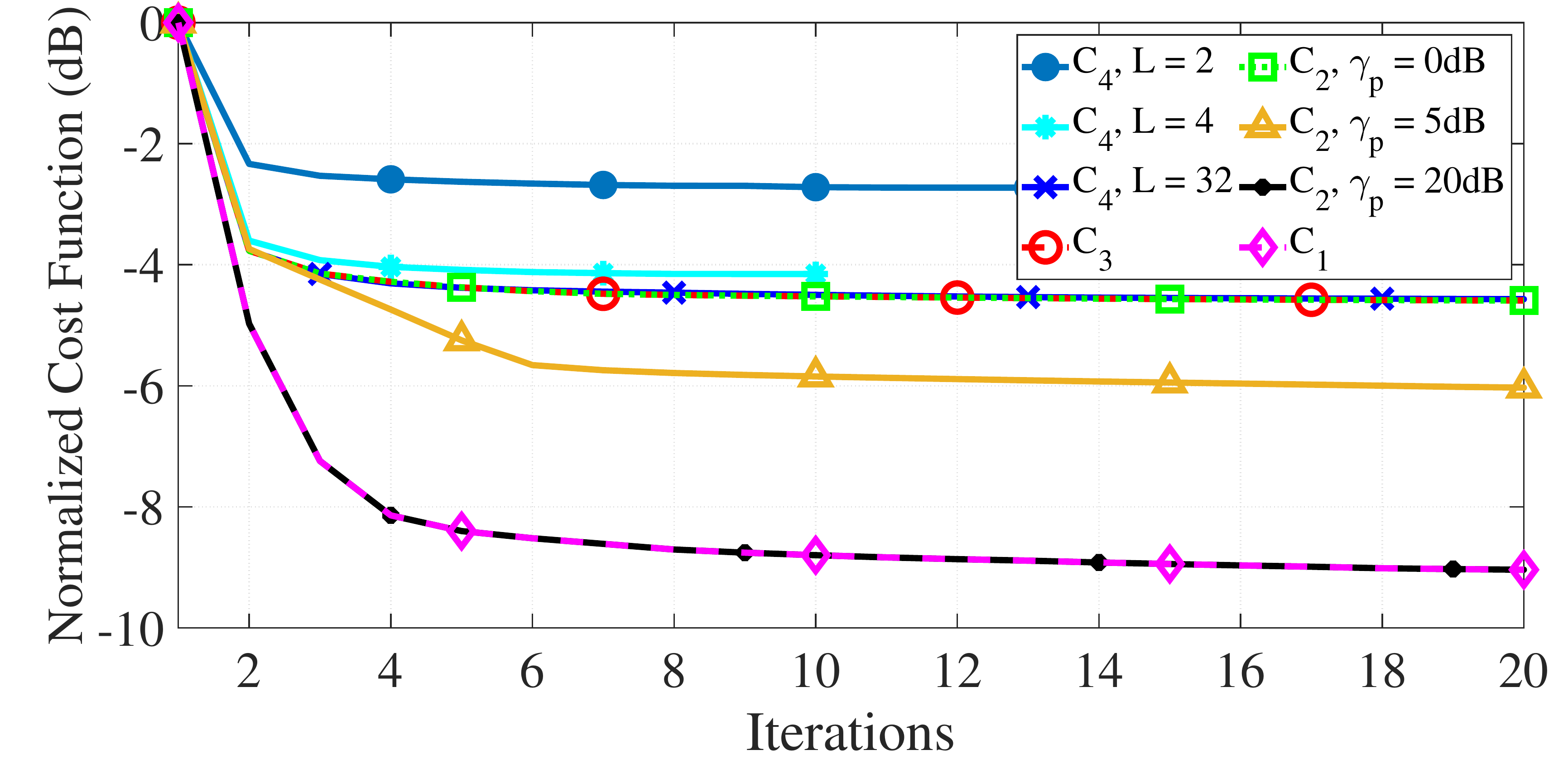}
		\caption[]{Convergence ($\eta = 0.95)$.}\label{fig:PAR_eta_impact_Convergence}
    \end{subfigure}
    \begin{subfigure}{.49\textwidth}
        \centering
		\includegraphics[width=1\linewidth]{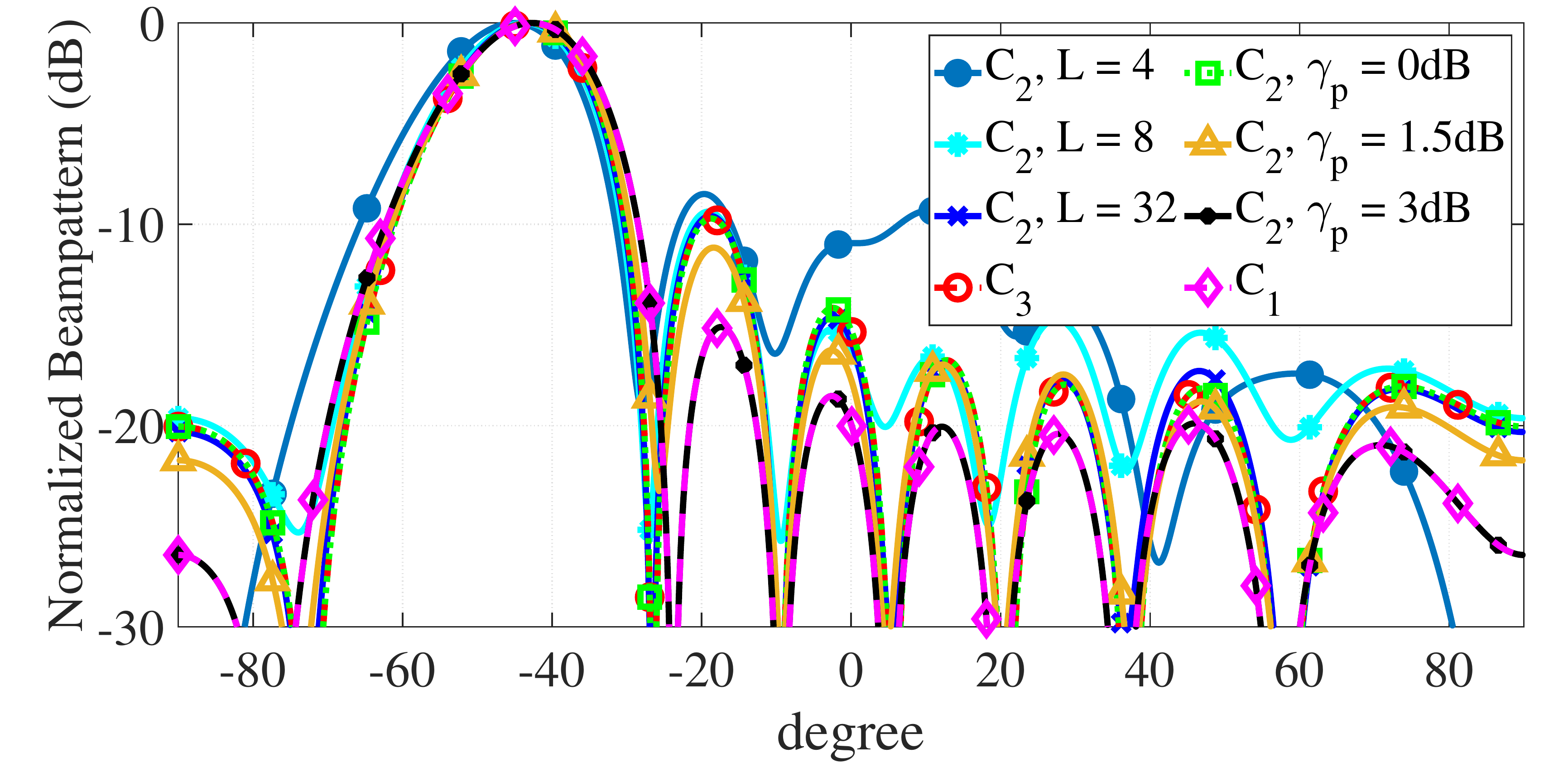}
		\caption[]{Beampattern ($\eta = 1$, $\Theta_d = [-55^o,-35^o]$ and $\Theta_u = [-90^o,-60^o] \cup [-30^o,90^o]$).}\label{fig:PAR_eta_Beam}
    \end{subfigure}
    \begin{subfigure}{.49\textwidth}
        \centering
		\includegraphics[width=1\linewidth]{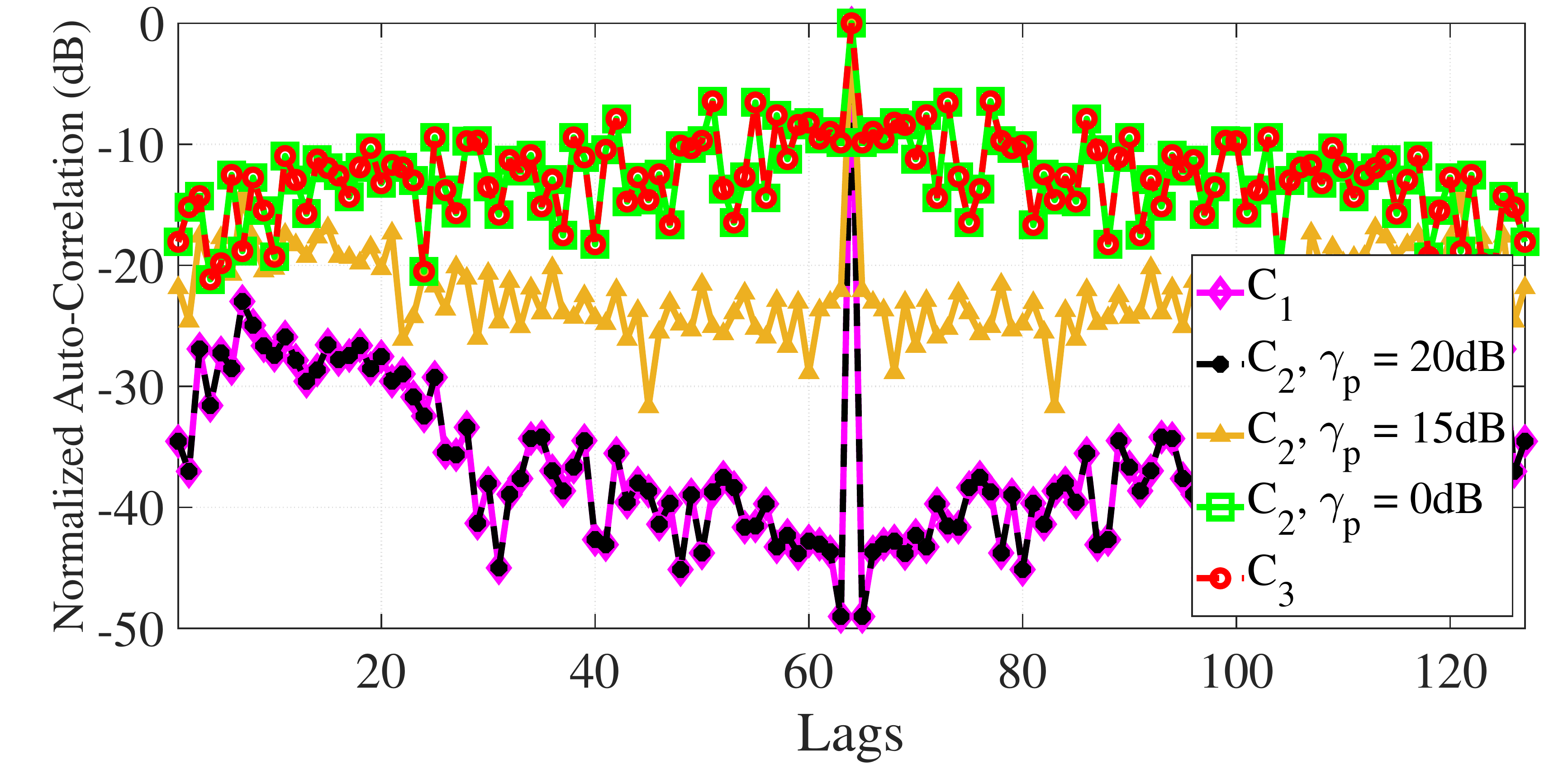}
		\caption[]{Auto-correlation ($\eta = 0$).}\label{fig:PAR_eta_autocorrelation}
    \end{subfigure}
    \begin{subfigure}{.49\textwidth}
        \centering
		\includegraphics[width=1\linewidth]{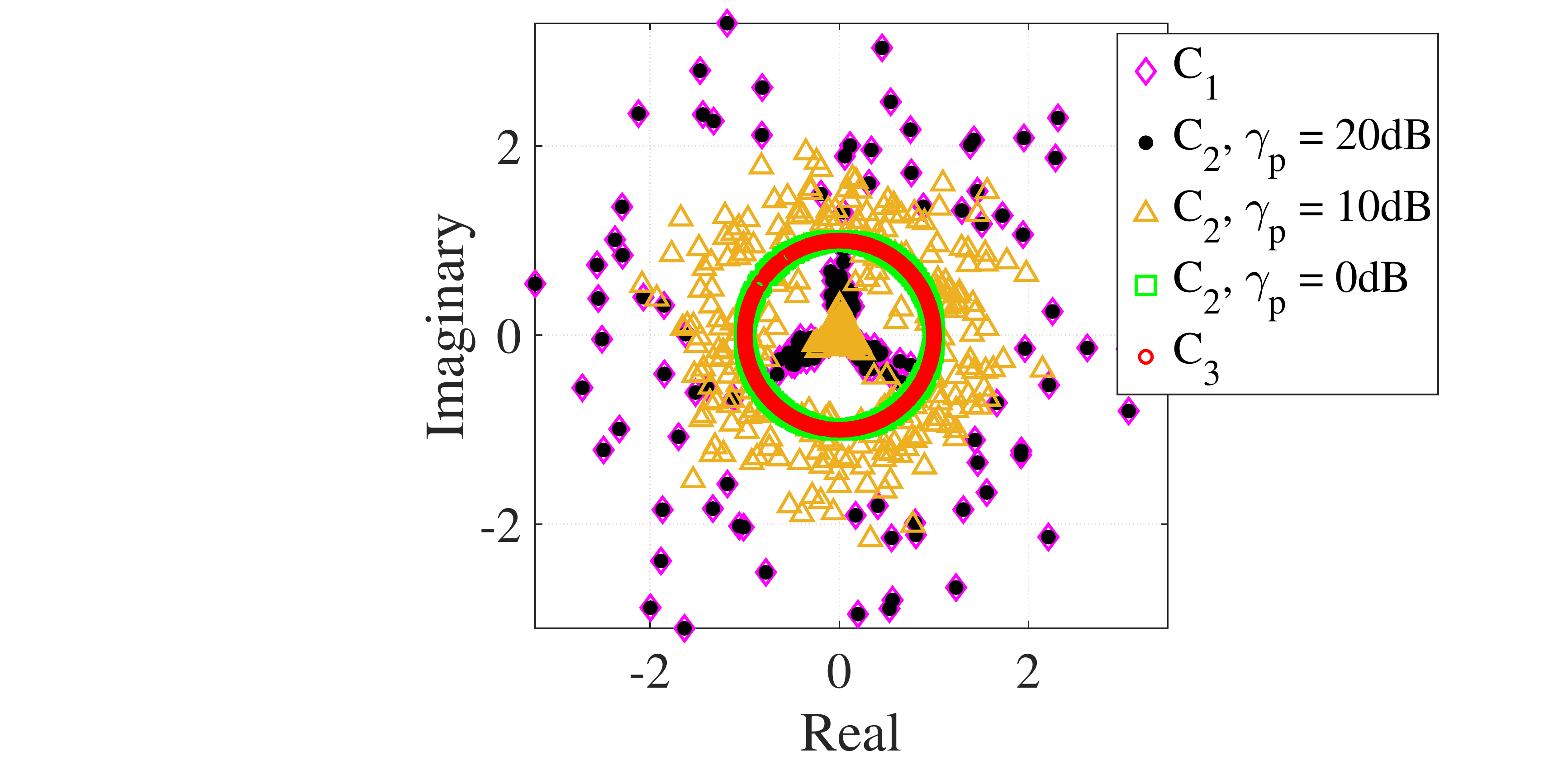}
		\caption[]{Constellation ($\eta = 1$).}\label{fig:PAR_eta_Constellation}
    \end{subfigure}
    \caption[]{The impact of alphabet size ((a) and (b)) and \gls{PAR} ((c) and (d)) ($M=8$ and $N=64$).}\label{fig:PAR_eta_impact}
\end{figure*}

\figurename{~\ref{fig:PAR_eta_impact}}a,b and \figurename{~\ref{fig:PAR_eta_impact}}c,d shows the impact of alphabet size and \gls{PAR} in several aspects respectively. As is evident, the solution of $C_4$ approaches that of $C_3$ for large alphabet sizes. This behavior is expected since the feasible set of $C_4$ will be close to that of $C_3$, and the optimized solutions will behave the same. On the other hand, by increasing \gls{PAR} threshold, the feasible set under $C_2$ constraint converges to $C_1$. By decreasing \gls{PAR} threshold to $1$, the feasible set will be limited to that specified in $C_3$.

\section{Conclusion}
In this paper, we aimed to effect a trade-off between  beampattern response and orthogonality using spatial- and range-\gls{ISLR} as representative figures of merit. Accordingly, we introduced a bi-objective Pareto framework to minimize the two metrics simultaneously for \gls{MIMO} radar systems, under power budget, \gls{PAR}, continuous and discrete phase constraints. The problem formulation led to a non-convex, multi-variable and NP-hard optimization problem. To tackle the problem, we proposed an iterative method based on \gls{CD}; in each of its steps we utilized an effective method to minimize the objective function. Specifically, we used a gradient based method under energy budget, \gls{PAR} and continuous phase constraints; a \gls{FFT}-based method was used under the discrete phase constraint.

Simulation results have illustrated the monotonicity of the proposed method in minimizing the objective function as well as the contradiction in minimizing the two \gls{ISLR}s. In this context, the proposed method is capable of effecting an optimal trade-off between the two. The paper also provided a Pareto curve aided cognitive radar system to decide on the operating levels of the two \gls{ISLR}s. Besides, the proposed framework also shows good performance in comparison to counterparts when  used for minimizing the spatial- and range-\gls{ISLR} individually; this indicates the flexibility offered by the framework.

Possible future research directions includes the consideration of Doppler filter bank and spectrum shaping for enhanced cognition in \gls{MIMO} radar systems. 

% \appendices

% % you can choose not to have a title for an appendix
% % if you want by leaving the argument blank
% \section{}

%\appendix
\appendices
% \section{}\label{app:1}
% The integrated auto and cross-correlation level of transmitted signals is defined as,
% \begin{equation}\label{eq:cross_correlation_level}
% \begin{aligned}
%     &\sum_{m=1}^{M}\sum_{l=1}^{M}\sum_{k=-N+1}^{N-1}|r_{m,l}(k)|^2 = \sum_{m=1}^{M}\sum_{\substack{{l=1}\\{l \neq m}}}^{M}\sum_{k=-N+1}^{N-1}|r_{m,l}(k)|^2 + \\
% 	& \sum_{m=1}^{M}\sum_{\substack{{k=-N+1}\\{k \neq 0}}}^{N-1}|r_{m,m}(k)|^2 + \sum_{m=1}^{M}|r_{m,m}(0)|^2.
% 	\end{aligned}
% \end{equation}
% The first and second terms of above equation represents the cross- and auto-correlation \gls{ISL} of the waveform. Therefore \gls{ISL} can be written as,
% \begin{equation}
% 	\text{ISL} = \sum_{m=1}^{M}\sum_{l=1}^{M}\sum_{k=-N+1}^{N-1}|r_{m,l}(k)|^2 - \sum_{m=1}^{M}|r_{m,m}(0)|^2. 
% \end{equation}

\section{}\label{app:2}
Writing \eqref{eq:sum_weighted} with respect to $s_{t,d}$ has the following parts.
\paragraph{Spatial-\gls{ISLR} coefficients} Beampattern of undesired angles can be written as,
\begin{equation*}
	\begin{aligned}                                  
	&\textstyle \sum_{n=1}^{N}\bar{\bs}_n^H\bA_u\bar{\bs}_n = \sum_{\substack{{n = 1}\\{n \neq d}}}^{N}\bar{\bs}_n^H\bA_u\bar{\bs}_n + \bar{\bs}_d^H\bA_u\bar{\bs}_d,
	\end{aligned}
\end{equation*}
where the second term can be expanded as,
\begin{equation*}
	\begin{aligned}                                  
	&\textstyle \sum_{m=1}^{M_t}\sum_{l=1}^{M_t} s_{m,d}^*a_{u_{m,l}}s_{l,d} = \textstyle \sum_{\substack{{m = 1}\\{m \neq t}}}^{M_t}\sum_{\substack{{l = 1}\\{l \neq t}}}^{M_t} s_{m,d}^*a_{u_{m,l}}s_{l,d} \\
	& \textstyle s_{t,d}\sum_{\substack{{m = 1}\\{m \neq t}}}^{M_t}s_{m,d}^*a_{u_{m,t}} + s_{t,d}^*\sum_{\substack{{l = 1}\\{l \neq t}}}^{M_t}a_{u_{t,l}}s_{l,d} + s_{t,d}^*a_{u_{t,t}}s_{t,d},
	\end{aligned}
\end{equation*}
with $a_{u_{m,l}}$ indicating $\{m,l\}$ entries of matrix $\bA_u$. Defining,
% \begin{equation}\label{eq:coeff_nom}
% 	\begin{aligned}  
% 	a_0 &\triangleq \sum_{\substack{{m = 1}\\{m \neq t}}}^{M}s_{m,d}^*a_{u_{m,t}}, \quad 
% 	a_3 \triangleq a_{u_{t,t}}, \quad
% 	a_2 \triangleq \sum_{\substack{{l = 1}\\{l \neq t}}}^{M}a_{u_{t,l}}s_{m,d},\\
% 	a_1 &\triangleq \sum_{\substack{{m = 1}\\{m \neq t}}}^{M}\sum_{\substack{{l = 1}\\{l \neq t}}}^{M}\sum_{n=1}^{N} s_{m,n}^*a_{u_{m,l}}s_{l,n}
% 	+ \sum_{\substack{{m = 1}\\{m \neq t}}}^{M}\sum_{\substack{{n = 1}\\{n \neq d}}}^{N} s_{m,n}^*a_{u_{m,t}}s_{t,n}\\
% 	&+ \sum_{\substack{{l = 1}\\{l \neq t}}}^{M}\sum_{\substack{{n = 1}\\{n \neq d}}}^{N} s_{t,n}^*a_{u_{t,l}}s_{l,n}
% 	+ \sum_{\substack{{n = 1}\\{n \neq d}}}^{N} s_{t,n}^*a_{u_{t,t}}s_{t,n}
% 	\end{aligned}
% \end{equation}
% \begin{equation*}\label{eq:coeff_nom}
% 	\begin{aligned}  
% 	a_0 &\triangleq \textstyle \sum_{\substack{{m = 1}\\{m \neq t}}}^{M}s_{m,d}^*a_{u_{m,t}}, \quad 
% 	a_3 \triangleq a_{u_{t,t}}, \quad
% 	a_2 \triangleq a_0^*,\\
% 	a_1 &\triangleq \textstyle \sum_{\substack{{m = 1}\\{m \neq t}}}^{M}\sum_{\substack{{l = 1}\\{l \neq t}}}^{M}\sum_{n=1}^{N} s_{m,n}^*a_{u_{m,l}}s_{l,n} + \sum_{\substack{{n = 1}\\{n \neq d}}}^{N} s_{t,n}^*a_{u_{t,t}}s_{t,n}
% 	\\
% 	&+ \textstyle\sum_{\substack{{l = 1}\\{l \neq t}}}^{M}\sum_{\substack{{n = 1}\\{n \neq d}}}^{N} s_{t,n}^*a_{u_{t,l}}s_{l,n}
% 	+ \sum_{\substack{{m = 1}\\{m \neq t}}}^{M}\sum_{\substack{{n = 1}\\{n \neq d}}}^{N} s_{m,n}^*a_{u_{m,t}}s_{t,n},
% 	\end{aligned}
% \end{equation*}
\begin{equation*}\label{eq:coeff_nom}
	\begin{aligned}  
	a_0 &\triangleq \textstyle \sum_{\substack{{m = 1}\\{m \neq t}}}^{M_t}s_{m,d}^*a_{u_{m,t}}, \quad 
	a_3 \triangleq a_{u_{t,t}}, \quad
	a_2 \triangleq a_0^*,\\
	a_1 &\triangleq \textstyle \sum_{\substack{{n = 1}\\{n \neq d}}}^{N}\bar{\bs}_n^H\bA_u\bar{\bs}_n + \sum_{\substack{{m = 1}\\{m \neq t}}}^{M_t}\sum_{\substack{{l = 1}\\{l \neq t}}}^{M_t} s_{m,d}^*a_{u_{m,l}}s_{l,d},
	\end{aligned}
\end{equation*}
the beampattern response on undesired angles is equivalent to,
\begin{equation}
	\textstyle \sum_{n=1}^{N}\bar{\bs}_n^H\bA_u\bar{\bs}_n = a_0s_{t,d} + a_1 + a_2s_{t,d}^* + a_3s_{t,d}^*s_{t,d}.
\end{equation}
Like wise the beampattern at desired angles is:
\begin{equation}
	\textstyle \sum_{n=1}^{N}\bar{\bs}_n^H\bA_d\bar{\bs}_n = b_0s_{t,d} + b_1 + b_2s_{t,d}^* + b_3s_{t,d}^*s_{t,d},
\end{equation}
%where,
% \begin{equation}\label{eq:coeff_den}
% 	\begin{aligned}  
% 	b_0 &\triangleq \sum_{\substack{{m = 1}\\{m \neq t}}}^{M}s_{m,d}^*a_{d_{m,t}}, \quad 
% 	b_3 \triangleq a_{u_{t,t}}, \quad
% 	b_2 \triangleq \sum_{\substack{{l = 1}\\{l \neq t}}}^{M}a_{d_{t,l}}s_{m,d},\\
% 	b_1 &\triangleq \sum_{\substack{{m = 1}\\{m \neq t}}}^{M}\sum_{\substack{{l = 1}\\{l \neq t}}}^{M}\sum_{n=1}^{N} s_{m,n}^*a_{d_{m,l}}s_{l,n}
% 	+ \sum_{\substack{{m = 1}\\{m \neq t}}}^{M}\sum_{\substack{{n = 1}\\{n \neq d}}}^{N} s_{m,n}^*a_{d_{m,t}}s_{t,n} \\
% 	&+ \sum_{\substack{{l = 1}\\{l \neq t}}}^{M}\sum_{\substack{{n = 1}\\{n \neq d}}}^{N} s_{t,n}^*a_{d_{t,l}}s_{l,n}
% 	+ \sum_{\substack{{n = 1}\\{n \neq d}}}^{N} s_{t,n}^*a_{d_{t,t}}s_{t,n}
% 	\end{aligned}
% \end{equation}
% \begin{equation*}\label{eq:coeff_den}
% 	\begin{aligned}  
% 	b_0 &\triangleq \textstyle \sum_{\substack{{m = 1}\\{m \neq t}}}^{M}s_{m,d}^*a_{d_{m,t}}, \quad 
% 	b_3 \triangleq a_{u_{t,t}}, \quad
% 	b_2 \triangleq b_0^*,\\
% 	b_1 &\triangleq \textstyle \sum_{\substack{{m = 1}\\{m \neq t}}}^{M}\sum_{\substack{{l = 1}\\{l \neq t}}}^{M}\sum_{n=1}^{N} s_{m,n}^*a_{d_{m,l}}s_{l,n} + \sum_{\substack{{n = 1}\\{n \neq d}}}^{N} s_{t,n}^*a_{d_{t,t}}s_{t,n} \\
% 	&+ \textstyle \sum_{\substack{{l = 1}\\{l \neq t}}}^{M}\sum_{\substack{{n = 1}\\{n \neq d}}}^{N} s_{t,n}^*a_{d_{t,l}}s_{l,n}
% 	+ \sum_{\substack{{m = 1}\\{m \neq t}}}^{M}\sum_{\substack{{n = 1}\\{n \neq d}}}^{N} s_{m,n}^*a_{d_{m,t}}s_{t,n}.
% 	\end{aligned}
% \end{equation*}
\begin{equation*}\label{eq:coeff_den}
	\begin{aligned}  
	b_0 &\triangleq \textstyle \sum_{\substack{{m = 1}\\{m \neq t}}}^{M_t}s_{m,d}^*a_{d_{m,t}}, \quad 
	b_3 \triangleq a_{d_{t,t}}, \quad
	b_2 \triangleq b_0^*,\\
	b_1 &\triangleq \textstyle \sum_{\substack{{n = 1}\\{n \neq d}}}^{N}\bar{\bs}_n^H\bA_d\bar{\bs}_n + \sum_{\substack{{m = 1}\\{m \neq t}}}^{M_t}\sum_{\substack{{l = 1}\\{l \neq t}}}^{M_t} s_{m,d}^*a_{d_{m,l}}s_{l,d},
	\end{aligned}
\end{equation*}
% Like wise as $\bA_d$ is a hermitian matrix, can be concluded that $b_2 = b^*_0$ and both  $b_1$ and $b_3$ are a real and positive value ($b_1, b_3 \in \mathbb{R^+}$) and the beampattern response on desired angles is a real and positive function. 
where $a_{d_{m,l}}$ are the $\{m,l\}$ entries of $\bA_d$. %Therefore $\bar{f}(\bS)$ can be written with respect to $s_{t,d}$ as
\eqref{eq:f_bar_std}. 
% follow,
% \begin{equation}
% 	\bar{f}(s_{t,d},\bS_{-(t,d)}) = \frac{ a_0s_{t,d} + \hat{a}_1 + a_2s_{t,d}^* + \hat{a}_3s_{t,d}^*s_{t,d} }{ b_0s_{t,d} + \hat{b}_1 + b_2s_{t,d}^* + \hat{b}_3s_{t,d}^*s_{t,d} }
% \end{equation}

\paragraph{Range-\gls{ISLR} coefficients} 
\eqref{eq:ISL2} can be written as,
\begin{comment}
\begin{equation*}
	\begin{aligned}
	&\text{ISL} = \textstyle \sum_{\substack{{m=1}\\{m \neq t}}}^{M_t}\sum_{\substack{{l=1}\\{l \neq t}}}^{M_t}\sum_{k=-N+1}^{N-1}|r_{m,l}(k)|^2 \\
	&+ \textstyle\sum_{\substack{{l=1}\\{l \neq t}}}^{M_t}\sum_{k=-N+1}^{N-1}|r_{t,l}(k)|^2 + \sum_{\substack{{m=1}\\{m \neq t}}}^{M_t}\sum_{k=-N+1}^{N-1}|r_{m,t}(k)|^2 \\
	&+ \textstyle \sum_{k=-N+1}^{N-1}|r_{t,t}(k)|^2 - \sum_{\substack{{m=1}\\{m \neq t}}}^{M_t}|r_{m,m}(0)|^2 + |r_{t,t}(0)|^2.
	\end{aligned}
\end{equation*}
\end{comment}
\begin{equation*}
	\begin{aligned}
	&\text{ISL} = \gamma_t + \textstyle \sum_{k=-N+1}^{N-1}|r_{t,t}(k)|^2 + |r_{t,t}(0)|^2 \\
	&+ \textstyle\sum_{\substack{{l=1}\\{l \neq t}}}^{M_t}\sum_{k=-N+1}^{N-1}|r_{t,l}(k)|^2 + \sum_{\substack{{m=1}\\{m \neq t}}}^{M_t}\sum_{k=-N+1}^{N-1}|r_{m,t}(k)|^2.
	\end{aligned}
\end{equation*}
where,
\begin{equation*}
    \begin{aligned}
	&\gamma_t \triangleq \textstyle \sum_{\substack{{m=1}\\{m \neq t}}}^{M_t}\sum_{\substack{{l=1}\\{l \neq t}}}^{M_t}\sum_{k=-N+1}^{N-1}|r_{m,l}(k)|^2 - \sum_{\substack{{m=1}\\{m \neq t}}}^{M_t}|r_{m,m}(0)|^2 \\ 
% 	= &\textstyle \sum_{\substack{{m=1}\\{m \neq t}}}^{M}\sum_{\substack{{l=1}\\{l \neq t}}}^{M}\sum_{k=-N+1}^{N-1}|(\tilde{\bs}_m \circledast \tilde{\bs}_l)_k|^2 - \sum_{\substack{{m=1}\\{m \neq t}}}^{M}|(\tilde{\bs}_m \circledast \tilde{\bs}_m)_0|^2
	\end{aligned}
\end{equation*}
%The cross-correlations $r_{m,t}(k)$, $r_{t,l}(k)$ and auto-correlation $r_{t,t}(k)$ can be written as,
Also,
\begin{equation*}
    \begin{aligned}
	r_{m,t}(k) &= \textstyle \sum_{\substack{{n=1}\\{n \neq d-k}}}^{N-k}s_{m,n}s_{t,n+k}^* + s_{m,d-k}s_{t,d}^*I_A(d-k)\\
	r_{t,l}(k) &= \textstyle \sum_{\substack{{n=1}\\{n \neq d}}}^{N-k}s_{t,n}s_{l,n+k}^* + s_{t,d}s_{l,d+k}^*I_A(d+k) \\
	r_{t,t}(k) &= \textstyle \sum_{\substack{{n=1}\\{n \neq d, n \neq d-k}}}^{N-k}s_{t,n}s_{t,n+k}^* + s_{t,d}s_{t,d+k}^*I_A(d+k)\\
	&+ s_{t,d}^*s_{t,d-k}I_A(d-k)
	\end{aligned}
\end{equation*}
where, $I_A(p)$ is the indicator function of set $A = \left \{1, \dots, N\right \}$, i.e, $I_A(p) \triangleq \begin{dcases}
	1, & p \in A\\
	0, & p \notin A
	\end{dcases}$. 
% \begin{equation*}
% 	I_A(p) \triangleq
% 	\begin{dcases}
% 	1, & p \in A\\
% 	0, & p \notin A
% 	\end{dcases}
% \end{equation*}
Let us define \footnote{By defining, $\tilde{\bs}_{\{t,m,l\}_{-d}} \triangleq \tilde{\bs}_m|_{s_{\{t,m,l\},d}=0}$, it can be shown that, $\gamma_{mtdk}$, $\gamma_{tldk}$ and $\gamma_{ttdk}$ can be considered as correlation of $\tilde{\bs}_{m_{-d}}$ and $\tilde{\bs}_{t_{-d}}$, $\tilde{\bs}_{t_{-d}}$ and $\tilde{\bs}_{l_{-d}}$, $\tilde{\bs}_{t_{-d}}$ and $\tilde{\bs}_{t_{-d}}$ respectively.},
\begin{equation*}
\begin{aligned}
	\gamma_{mtdk} &\triangleq \textstyle \sum_{\substack{{n=1}\\{n \neq d-k}}}^{N-k}s_{m,n}s_{t,n+k}^*, \beta_{mtdk} \triangleq s_{m,d-k}I_A(d-k)\\
	\gamma_{tldk} &\triangleq \textstyle \sum_{\substack{{n=1}\\{n \neq d}}}^{N-k}s_{t,n}s_{l,n+k}^*, \alpha_{tldk} \triangleq s_{l,d+k}^*I_A(d+k)\\
	\gamma_{ttdk} &\triangleq \textstyle \sum_{\substack{{n=1}\\{n \neq d, n \neq d-k}}}^{N-k}s_{t,n}s_{t,n+k}^*, \alpha_{ttdk} \triangleq s_{t,d+k}^*I_A(d+k) \\
	\beta_{ttdk} &\triangleq s_{t,d-k}I_A(d-k)
\end{aligned}
\end{equation*}
% {\color{red} What do you mean?
% \textit{\underline{Note}}: Defining, $\tilde{\bs}_{\{t,m,l\}_{-d}} \triangleq \tilde{\bs}_m|_{s_{\{t,m,l\},d}=0}$, it can be shown that, $\gamma_{mtdk}$, $\gamma_{tldk}$ and $\gamma_{ttdk}$ can be considered as correlation of $\tilde{\bs}_{m_{-d}}$ and $\tilde{\bs}_{t_{-d}}$, $\tilde{\bs}_{t_{-d}}$ and $\tilde{\bs}_{l_{-d}}$, $\tilde{\bs}_{t_{-d}}$ and $\tilde{\bs}_{t_{-d}}$ respectively.
% }
Thus, we obtain,
\begin{equation}\label{eq:ApISL}
	\text{ISL} = c_0s_{t,d}^2 + c_1s_{t,d} + c_2 + c_3s_{t,d}^* + c_4{s_{t,d}^*}^2 + c_5|s_{t,d}|^2
\end{equation}
with,
\begin{fleqn}
\begin{equation*}
	\begin{aligned}
	c_0 &\triangleq \textstyle \sum_{\substack{{k = -N+1}\\{k \neq 0}}}^{N-1}\alpha_{ttdk}\beta_{ttdk}^*, \quad c_4 \triangleq c_0^*,
	\end{aligned}
\end{equation*}
\end{fleqn}
\begin{fleqn}
\begin{equation*}
	\begin{aligned}
	&c_3 \triangleq c_1^*, \quad c_1 \triangleq \textstyle \sum_{\substack{{k=-N+1}\\{k \neq 0}}}^{N-1}(\gamma_{ttdk}^*\alpha_{ttdk}+\gamma_{ttdk}\beta_{ttdk}^*) +\\ 
	& \textstyle \sum_{\substack{{l=1}\\{l \neq t}}}^{M_t}\sum_{k=-N+1}^{N-1}\gamma_{tldk}^*\alpha_{tldk} + \sum_{\substack{{m=1}\\{m \neq t}}}^{M_t}\sum_{k=-N+1}^{N-1}\gamma_{mtdk}\beta_{mtdk}^*,
	\end{aligned}
\end{equation*}
\end{fleqn}
\begin{fleqn}
\begin{equation*}
	\begin{aligned}
	c_2 & \textstyle \triangleq \sum_{\substack{{k=-N+1}\\{k \neq 0}}}^{N-1}|\gamma_{ttdk}|^2 + \sum_{\substack{{l=1}\\{l \neq t}}}^{M_t}\sum_{k=-N+1}^{N-1}|\gamma_{tldk}|^2 \\
	&+ \textstyle \sum_{\substack{{m=1}\\{m \neq t}}}^{M_t}\sum_{k=-N+1}^{N-1}|\gamma_{mtdk}|^2 + \gamma_t,
	\end{aligned}
\end{equation*}
\end{fleqn}
\begin{fleqn}
\begin{equation*}
	\begin{aligned}
	c_5 &\triangleq \textstyle \sum_{\substack{{k=-N+1}\\{k \neq 0}}}^{N-1}(|\alpha_{ttdk}|^2+|\beta_{ttdk}|^2) \\
	&+ \textstyle \sum_{\substack{{l=1}\\{l \neq t}}}^{M_t}\sum_{k=-N+1}^{N-1}|\alpha_{tldk}|^2 + \sum_{\substack{{m=1}\\{m \neq t}}}^{M_t}\sum_{k=-N+1}^{N-1}|\beta_{mtdk}|^2.
	\end{aligned}
\end{equation*}
\end{fleqn}
Since $c_0 = c_4^*$, $c_1 = c_3^*$ and $c_1$, $c_5$ are real coefficient, \eqref{eq:ApISL} is a real and non-negative function. Also for the mainlobe,  
\begin{equation*}
	\begin{aligned}  
	& \textstyle \sum_{m=1}^{M_t}|r_{m,m}(0)|^2 = \textstyle \sum_{\substack{{m=1}\\{m \neq t}}}^{M_t}\left(\sum_{n=1}^{N} |s_{m,n}|^2\right)^2 + \\ 
	&\textstyle \left(\sum_{{\substack{{n=1}\\{n \neq d}}}}^{N}|s_{t,n}|^2\right)^2 + 2|s_{t,d}|^2\sum_{{\substack{{n=1}\\{n \neq d}}}}^{N}|s_{t,n}|^2 + |s_{t,d}|^4.
	\end{aligned}  
\end{equation*}
Defining, $d_2 \textstyle \triangleq \sum_{\substack{{m=1}\\{m \neq t}}}^{M_t}\left(\sum_{n=1}^{N} |s_{m,n}|^2\right)^2 + \left(\sum_{{\substack{{n=1}\\{n \neq d}}}}^{N}|s_{t,n}|^2\right)^2$ and $d_1 \textstyle \triangleq 2\sum_{{\substack{{n=1}\\{n \neq d}}}}^{N}|s_{t,n}|^2$,
% \begin{equation}\label{eq:d_coeff}
% 	d_2 \textstyle \triangleq \sum_{\substack{{m=1}\\{m \neq t}}}^{M}\left(\sum_{n=1}^{N} |s_{m,n}|^2\right)^2 + \left(\sum_{{\substack{{n=1}\\{n \neq d}}}}^{N}|s_{t,n}|^2\right)^2,
% \end{equation}
we have, 
\begin{equation}
	\textstyle \sum_{m=1}^{M_t}|r_{m,m}(0)|^2 = |s_{t,d}|^4 + d_1|s_{t,d}|^2 + d_2.
\end{equation}
%Therefore $\tilde{f}(\bS)$ can be written with respect to $s_{t,d}$ as \eqref{eq:f_tld_std}.
% \begin{equation}
%     \begin{aligned}
%     &\tilde{f}\left(s_{t,d}, \bS^{(i)}_{-(t,d)}\right) = \\ &\frac{c_0s_{t,d}^2+c_1s_{t,d}+c_2+c_3s_{t,d}^*+c_4{s_{t,d}^*}^2+c_5|s_{t,d}|^2}{|s_{t,d}|^4+d_1|s_{t,d}|^2+d_2}
%     \end{aligned}
% \end{equation}
\paragraph{$C_1$ Constraint} It is straight-forward to show that,
\begin{equation}
	\gamma_e \triangleq M_t N - \textstyle \sum_{\substack{{m=1}\\{m \neq t}}}^{M_t}\sum_{n=1}^{N}|s_{m,n}|^2 - \sum_{\substack{{n=1}\\{n \neq d}}}^{N}|s_{t,n}|^2.
\end{equation}
\underline{Note}: $\norm{\bS}_F^2 = \textstyle \sum_{\substack{{m=1}\\{m \neq t}}}^{M_t}\sum_{n=1}^{N}|s_{m,n}|^2 + \sum_{\substack{{n=1}\\{n \neq d}}}^{N}|s_{t,n}|^2 + |s_{t,d}|^2$.
\paragraph{$C_2$ Constraint}
The \gls{PAR} constraint can be written as, $M_t N\max {|s_{m,n}|}^2 \leqslant \gamma_p\norm{\bS}_F^2$. Defining $P_{-(t,d)} \triangleq \max\{|s_{m,n}|^2; (m,n) \neq (t,d)\}$, we obtain
\begin{equation*}
M_t N\max \{|s_{t,d}|^2, P_{-(t,d)}\} \leqslant \gamma_p \left(|s_{t,d}|^2 + \norm{\bS_{-(t,d)}}_F^2 \right)
\end{equation*}
Defining, 
\begin{equation*}
    \gamma_l \triangleq \frac{M_t NP_{-(t,d)} - \gamma_p\norm{\bS_{-(t,d)}}_F^2}{\gamma_p}, \quad \gamma_u \triangleq \frac{\gamma_p\norm{\bS_{-(t,d)}}_F^2}{M_t N-\gamma_p},
\end{equation*}
Hence, $|s_{t,d}|^2 \geqslant \gamma_l$ when $|s_{t,d}|^2 \leqslant P_{-(t,d)}$, and $|s_{t,d}|^2 \leqslant \gamma_u$ when $|s_{t,d}|^2 \geqslant P_{-(t,d)}$.

\section{}\label{app:3}
Considering $a_2 = a_0^*$, $b_2 = b_0^*$, $c_4 = c_0^*$ and $c_3 = c_1^*$, the \eqref{eq:f_bar_rphi} and \eqref{eq:f_tld_rphi} can be written as \footnote{It is possible to consider $e^{j\phi}$ as the variable and solve the problem. However, we reformulate the problem in the real variable to enable computations in real domain to be closer to practical implementation.},
\begin{equation}
    \begin{aligned}
	\bar{f}\left(r, \phi \right) &= \frac{ 2\Re\{a_0re^{j\phi}\} + a_1 + a_3r^3 }{ 2\Re\{b_0re^{j\phi}\} + b_1 + b_3r^3 } \\
	&= \frac{a_3r^2 + 2(a_{0r}\cos{\phi}-a_{0i}\sin{\phi})r + a_1}{b_3r^2 + 2(b_{0r}\cos{\phi}-b_{0i}\sin{\phi})r + b_1}
	\end{aligned}
\end{equation}
\begin{equation}
\begin{aligned}
	&\tilde{f}\left(r, \phi \right) = \frac{ 2\Re\{c_0r^2e^{j2\phi}\} + 2\Re\{c_1re^{j\phi}\} + c_2 +c_5r^2 }{ r^4 + d_1r^2 + d_2 }\\
	&= [(2c_{0r}\cos{2\phi}-2c_{0i}\sin{2\phi}+c_5)r^2\\
	&+ 2(c_{1r}\cos{\phi}-c_{1i}\sin{\phi})r + c_2]\frac{1}{r^4 + d_1r^2 + d_2},
	\end{aligned}
\end{equation}
where, $a_{0r} = \Re(a_0)$, $a_{0i} = \Im(a_0)$, $b_{0r} = \Re(b_0)$, $b_{0i} = \Im(b_0)$, $c_{0r} = \Re(c_0)$, $c_{0i} = \Im(c_0)$, $c_{1r} = \Re(c_1)$ and $c_{1i} = \Im(c_1)$.

\section{}\label{app:4}
As $f_o(r,\phi_0)$ is a fractional function, $\frac{\partial f_o(r,\phi_0)}{\partial r}$ is also a fractional function. Hence to find the roots of $\frac{\partial f_o(r,\phi_0)}{\partial r} = 0$ it is sufficient to find the roots of the numerator. By some mathematical manipulation it can be shown that the numerator can be written as \eqref{eq:dfr_root}, and the coefficients are,
% To this end it can be shown that the roots of nominator is equivalent by finding the roots of the following polynomial,
% \begin{equation}
% 	\sum_{k=0}^{10} p_k r^k = 0,
% \end{equation}
% where,

\begin{fleqn}
\begin{equation*}
	p_0 \triangleq 2\eta\Re\{\rho_0e^{j\phi0}\}, \quad p_1 \triangleq 2(\eta\rho_1 + (\eta-1)b_3^2\rho_2),
\end{equation*}
\end{fleqn}

% \begin{fleqn}
% 	\begin{equation*}
% 	p_1 \triangleq 2(\eta\rho_1 + (\eta-1)b_3^2\rho_2),
% 	\end{equation*}
% \end{fleqn}

\begin{fleqn}
	\begin{equation*}
	\begin{aligned}
	p_2 &\triangleq 2(\eta\Re\{(\rho_3 + 2d_1\rho_0)e^{j\phi_0}\} + (\eta-1)(3b_3^2\rho_4 + 4b_3\rho_5\rho_2)),
	\end{aligned}
	\end{equation*}
\end{fleqn}
\begin{fleqn}
	\begin{equation*}
	\begin{aligned}
	p_3 &\triangleq 4(\eta d_1\rho_1 + (\eta-1)((2\rho_5^2 + b_1b_3)\rho_2 + c_2b_3^2 + 6b_3\rho_5\rho_4)),
	\end{aligned}
	\end{equation*}
\end{fleqn}
\begin{fleqn}
	\begin{equation*}
	\begin{aligned}
	    p_4 &\triangleq 2(\eta\Re\{(\rho_6\rho_0 + 2d_1\rho_3)e^{j\phi_0}\} + \\
	    & (\eta-1)(\rho_4(12\rho_5^2 + 6b_1b_3 + b_3^2d_1) + 4\rho_5(b_1\rho_2 + 2b_3c_2))),	\end{aligned}
	\end{equation*}
\end{fleqn}
\begin{fleqn}
	\begin{equation*}
	\begin{aligned}
	p_5 &\triangleq 2(\eta\rho_6\rho_1 + (\eta-1)(\rho_2(b_1^2 - d_2b_3^2) + b_3^2c_2d_1 + \\
	& 4c_2(2\rho_5^2 + b_1b_3) + 4\rho_5\rho_4(3b_1 + b_3d_1))),
	\end{aligned}
	\end{equation*}
\end{fleqn}
\begin{fleqn}
	\begin{equation*}
	\begin{aligned}
	p_6 &\triangleq 2(\eta\Re\{(\rho_6\rho_3 + 2d_1d_2\rho_0)e^{j\phi_0}\} +  (\eta-1)(\rho_4(3b_1^2 - b_3^2d_2 \\
	&+ 2d_1(2\rho_5^2 + b_1b_3)) + 4\rho_5(2b_1c_2 - b_3(d_2\rho_2 - c_2d_1)))),
	\end{aligned}
	\end{equation*}
\end{fleqn}
\begin{fleqn}
	\begin{equation*}
	\begin{aligned}
	p_7 &\triangleq 4(\eta d_1d_2\rho_1 + (\eta-1)(b_1^2c_2 + 2(b_1d_1 - b_3d_2)\rho_5\rho_4 \\
	& - (d_2\rho_2 - c_2d_1)(2\rho_5^2 + b_1b_3))),
	\end{aligned}
	\end{equation*}
\end{fleqn}
\begin{fleqn}
	\begin{equation*}
	\begin{aligned}
	p_8 &\triangleq 2(\eta\Re\{(d_2^2\rho_0 + 2d_1d_2\rho_3)e^{j\phi_0}\} + (\eta-1)(\rho_4(b_1^2d_1 \\
	&- 2d_2(2\rho_5^2 + b_1b_3)) - 4b_1\rho_5(d_2\rho_2 - c_2d_1))),
	\end{aligned}
	\end{equation*}
\end{fleqn}
\begin{fleqn}
	\begin{equation*}
	\begin{aligned}
	p_9 \triangleq& 2(\eta d_2^2\rho_1 - (\eta-1)(b_1^2(d_2\rho_2 - c_2d_1) + 4b_1d_2\rho_5\rho_4)),
	\end{aligned}
	\end{equation*}
\end{fleqn}
\begin{fleqn}
	\begin{equation*}
	\begin{aligned}
	p_{10} \triangleq& 2(\eta d_2^2\Re\{\rho_3e^{j\phi_0}\} - (\eta-1)b_1^2d_2\rho_4),
	\end{aligned}
	\end{equation*}
\end{fleqn}
where, $\rho_0 \triangleq a_3b_0 - b_3a_0$, $\rho_1 \triangleq a_3b_1 - a_1b_3$, $\rho_2 \triangleq c_5 + 2\Re\{c_0e^{j2\phi_0}\}$, $\rho_3 \triangleq b_1a_0 - a_1b_0$, $\rho_4 \triangleq \Re\{c_1e^{j\phi_0}\}$, $\rho_5 \triangleq \Re\{b_0e^{j\phi_0}\}$ and $\rho_6 \triangleq d_1^2 + 2d_2$.

\section{}\label{app:5}
After substituting $\cos(\phi) = {(1-\tan^2(\frac{\phi}{2}))}/{(1+\tan^2(\frac{\phi}{2}))}$, $\sin(\phi) = {2\tan(\frac{\phi}{2})}/{(1+\tan^2(\frac{\phi}{2}))}$ in $\frac{\partial f_o(r_e^{\star},\phi)}{\partial \phi}$ and considering $z \triangleq \tan(\frac{\phi}{2})$, we encounter with a fractional function. In this case it is sufficient to find the roots of nominator. It can be shown that the nominator can be written as, \eqref{eq:dfz_root}, where,
\begin{fleqn}
	\begin{equation*}
	\begin{aligned}
	q_0 &\triangleq 2r_e^{\star}(\eta\xi_0(2\xi_3 - \xi_2) \\
	&+ (1-\eta)(c_{1i} - 2\xi_9)(\xi_4^2 - 4\xi_6(\xi_4 - \xi_6))),
	\end{aligned}
	\end{equation*}
\end{fleqn}
\begin{fleqn}
	\begin{equation*}
	\begin{aligned}
	q_1 &\triangleq 4r_e^{\star}(\eta\xi_0\xi_1 + (1-\eta)(4\xi_7(2\xi_9 - c_{1i})(\xi_4 - 2\xi_6) \\
	&+ (4\xi_8 - c_{1r})(\xi_4^2 - 4\xi_6(\xi_4 - \xi_6)))),
	\end{aligned}
	\end{equation*}
\end{fleqn}
\begin{fleqn}
	\begin{equation*}
	\begin{aligned}
	q_2 &\triangleq 4r_e^{\star}(\eta\xi_0(4\xi_3 - \xi_2) + (1-\eta)(- 8\xi_7(4\xi_8 - c_{1r})(\xi_4 - 2\xi_6) \\
	&+ \xi_4^2(4\xi_9 + c_{1i}) + 4({r_e^{\star}}^2\xi_5(2\xi_9 - c_{1i}) - 6\xi_6\xi_9(\xi_4 - \xi_6)))),
	\end{aligned}
	\end{equation*}
\end{fleqn}
\begin{fleqn}
	\begin{equation*}
	\begin{aligned}
	&q_3 \triangleq 4r_e^{\star}(3\eta\xi_0\xi_1 + (1-\eta)(\xi_4^2(4\xi_8 - 3c_{1r}) + 8\xi_{10} + 4\xi_{11} + \\
	&4(\xi_5{r_e^{\star}}^2(c_{1r} - 8\xi_8) - 2\xi_7^2c_{1r} - 2\xi_6(2\xi_6\xi_8 - \xi_7(14\xi_9 - c_{1i}))) )),
	\end{aligned}
	\end{equation*}
\end{fleqn}
\begin{fleqn}
	\begin{equation*}
	\begin{aligned}
	q_4 &\triangleq 8r_e^{\star}(3\eta\xi_0\xi_3 + (1-\eta)(\xi_9(5\xi_4^2 - 24{r_e^{\star}}^2\xi_5) \\
	&+ 2\xi_4(4\xi_7c_{1r} + \xi_6c_{1i}) - 4\xi_6(16\xi_7\xi_8 + \xi_9\xi_6))),
	\end{aligned}
	\end{equation*}
\end{fleqn}
\begin{fleqn}
	\begin{equation*}
	\begin{aligned}
	&q_5 \triangleq 4r_e^{\star}(3\eta\xi_0\xi_1 + (1-\eta)(- \xi_4^2(4\xi_8 + 3c_{1r}) + 8\xi_{10} - 4\xi_{11} +\\
	&4(\xi_5{r_e^{\star}}^2(c_{1r} + 8\xi_8) - 2\xi_7^2c_{1r} + 2\xi_6(2\xi_6\xi_8 - \xi_7(14\xi_9 + c_{1i}))))),
	\end{aligned}
	\end{equation*}
\end{fleqn}
\begin{fleqn}
	\begin{equation*}
	\begin{aligned}
	q_6 &\triangleq 4r_e^{\star}(\eta\xi_0(4\xi_3 + \xi_2) + (1-\eta)(8\xi_7(4\xi_8 + c_{1r})(\xi_4 + 2\xi_6) \\
	&+ \xi_4^2(4\xi_9 - c_{1i}) + 4({r_e^{\star}}^2\xi_5(2\xi_9 + c_{1i}) + 6\xi_6\xi_9(\xi_4 + \xi_6)))),
	\end{aligned}
	\end{equation*}
\end{fleqn}
\begin{fleqn}
	\begin{equation*}
	\begin{aligned}
	q_7 &\triangleq 4r_e^{\star}(\eta\xi_0\xi_1 + (1-\eta)(4\xi_7(2\xi_9 + c_{1i})(\xi_4 + 2\xi_6) \\
	&- (4\xi_8 + c_{1r})(\xi_4^2 + 4\xi_6(\xi_4 + \xi_6)))),
	\end{aligned}
	\end{equation*}
\end{fleqn}

\begin{fleqn}
	\begin{equation*}
	\begin{aligned}
	q_8 &\triangleq 2r_e^{\star}(\eta\xi_0(2\xi_3 + \xi_2) \\
	&- (1-\eta)(c_{1i} + 2\xi_9)(\xi_4^2 + 4\xi_6(\xi_4 + \xi_6))),
	\end{aligned}
	\end{equation*}
\end{fleqn}
where, $\xi_0 \triangleq {r_e^{\star}}^4 + {r_e^{\star}}^2d_1 + d_2$, $\xi_1 \triangleq {r_e^{\star}}^2(a_3b_{0r} - a_{0r}b_3) + (a_1b_{0r} - a_{0r}b_1)$, $\xi_2 \triangleq {r_e^{\star}}^2(a_3b_{0i} - a_{0i}b_3) + (a_1b_{0i} - a_{0i}b_1)$, $\xi_3 \triangleq r_e^{\star}(a_{0r}b_{0i} - a_{0i}b_{0r})$, $\xi_4 \triangleq {r_e^{\star}}^2b_3 + b_1$, $\xi_5 \triangleq b_{0r}^2 - 2b_{0i}^2$, $\xi_6 \triangleq r_e^{\star}b_{0r}$, $\xi_7 \triangleq r_e^{\star}b_{0i}$, $\xi_8 \triangleq r_e^{\star}c_{0r}$, $\xi_9 \triangleq r_e^{\star}r_{0i}$, $\xi_{10} \triangleq \xi_4(2\xi_6\xi_8 - 5\xi_7\xi_9)$ and $\xi_{11} \triangleq \xi_4(\xi_6c_{1r} - \xi_7c_{1i})$.

\section{}\label{app:6}
% If $r=1$ the integrated mainlobe level of auto-correlation is equal to, $\sum_{m=1}^{M}\norm{\tilde{\bs}_m^H\tilde{\bs}_m}_2^2 = MN^2$. In this case the objective function of $\mathcal{P}_{r,\phi}$ can be written as,
% \begin{equation}
% 	\bar{f}\left(r, \phi \right) = \frac{a_0e^{j\phi}+a_1+a_2e^{-j\phi}+a_3}{b_0e^{j\phi}+b_1+b_2e^{-j\phi}+b_3},
% \end{equation}
% \begin{equation}
% \begin{aligned}
% 	&\tilde{f}\left(r, \phi \right) = \frac{c_0e^{j2\phi}+c_1e^{j\phi}+c_2+c_3e^{-j\phi}+c_4e^{-j2\phi}+c_5}{NM^2}.
% 	\end{aligned}
% \end{equation}
% Therefore
% \begin{equation}
% \begin{aligned}
% f_d(\phi) = \frac{e^{j3\phi}\sum_{k=0}^{6} g_ke^{-jk\phi}}{e^{j\phi}\sum_{k=0}^{2} h_k^{-jk\phi}}
% \end{aligned}
% \end{equation}
By substituting $r=1$ in \eqref{eq:f_bar_rphi2} and \eqref{eq:f_tld_rphi2}, the objective function under $C_4$ constraint can be written as, \eqref{eq:P_d}, where, 
\begin{equation}\label{eq:coeff_f_o(phi)}
\begin{aligned}  
h_0 &\triangleq b_0, \: h_1 \triangleq b_1 + b_3, \: h_2 \triangleq b_2, \: 
g_0 \triangleq c_0b_0\frac{1-\eta}{M_t N^2}, \: g_6 \triangleq g_0^*\\
g_1 &\triangleq (c_0b_1+c_1b_0)\frac{1-\eta}{M_t N^2}, \quad g_5 \triangleq g_1^*\\ 
g_2 &\triangleq (c_0b_2+c_1b_1+c_2b_0)\frac{1-\eta}{M_t N^2} + a_0\eta, \quad g_4 \triangleq g_2^*,\\
g_3 &\triangleq (c_1b_2+c_2b_1+c_3b_0)\frac{1-\eta}{M_t N^2} + a_1\eta.
\end{aligned}
\end{equation}
% \begin{equation}\label{eq:coeff_f_o(phi)}
% \begin{aligned}  
% h_0 &\triangleq b_0, \quad h_1 \triangleq b_1 + b_3, \quad h_2 = b_2\\
% g_0 &\triangleq c_0b_0\frac{1-\eta}{MN^2},\quad g_1 \triangleq (c_0b_1+c_1b_0)\frac{1-\eta}{MN^2},\\ 
% g_2 &\triangleq (c_0b_2+c_1b_1+c_2b_0)\frac{1-\eta}{MN^2} + a_0\eta ,\\
% g_3 &\triangleq (c_1b_2+c_2b_1+c_3b_0)\frac{1-\eta}{MN^2} + a_1\eta ,\\
% g_4 &\triangleq (c_4b_0+c_3b_1+c_2b_2)\frac{1-\eta}{MN^2} + a_0\eta ,\\ 
% g_5 &\triangleq (c_3b_2+c_4b_1)\frac{1-\eta}{MN^2}, \quad g_6 \triangleq c_4b_2\frac{1-\eta}{MN^2}.
% \end{aligned}
% \end{equation}
% Considering $a_2 = a_0^*$, $b_2 = b_0^*$, $c_4 = c_0^*$ and $c_3 = c_1^*$ it can be concluded that, $g_6 = g_0^*$, $g_5 = g_1^*$ and $g_4 = g_2^*$.

% \section*{Acknowledgment}

\ifCLASSOPTIONcaptionsoff
  \newpage
\fi



%\bibliographystyle{IEEEtran}



% \end{thebibliography}

% \bibliographystyle{ieeetr}
% \bibliography{ref1}
\bibliographystyle{IEEEtran}
\bibliography{IEEEabrv,J1_TSP_2020_arxiv}


% \begin{IEEEbiography}{Ehsan Raei}
% \end{IEEEbiography}

% \begin{IEEEbiographynophoto}{Mohammad Alaee-Kerahroodi}

% % insert where needed to balance the two columns on the last page with
% %\newpage

% \begin{IEEEbiographynophoto}{M.R. Bhavani Shankar}


%\vfill


\end{document}